\documentclass{aa}  

\usepackage{graphicx, array}
\usepackage[varg]{txfonts}
\usepackage{natbib,twoopt}
\usepackage{xcolor}
\usepackage[breaklinks=true]{hyperref} 

\newcommand{\teff}  {\mbox{T$_\mathrm{eff}$}}
\newcommand{\feh}   {\mbox{[Fe/H]}}
\newcommand{\logg}   {\mbox{$\log$ g}}
\newcommand{\tgm}   {\mbox{(T$_\mathrm{eff}$, $\log$g, [Fe/H])}}
\newcommand{\pastel}{{\small PASTEL}}
\newcommand{\apo}{{\small APOGEE}}
\newcommand{\rave}{{\small RAVE}}
\newcommand{\lamost}{{\small LAMOST}}
\newcommand{\galah}{{\small GALAH}}
\newcommand{\ges}{{\small Gaia-ESO Survey}}
\newcommand{\segue}{{\small SEGUE}}

\begin{document}

   \title{Assessment of \feh\ determinations for FGK stars in spectroscopic surveys
   \thanks{Tables are only available in electronic form at the CDS via anonymous ftp to cdsarc.u-strasbg.fr (130.79.128.5) 
or via http://cdsarc.u-strasbg.fr/viz-bin/qcat?J/A+A/?/?}}

    \author{
C. Soubiran\inst{\ref{LAB}}, N. Brouillet\inst{\ref{LAB}}, L. Casamiquela\inst{\ref{LAB}}
}

\institute{
Laboratoire d'Astrophysique de Bordeaux, Univ. Bordeaux, CNRS, B18N, all\'ee Geoffroy Saint-Hilaire, 33615 Pessac, France\label{LAB}
\email{caroline.soubiran@u-bordeaux.fr}
}
  
 \date{Received \today, accepted }

  \abstract
   {The iron abundance \feh\ in the atmosphere of FGK-type stars is crucial in stellar and galactic physics. The number of stars with a measured value of \feh\ is considerably increasing  thanks to spectroscopic surveys. However different methodologies, inputs and assumptions used in spectral analyses lead to different precisions in \feh\ and possibly to systematic differences that need to be evaluated. It is essential to understand the characteristics of each survey to fully exploit their potential, in particular if the surveys are combined to probe a larger galactic volume and to improve statistics.}
   {The purpose of this study is to compare \feh\ determinations from the largest spectroscopic surveys to other catalogues taken as reference.  Offsets and dispersions of the residuals are examined as well as their trends with other parameters. The investigated surveys are the latest public releases of \apo, \galah, \rave, \lamost, \segue\ and the \ges.  }
  {We use reference samples providing independent determinations of \feh\ which are compared to those from the surveys for common stars. The distribution of the residuals is assessed through simple statistics that measures the offset between two catalogues and the dispersion representative of the precision of both catalogues.  When relevant, linear fits are performed. A large sample of FGK-type stars with \feh\ based on high-resolution, high signal to noise spectroscopy was built from the \pastel\ catalogue to provide a reference sample. We also use  FGK members in open and globular clusters to assess the internal consistency of \feh\ of each survey. The agreement of median \feh\ values for clusters observed by different surveys is discussed.}
 {All the surveys overestimate the low metallicities, and some of them also underestimate the high metallicities. They perform well in the most populated intermediate metallicity range, whatever the resolution. In most cases the typical precision that we deduce from the comparisons is in good agreement with the  uncertainties quoted in the catalogues. Some exceptions to this general behaviour are discussed.}
   {}


   \maketitle
%

\section{Introduction}
We are in the middle of a new era where stellar atmospheric parameters (APs) and abundances are massively produced by spectroscopic surveys. 
The iron abundance \feh\ is an essential stellar property which has to be known for the determination of other parameters through stellar models, like the mass and the age. Iron abundances are also needed in galactic archeology in order to understand how the different stellar populations have formed and evolved. 

The pioneer of spectroscopic surveys, the Radial Velocity Experiment (\rave), produced its first data release (DR) fifteen years ago \citep{ste06} and its final one, DR6,  last year  \citep{rave, ste20}. In the meantime, other surveys have been operated and developed new methodologies, learning progressively from their own and each other experience how to reduce biases and uncertainties in the automated determination of APs and abundances from massive datasets of stellar spectra with various resolutions and spectral coverages. Besides \rave, other surveys have published successive DRs available for a public use. At the time of writing, everyone has access to \apo\ \citep[DR16,][]{jon20}, \galah\ \citep[DR3,][]{galah}, \rave\ \citep[DR6,][]{rave, ste20}, \lamost\ \citep[DR5,][]{luo19}, \segue\ \citep{yan09},  \ges\ \citep[DR3,][]{gil12,ran13}. Additional versions of APs,  based on different methods, are also provided for \rave\ DR6 \citep{gui20} and for \lamost\ DR5 \citep{xia19}. The next generation of optical and near-infrared spectrographs, wide-field and massively multiplexed, is in preparation and will soon provide even larger catalogues of APs and abundances, such as {\small WEAVE} \citep{weave}, the Multi-Object Optical and Near-infrared Spectrograph on  ESO's Very Large Telescope \citep[][{\small MOONS}]{moons}, the 4-metre Multi-Object Spectroscopic Telescope  \citep[{\small 4MOST},][]{dej19}, the Prime Focus Spectrograph \citep[{\small PFS},][]{pfs}, the Maunakea Spectroscopic Explorer \citep[{\small MSE},][]{mse}. In terms of number of stars, the most revolutionary survey will certainly be that of the Gaia  space mission \citep{gaia} which will deliver in its DR3 in 2022\footnote{The Gaia data release scenario and the DR3 content can be found at \url{https://www.cosmos.esa.int/web/gaia}} estimates of the physical properties, including metallicities, for millions of stars, obtained with various methods through an astrophysical parameters inference system \citep{bai13}. 

Each survey has its own strategy for the calibration and validation of APs and abundances. The term calibration usually invokes standard stars with {\it true} APs. If effective temperatures and surface gravities, \teff\ and \logg, can be determined independently of atmospheric models thanks to fundamental relations \citep{hei15}, this is not the case for the metallicity \feh\ which has therefore no absolute zero point. Abundances are expressed relatively to the Sun, the chemical composition of which is still subject to debate \citep{asp09,asp21}. It is thus impossible to measure the typical accuracy of a given catalogue of metallicities, but instead the zero point agreement between two catalogues can be assessed. It is also possible to evaluate the relative precisions of different catalogues by comparing them to another independent source. In classical spectroscopy, the assessment of uncertainties usually evaluates the random errors due to the characteristics of the input spectra and to the line selection, and the systematic errors due to the adopted assumptions (e.g. local thermodynamic equilibrium) or to the method itself (e.g. equivalent width versus synthetic spectrum fitting). Comparisons to independent reference datasets and inter-comparisons of surveys are mandatory to track systematic differences, although these comparisons are limited by the number of stars in common and their range of properties. The strategies adopted by the ongoing surveys for calibrations and validations are reviewed in \cite{jof19}. 

The validation of the Gaia's APs is very challenging due to the size of the dataset, the large magnitude range and the observing mode which collects all the objects down to a limiting magnitude, including stars with properties that prevent a reliable determination of APs by automated methods (e.g. rotation, emission, binarity). All the information from ground-based surveys and catalogues of APs is being used to assess the accuracy and precision of Gaia's APs. In this context, an important task is to deepen our knowledge and understanding of the AP uncertainties of ground-based surveys. Any systematic difference between large surveys  has potentially important implications for the study of stellar populations in the Milky Way and the galactic chemical evolution.  It is also important to make these comparisons in the perspective of combining different surveys to probe a larger galactic volume and to improve the statistics, as attempted by \cite{nan20} for instance.

In this study, we focus on \feh\ of FGK-type stars in the effective temperature range 4000-6500 K. The upper limit avoids hot stars the metallicity of which can be affected by rotation and chemical peculiarities. The lower limit avoids cool giants and dwarfs the metallicity of which is reputed difficult to measure because of many blended lines in the spectra.  FGK stars span the full age range of the Galaxy, with their chemical composition reflecting the chemical composition of the interstellar matter from which they formed, from very low to very high metallicity. FGK members in open clusters (OCs) and globular clusters (GCs) are supposed to share the same iron abundance. Indeed high precision differential studies have shown that the chemical homogeneity is at the level of 0.02 dex in OCs  \citep{liu16,cas20} and of 0.03 in most GCs \citep[e.g.][]{yon13}. This property offers the possibility to assess the precision of a given catalogue by measuring the typical dispersion of \feh\ among members of a given cluster. The consistency of the metallicity can be tested all over the temperature range of FGK stars and among giants and dwarfs.  Many OCs and GCs have been observed for decades in spectroscopy so that their chemical composition is reasonably known. Clusters are ideal for multi-object spectroscopy and a significant observing time is dedicated to them by  spectroscopic surveys. Gaia DR2 \citep{gdr2} and EDR3 \citep{edr3} have considerably enlarged the number of stars and clusters for which membership probabilities are available \citep{bab18,  can18, can20, vas21}. This offers an opportunity to revise the metallicity of clusters. 

In order to  evaluate the precision and zero-point agreement of  \feh\  determinations in surveys we constructed three reference samples, based on: (1) the \pastel\ catalogue \citep{sou16}, (2) OC members , (3) GC members. \pastel, updated in 2020,  compiles \feh\ determinations based on high resolution, high S/N ratio spectroscopy, with a significant fraction of metal-poor stars.  Together with GCs, it provides a mean to test the poorly constrained low metallicity part of the AP space.  Our procedure measures the dispersion of the residuals resulting from the \feh\ comparison between the different surveys and the reference catalogues, and looks for trends with magnitude and APs. 

In this paper we first present our compilation of cluster members and the \pastel\ catalogue. We then briefly present the investigated surveys, and the selections applied on \teff, \feh\ errors,  flags or other criteria, depending on the survey, to retain the atmospheric parameters of the best quality.  We cross-match the resulting samples to the reference catalogues to evaluate the \feh\ residuals and their dependency to other parameters. We discuss the results in terms of typical precision of the surveys in the metal-rich and metal-poor range, and in the giant and dwarf subspaces. We also compare the surveys to \apo.
In the whole paper we use median values denoted MED to measure offsets. We evaluate the dispersion of the various distributions measured through their median absolute deviation denoted MAD. When relevant we fit a line to highlight a trend.

\section{Reference catalogues}

\subsection{Cluster members}
For the purpose of testing the metallicity precision of the surveys, we select clusters with reasonably known metallicity and their members of highest probability. 

For OCs we adopt the mean metallicities per cluster compiled by \cite{net16} for 172 OCs. In that paper, the metallicities are either derived from spectroscopy at high resolution and high S/N (88 clusters) or lower resolution (12 clusters) or from photometry (72 clusters).  The spectroscopic metallicities are updated from \cite{hei14}. We complete this compilation by the high-precision and homogeneous mean \feh\ determined by \cite{cas21} for 47 OCs, based on clump giants only. This adds 18 clusters. We adopt the metallicity from \cite{cas21}  over that of \cite{net16}  for three clusters (NGC 7245, NGC 6940, King 1) because of a photometric determination and for  two other clusters (NGC 2266, NGC 2639) because of a spectroscopic metallicity relying on one star only.  The resulting compilation of mean \feh\ per cluster is not of homogeneous quality but our purpose is to use clusters to test the precision of surveys, not their accuracy, so that an absolute reference value is not mandatory. The most important is to have a large number of reliable members per cluster giving a significant intersection with the surveys.  In the following we will analyse the distribution of the \feh\ residuals in terms of dispersion per cluster since an offset seen for a given OC could be due to an erroneous mean metallicity from the literature. Despite different precisions of the reference metallicities of OCs, trends can still be observed.

We adopt the list of members with a probability higher than 70\% to belong to their parent cluster that \cite{can20b} used to determine the physical properties of $\sim$2000 OCs, based on Gaia DR2 data. All the stars have their Gaia magnitude  Gmag $<$ 18. Three clusters from \cite{net16} were not found in \cite{can20b} (Saurer 1, Loden 807, Collinder 173). This gives  77\,899 stars in 187 OCs, spanning metallicities from -0.50 to +0.43. The number of members per cluster ranges from 11 to nearly 3000.

For GCs, we adopt the catalogue of \cite{har10} which provides the metallicity of 152 GCs in the Milky Way and
the membership probabilities recently computed by \cite{vas21} using Gaia EDR3 for 170 galactic GCs.  We select the most reliable members having  Gmag $<$ 18 and a membership probability higher than 70\%, and we keep only the GCs which have at least 10 members remaining after these cuts. It is well established that most globular clusters have a scatter in metallicity lower than 0.05 dex \citep{car09} with a few exceptions that show a larger dispersion possibly related to multiple populations \citep{gra12}. These objects, NGC 5139 (Omega Cen), NGC 6715 (M 54), NGC 6656 (M 22), Terzan 5, NGC 1851 and NGC 2419, have been removed from our reference sample. 
This gives 146\,147 stars in 134 GCs spanning metallicities from -2.37 to 0 dex.

\subsection{The \pastel\ catalogue}
\label{s:pastel}

The compilation of  atmospheric parameters started in the eighties with the so-called \feh\ catalogue \citep{cay80, cay81,cay85, cay92, cay97, cay01}, and continued in 2010 with the \pastel\ catalogue which was regularly updated  until 2020 \citep{sou10, sou16}.  Only \feh\ determinations based on high-resolution (R $\geq$ 25\,000), high S/N (S/N$\geq$50) spectra are recorded in \pastel\ with a few exceptions explained in \cite{sou16}.  \pastel\ also provides effective temperatures and surface gravities determined from various methods.  \pastel\  does not include AP determinations from spectroscopic surveys that have a resolution and S/N fitting the criteria (e.g. \galah\ and the UVES part of the \ges).

As of January 2020, \pastel\ has 81\,362 records, including   42\,932 determinations of the 3 atmospheric parameters \tgm\ for 18\,119 different stars.  The \feh\ determinations  range from -4.80 to +2.40 dex. The solar metallicity is by far the most frequent value. More than 80\% of the \feh\ determinations are between -0.50 dex and +0.50 dex. In the literature monitoring, a particular effort was put on trying to be as complete as possible for metal-poor stars. Such stars are rare in the solar neighbourhood  so that observers have to consider targets at larger distances, which are fainter and challenging for spectroscopic observations at high-resolution, high S/N. Nevertheless \pastel\ includes a significant number of AP determinations for metal-poor stars: 5\,544 values with \feh$\le$-1.0 dex ($\sim$2\,000 different stars), 1\,973 values with \feh$\le$-2.0 dex ($\sim$850 different stars), 418 values with \feh$\le$-3.0 dex ($\sim$240 different stars). The top 5 most studied metal-poor stars are HD140283 (weighted mean \feh=-2.47$\pm$0.03 dex, for N=58 determinations), HD103095 (\feh=-1.34$\pm$0.02 dex, N=44), HD19445 (\feh=-1.98$\pm$0.03 dex, N=43), \* tau Cet (\feh=-0.51$\pm$0.01 dex, N=43), HD122563 (\feh=-2.67$\pm$0.02 dex, N=42). \pastel\ includes 18 extremely metal-poor stars with \feh$\le$-4.0 dex. Most have been studied only once at high-resolution. The most studied one is CD-38 245 (\feh=-4.10$\pm$0.08 dex, N=12).



For a practical use of the \pastel\ catalogue in the comparison to other catalogues, it is needed to have a single value of AP per star. Among the different strategies we adopt the weighted average of the parameters, although we are aware that systematics make this procedure theoretically improper. But \pastel\ compiles more than 1200 papers with very different numbers of targets, making impossible any kind of homogenization.

The mean APs and uncertainties were computed for each star with a weighting scheme based on the uncertainty of the individual measurements, following the method described in \cite{sou13}. Errors on \tgm\  listed in \pastel\ have median values of 50 K, 0.1 dex and 0.06 dex respectively, for FGK stars with 4000 $\le$ \teff $\le$ 6500 K, but not all the individual AP determinations are given with an error. Therefore these median values are adopted as default errors for each AP determination, doubled when \feh$<$-1.0 dex, and doubled again if the year of the publication is before 1990.  The error adopted for the weighting scheme is the maximum between this default value and that provided in \pastel, when available. Flagged values of \feh, corresponding to global metallicities [M/H] or to non-LTE abundances or based on ionised iron lines, are not considered in the average.

\begin{figure}[ht!]
\centering
\includegraphics[width=0.45\columnwidth]{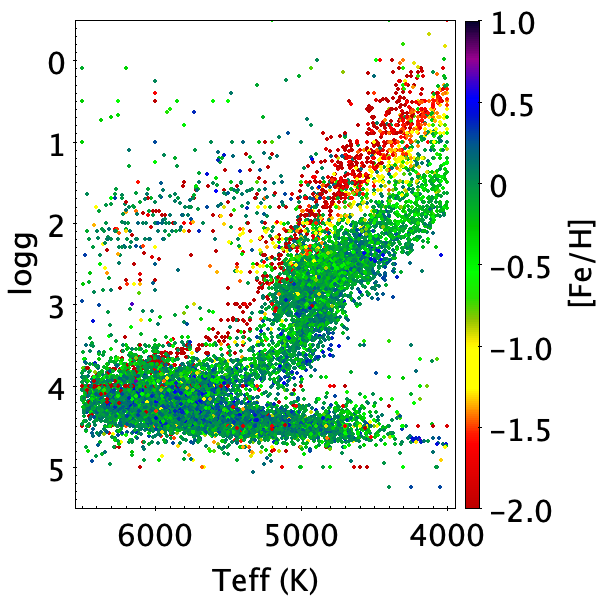}      
\includegraphics[width=0.45\columnwidth]{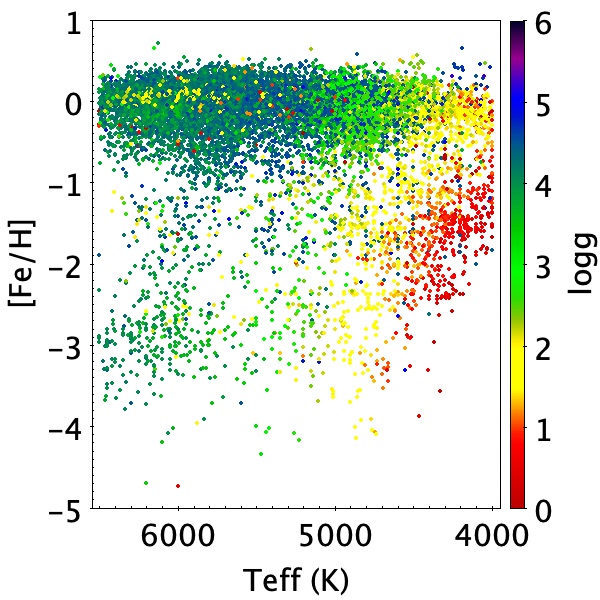}      
\caption{AP distribution for the mean \pastel\ catalogue limited to the FGK regime. Left: Kiel diagram coloured according to \feh\, right: \feh\ vs \teff\  coloured according to \logg.}
\label{f:kiel}
\end{figure}

The resulting mean \pastel\ catalogue, available in VizieR, provides the three parameters \tgm\ for 14\,181 FGK stars of which 13\,506 are in Gaia DR2. For the 7\,400 stars that have at least two \feh\  determinations, the median  uncertainty of the mean is 0.04 dex (0.06 dex for metal-poor stars with \feh$<$-0.50 dex).  The Kiel diagram of the mean \pastel\ APs with 4000 K $\leq$ \teff $\leq$ 6500 K is shown in Fig.~\ref{f:kiel} together with the metallicity distribution as a function of \teff.

\subsection{Clusters in \pastel}

\begin{figure*}[ht!]
\centering
\includegraphics[width=0.48\columnwidth]{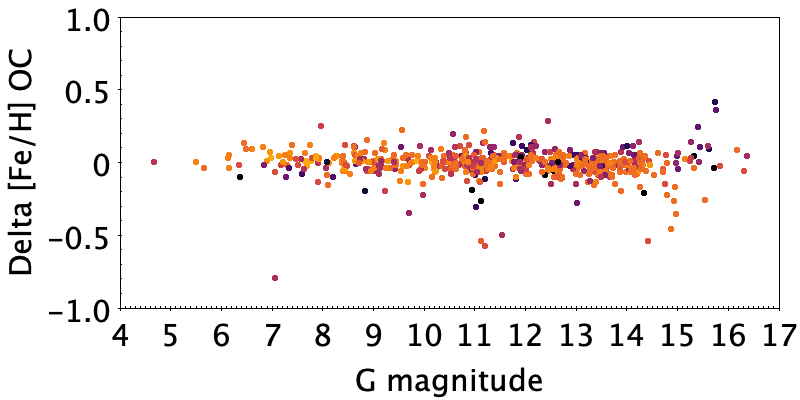}      
\includegraphics[width=0.48\columnwidth]{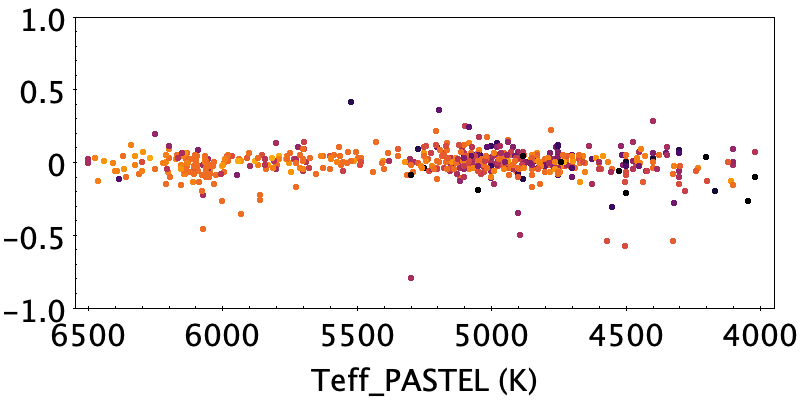}      
\includegraphics[width=0.48\columnwidth]{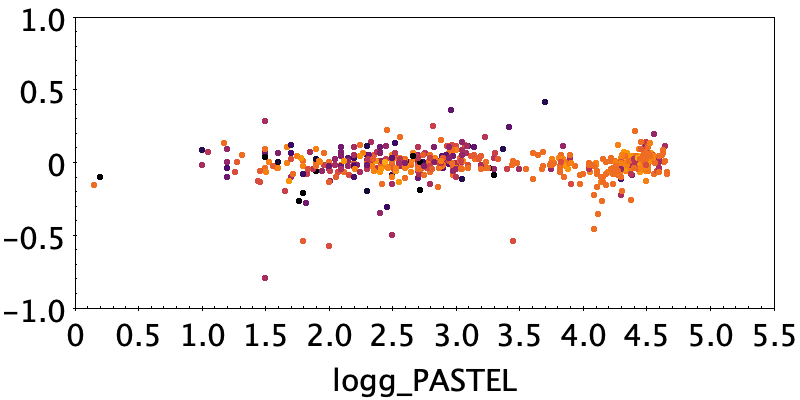}      
\includegraphics[width=0.48\columnwidth]{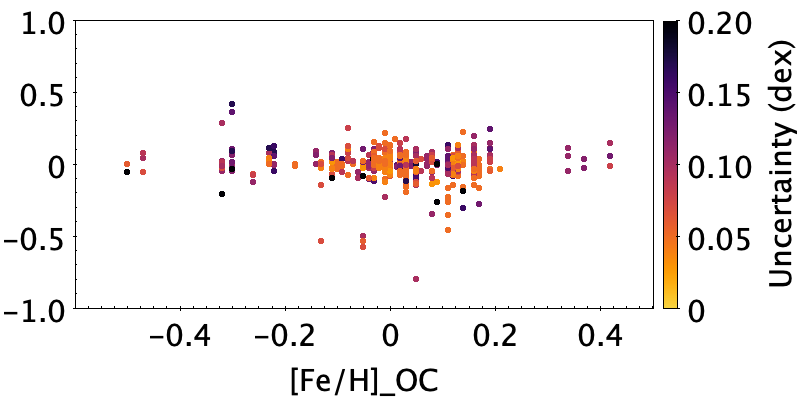}      

\includegraphics[width=0.48\columnwidth]{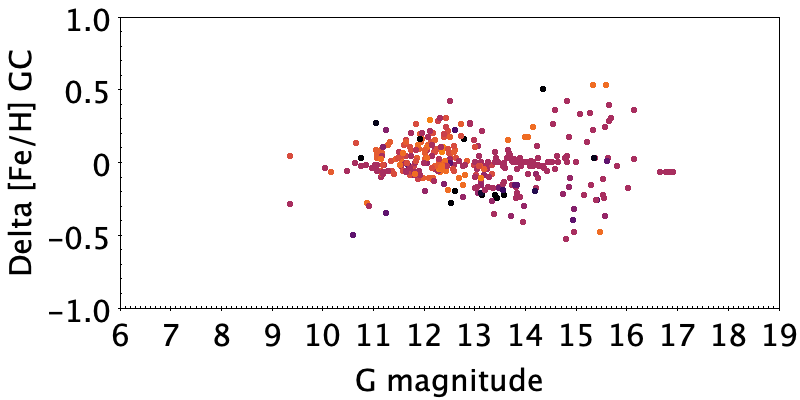}      
\includegraphics[width=0.48\columnwidth]{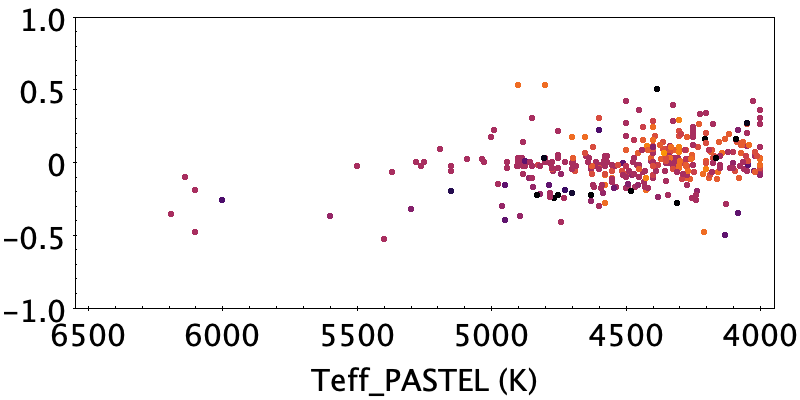}      
\includegraphics[width=0.48\columnwidth]{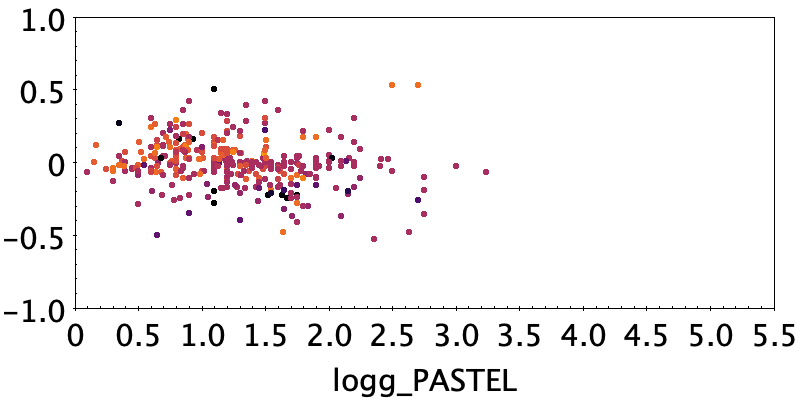}      
\includegraphics[width=0.48\columnwidth]{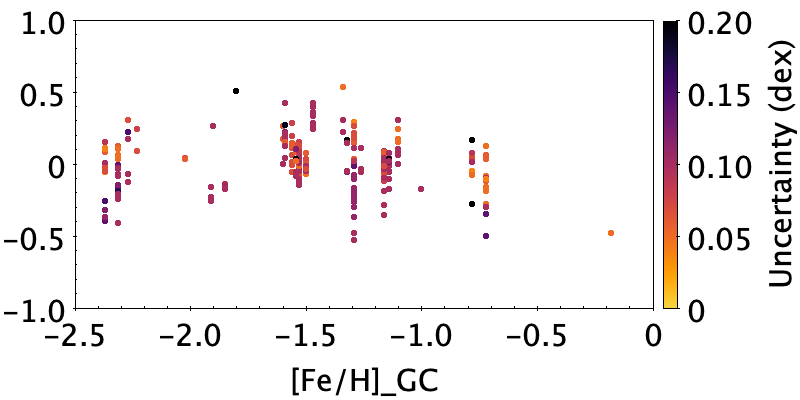}      
\caption{Difference between mean \pastel\  determinations of \feh\ for individual members and the mean value per cluster from \cite{net16}  for OCs (upper panel) and from \cite{har10} for GCs (bottom panel) versus G magnitude, \teff, \logg\ from \pastel\ and the  mean \feh\ per cluster. The colour is related to the \feh\ uncertainty from \pastel.}
\label{f:pastel_clusters}
\end{figure*}

The cross-match between the mean \pastel\ catalogue and the list of reference OC members gives 590 common FGK stars in 87 clusters.  
Figure~\ref{f:pastel_clusters} (upper panel) shows the metallicity difference between individual stars and the literature mean value  as a function of the G magnitude, of \teff\ and \logg\ from \pastel, and of the literature \feh\ of the cluster. The distribution of residuals is flat, there is no trend. The median difference is null, and the dispersion is MAD=0.04, in perfect agreement with the typical uncertainty of the mean \pastel\ \feh. The \feh\ value of individual stars agree well in general with the mean OC metallicity from the literature which is not surprising since the previous version of \pastel\ has been used by \cite{hei14} and \cite{net16} to build the compilation of OC metallicities. The two most represented OCs in \pastel\ are M67 and the Hyades with respectively 91 and 55 FGK stars giving a metallicity of 0.0$\pm$0.04 dex and +0.14$\pm$0.03 dex (MED$\pm$MAD).  

The cross-match between the mean \pastel\ catalogue and the list of reference GC members gives 350 common FGK stars in 29 clusters. The metallicity residuals are shown in Fig.~\ref{f:pastel_clusters} (bottom panel). The median difference is -0.01 dex, and the dispersion is MAD=0.08 dex, slightly larger than for the metal-poor field stars mentioned in the previous section. Four GCs show a remarkably low dispersion in metallicity with a MAD lower or equal to 0.01 dex: NGC 2808 (MED \feh=-1.12 dex for 23 stars), NGC 4833 (MED \feh=-2.02 dex for 12 stars), NGC 7078 (MED \feh=-2.37 dex for 38 stars), NGC 7078 (MED \feh=-2.34 dex for 9 stars). The other GCs with at least five members in \pastel\ have also low dispersions with their MAD ranging from 0.01 to 0.08 dex. The only exception is NGC 5904 (M5) which exhibits a large dispersion of 0.19 dex (MED \feh=-1.31 dex for 33 stars), clearly visible in Fig.~\ref{f:pastel_clusters}.  The chemical composition of this cluster has been extensively studied in the past. The dispersion that we observe in \pastel\ reflects the fact that the authors of different analyses do not agree on the metallicity of this cluster. \cite{sne92} and \cite{car97} report metallicities of individual members giving on average \feh=-1.17$\pm$0.01 dex and -1.11$\pm$0.03 dex for the cluster, respectively, while \cite{lai11}  determine metallicities ranging from -1.82 to -1.33 dex. 

In the next sections we cross-match the mean \pastel\ catalogue with the most recent versions of spectroscopic surveys,  in order to compare \feh\ determinations. Although the results of each comparison encompass the systematic errors and uncertainties of both \pastel\ and the compared survey, they provide relevant information on the strengths and weaknesses of the datasets. In addition we apply to each survey the same tests on stellar clusters that reveal typical dispersions in different ranges of magnitude, \teff, \logg\ and \feh, independently of any other catalogue.

\section{Surveys vs reference catalogues}
\subsection{APOGEE}
\apo\ {\small DR16} \citep{jon20} includes about 430\,000 stars with APs. \apo\ spectra have a resolution of  $\sim$22\,500  and cover the near-infrared range from 15\,140 \AA\ to  16\,940 \AA. The APOGEE Stellar Parameters and Abundances Pipeline \citep[ASCAP][]{gar16} compares APOGEE observations to a large library of synthetic spectra (from MARCS models for FGK stars considered here) and determines the best matching synthetic spectrum using the code FERRE \citep{all06}, which additionally allows for interpolation within the library. The spectroscopic \teff\ and \logg\ are then calibrated. \feh\ is measured in a second step by tuning the fit around iron lines. The uncertainty on \feh\ is set through a function of \teff, global metallicity and S/N of the spectrum, the coefficients of which are deduced from repeat observations.  We use here the cleaned and calibrated \feh\, as recommended, which verify FE\_ H\_FLAG = 0. Together with the FGK selection 4000 $\le$ \teff $\le$ 6500 K, this gives 236,966 stars with a median uncertainty of 0.01 dex.

\begin{figure*}[ht!]
\centering
\includegraphics[width=0.48\columnwidth]{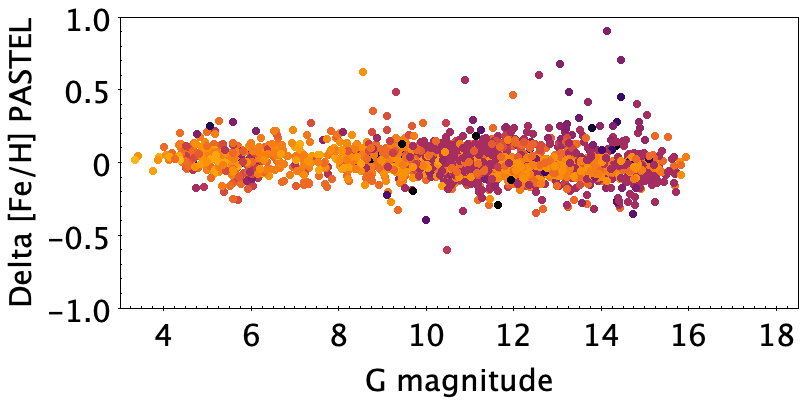}      
\includegraphics[width=0.48\columnwidth]{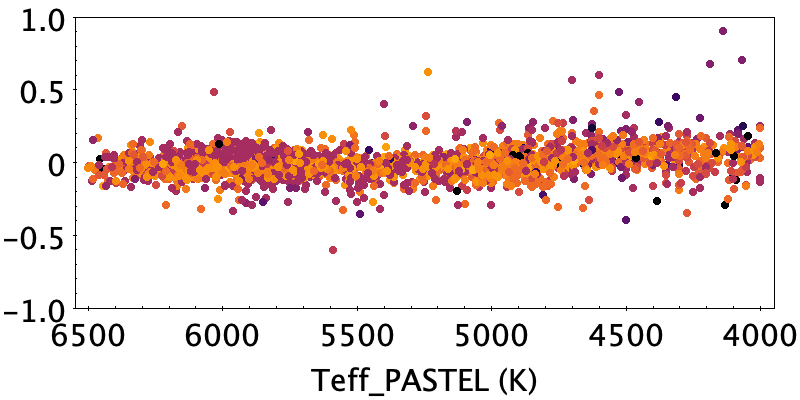}      
\includegraphics[width=0.48\columnwidth]{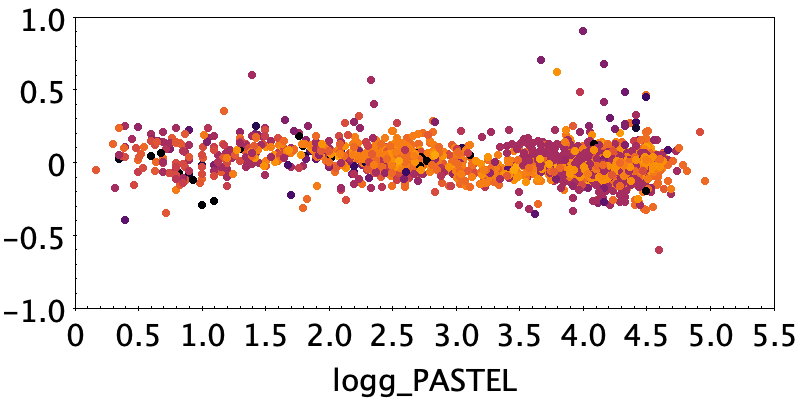}      
\includegraphics[width=0.48\columnwidth]{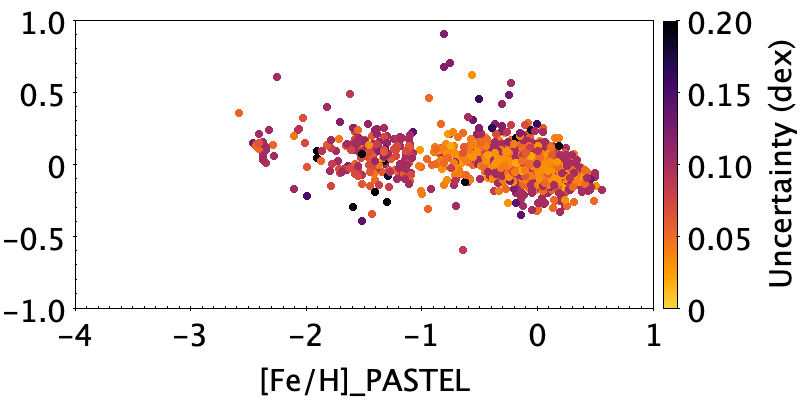}      
\includegraphics[width=0.48\columnwidth]{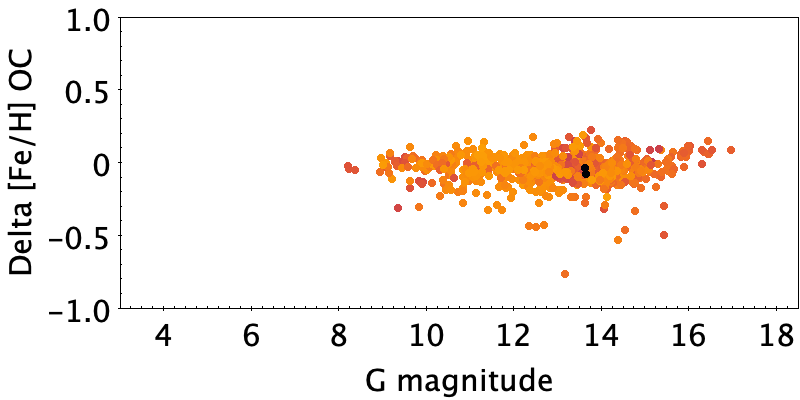}      
\includegraphics[width=0.48\columnwidth]{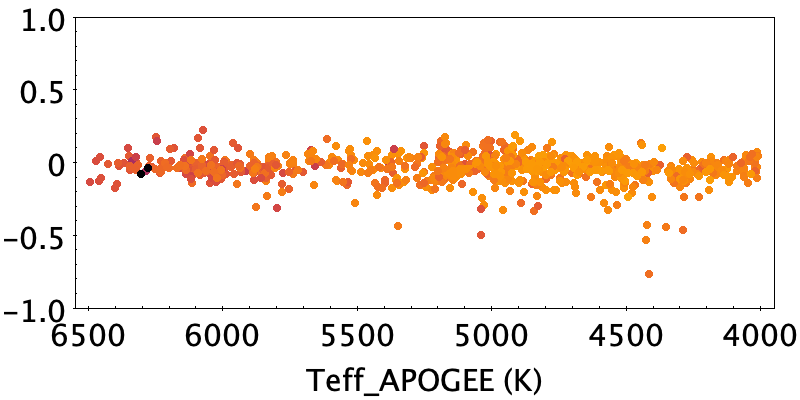}      
\includegraphics[width=0.48\columnwidth]{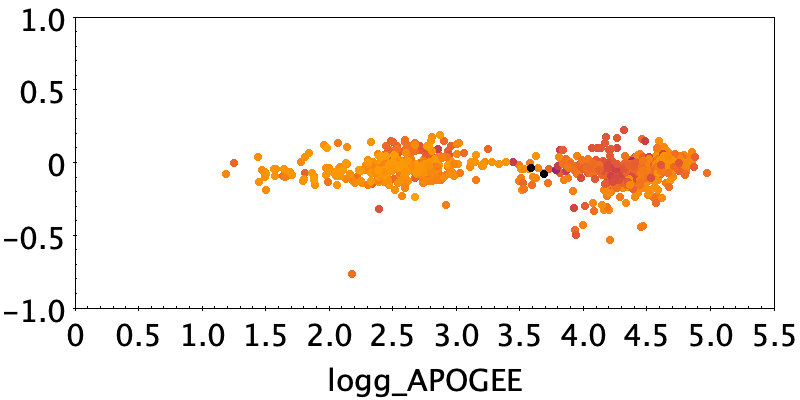}      
\includegraphics[width=0.48\columnwidth]{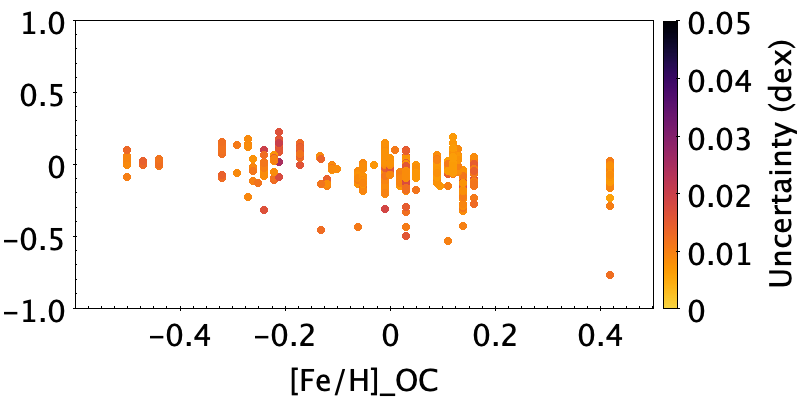}      
\includegraphics[width=0.48\columnwidth]{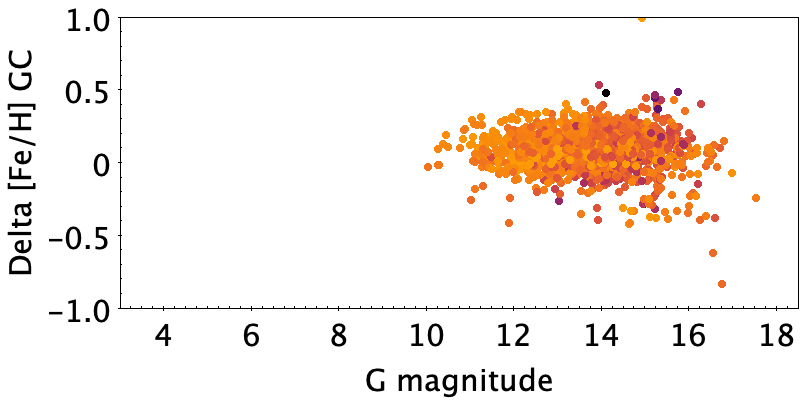}      
\includegraphics[width=0.48\columnwidth]{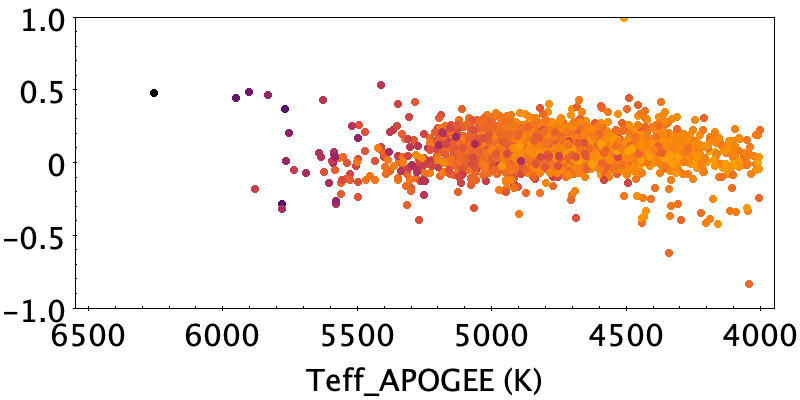}      
\includegraphics[width=0.48\columnwidth]{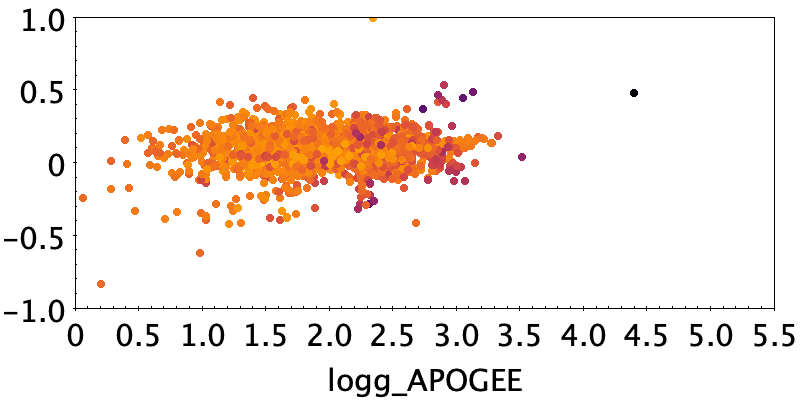}      
\includegraphics[width=0.48\columnwidth]{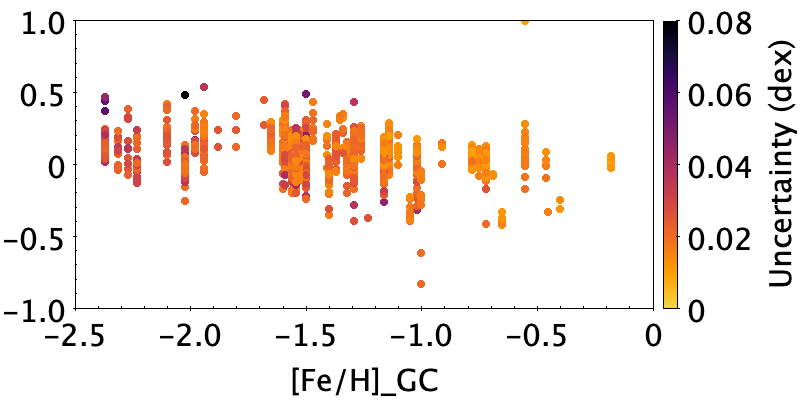}      
\caption{Difference (Delta) between \apo\  determinations of \feh\ and those from the reference catalogues versus magnitude, \teff, \logg\ and \feh. Upper panel is for literature mean value from \pastel, middle panel for OC members with cluster mean metallicities from \cite{net16}, bottom panel for GC members with cluster mean metallicities from  \cite{har10}. The colour code reflects the metallicity uncertainty, quadratically combined for \apo\ vs \pastel, from \apo\ only for the clusters (note the different scales). }
\label{f:apo_ref}
\end{figure*}

There are 2\,155 stars in common between \pastel\ and this \apo\ sample, 844 FGK cluster members in 43 OCs and 1958 FGK cluster members in 48 GCs. The residuals are shown in Fig.~\ref{f:apo_ref}  as a function of magnitude and of \tgm.

For \pastel\ the residuals exhibit a rather low dispersion (MAD=0.05), very similar to that computed  by \cite{jon18} when comparing previous \apo\ data releases to independent studies. The few outliers correspond to stars with larger uncertainties. There is a trend in \feh\ showing that \apo\ systematically overestimates metallicities of metal-poor stars  (\feh$<$-0.50 dex) and underestimates that of the most metal-rich stars (\feh$>$0 dex) compared to \pastel. The offset is +0.06 dex in the metal-poor regime while there is a decreasing trend with \feh\ in the metal-rich regime (a linear fit gives a slope of -0.18 dex$^{-1}$). These metallicity trends seem to be also there in the residuals of clusters. \cite{nid20} reported an offset of +0.08 dex for GC metallicities from \apo\ DR16 compared to high quality determinations from \cite{car09}, that they considered consistent with uncertainties. Here, with a larger sample of GCs and with \pastel, we confirm this positive offset of the metallicity scale in the metal-poor regime. The internal scatter (MAD) for clusters ranges from 0.007 to 0.05 dex for individual OCs and from 0.02 to 0.1 dex for GCs.  

\subsection{GALAH}
The current public version is GALAH+ DR3  \citep{galah}  which contains 588\,571 stars. The data for GALAH consist of spectra at a resolution of R$\sim$28\,000 in four wavelength ranges covering 4713--4903, 5648--5873, 6478--6737 and 7585--7887 \AA.  The determination of atmospheric parameters and abundances is performed with the code Spectroscopy Made Easy \citep[SME,][]{sme} through spectrum synthesis with MARCS models. Non-LTE corrections for Fe lines from \citet{ama16} are applied. The gravity \logg\ is constrained from Gaia astrometry and 2MASS photometry. The precision of \feh\ is estimated with both
the internal SME covariance errors and the standard deviation of repeat observations of the same
star resulting in an exponential function of S/N. The validation of iron abundances is made by comparison to the Gaia Benchmark Stars \citep{hei15, jof15} and to clusters. We considered only the FGK stars in the version 2 of the catalogue (GALAH\_DR3\_main\_allstar\_v2.fits) for which flag\_fe\_h = 0 and flag\_sp = 0, as recommended. The median \feh\ uncertainty for this selection of 407,276 stars  is 0.08 dex.

\begin{figure*}[ht!]
\centering
\includegraphics[width=0.48\columnwidth]{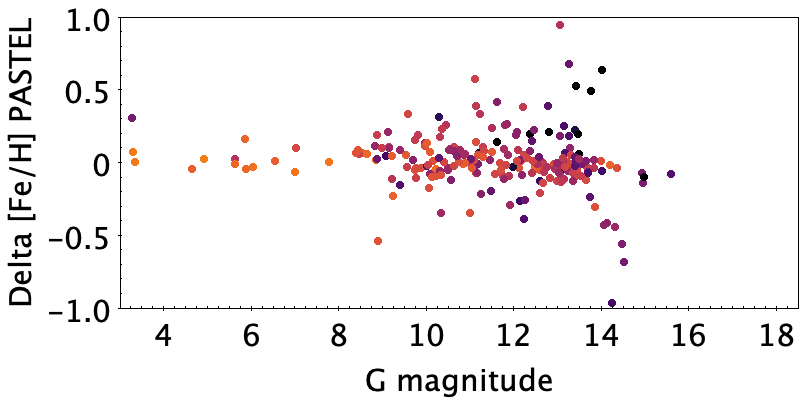}      
\includegraphics[width=0.48\columnwidth]{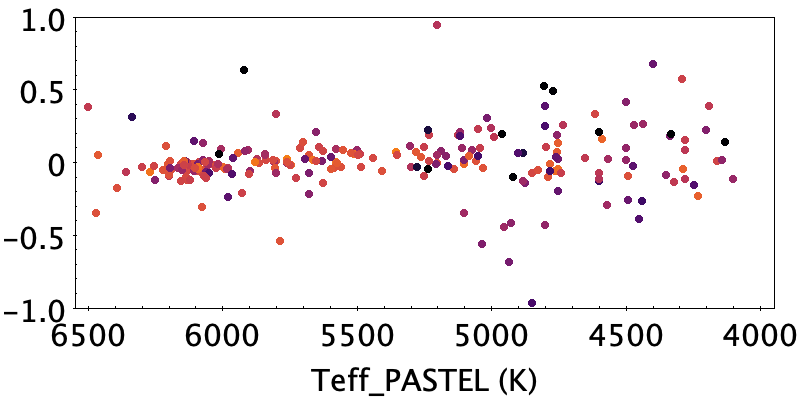}      
\includegraphics[width=0.48\columnwidth]{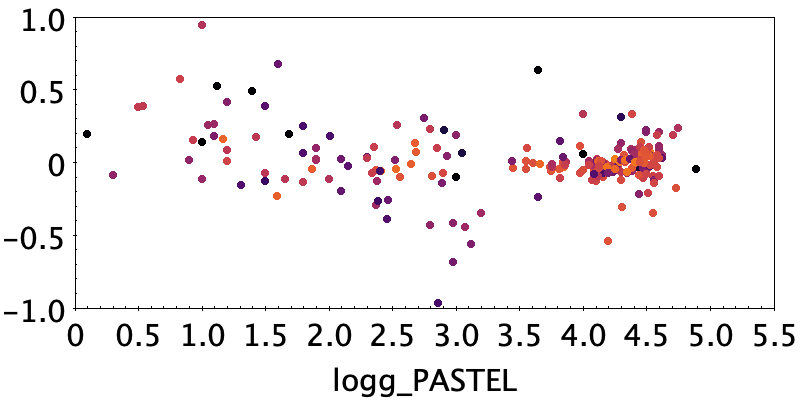}      
\includegraphics[width=0.48\columnwidth]{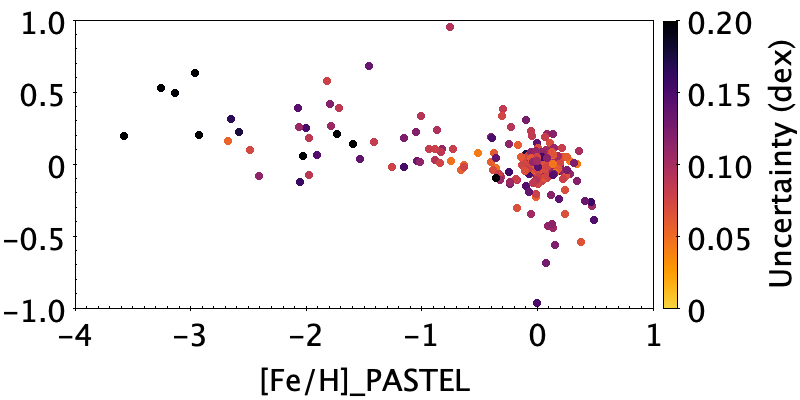}      

\includegraphics[width=0.48\columnwidth]{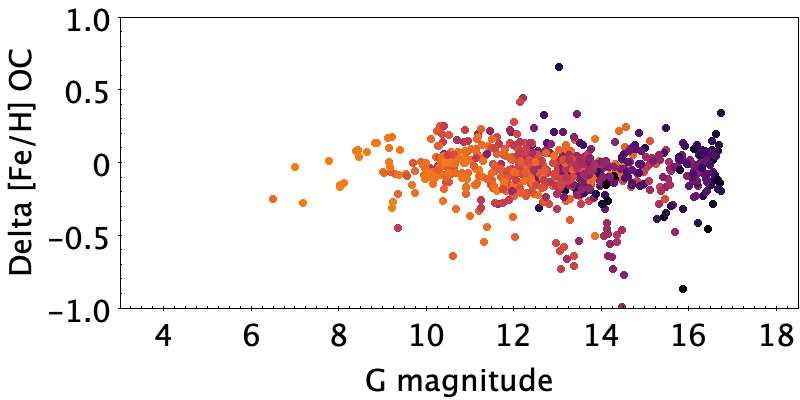}      
\includegraphics[width=0.48\columnwidth]{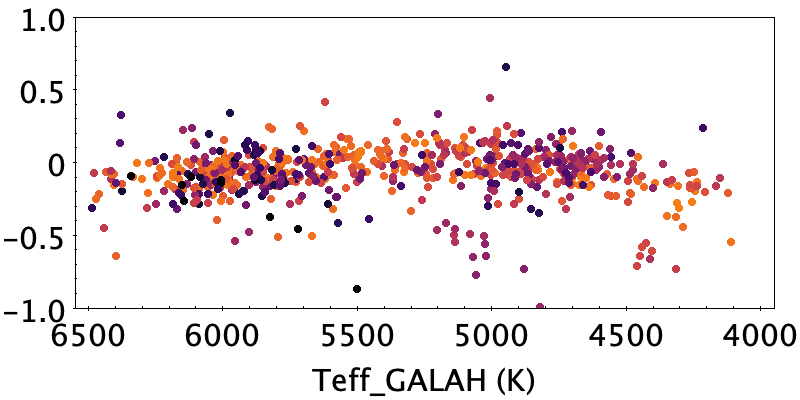}      
\includegraphics[width=0.48\columnwidth]{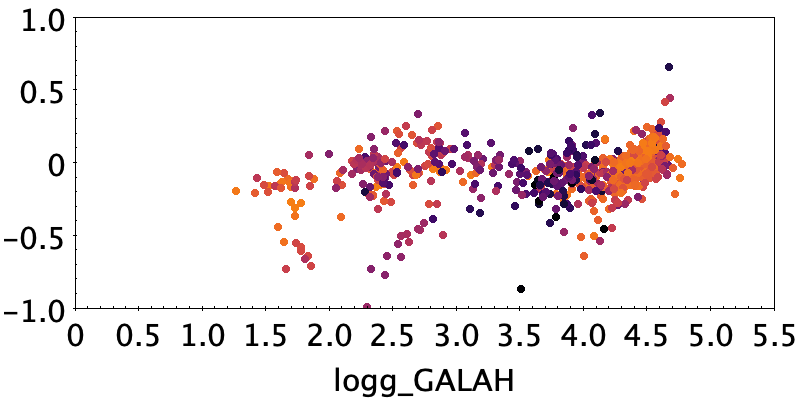}      
\includegraphics[width=0.48\columnwidth]{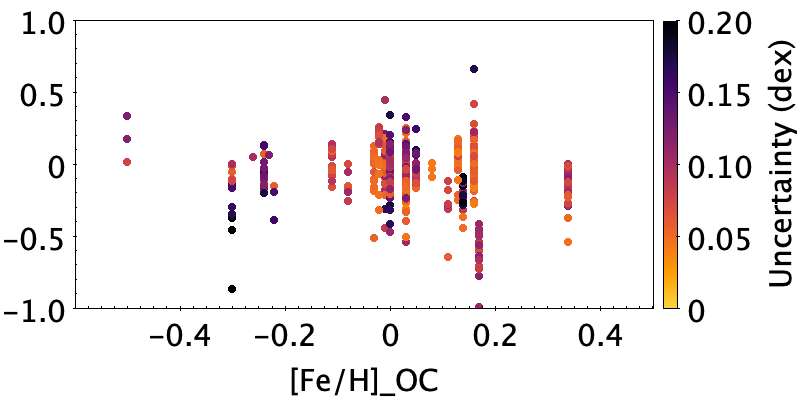}      

\includegraphics[width=0.48\columnwidth]{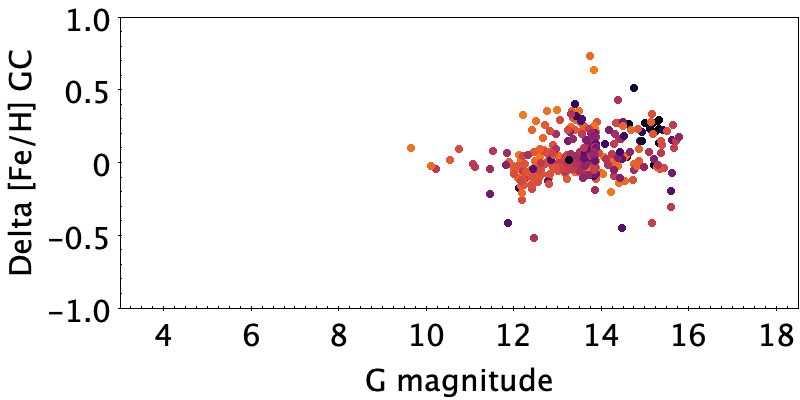}      
\includegraphics[width=0.48\columnwidth]{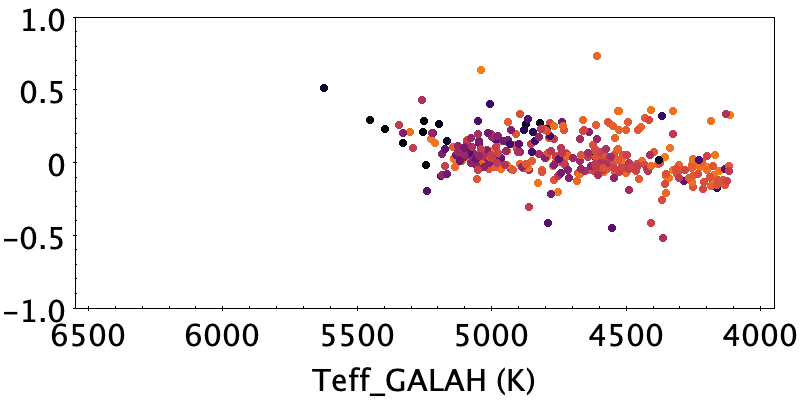}      
\includegraphics[width=0.48\columnwidth]{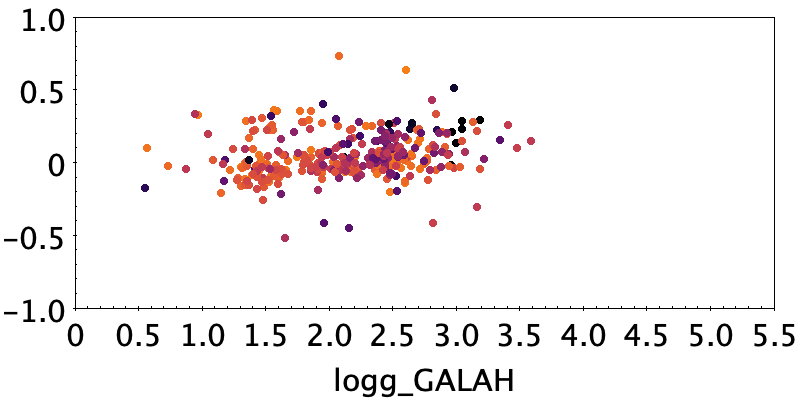}      
\includegraphics[width=0.48\columnwidth]{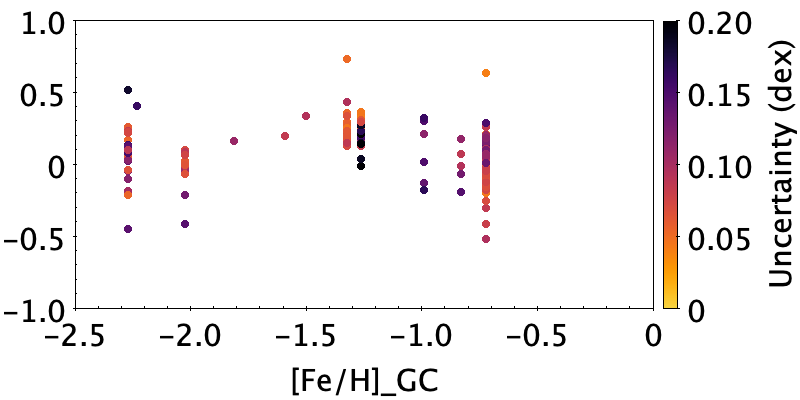}      
\caption{Same as Fig. \ref{f:apo_ref} for \galah.}
\label{f:gal_ref}
\end{figure*}

There are 232 stars in common between \pastel\ and this \galah\ sample, 682 FGK cluster members in 25 OCs and 363 FGK cluster members in 11 GCs. The residuals are shown in Fig.~\ref{f:gal_ref}  as a function of magnitude and of \tgm.

In the comparison to \pastel\ we note a larger dispersion for giants (\logg$<$3.8) than for dwarfs (\logg$\geq$3.8) with MAD=0.12 dex and MAD=0.05 dex respectively. With the same MAD values, the dispersion for metal-poor stars (\feh$<$-0.5 dex) is larger than that for  metal-rich stars (\feh$>$0 dex) although there are some outliers.  In addition, there is an offset of +0.14 dex for the metal-poor stars.  In the comparison to OCs, there are trends giving a negative offset at the two extrema of the  \teff\ range, and a pronounced oscillation along the \logg\ axis. For dwarfs there seems to be a positive slope of with \logg. Some outliers are visible, mainly on the negative side of the residuals. The largest uncertainties correspond to the faintest stars but those do not exhibit large deviations. For the GCs, there are only giants and there is no obvious trend.

\subsection{Gaia-ESO Survey}

The current public version of GES is the DR3, available in the ESO science archive since 2016. It includes 25\,533 stars observed with the FLAMES instrument on the Very Large Telescope, either at  medium-resolution (R$\sim$20\,000) with GIRAFFE setups or at high-resolution (R$\sim$47\,000) with UVES. The APs have been determined by different groups using a variety of parametrisation methodologies with common inputs (synthetic spectra, line list, solar abundance, LTE regime) from which recommended parameters and their errors were provided. 
The parameter homogenisation was performed with a weighting scheme that takes into account the performances of each group, after outlier rejection.
The parameters derived by the different groups were put on the same scale based on calibrators analysed by all. Our selection of FGK stars with valid \tgm\ in the \teff\ range 4000-6500 K gives 11\,638 stars with a median \feh\ uncertainty of 0.09 dex.

\begin{figure*}[ht!]
\centering
\includegraphics[width=0.48\columnwidth]{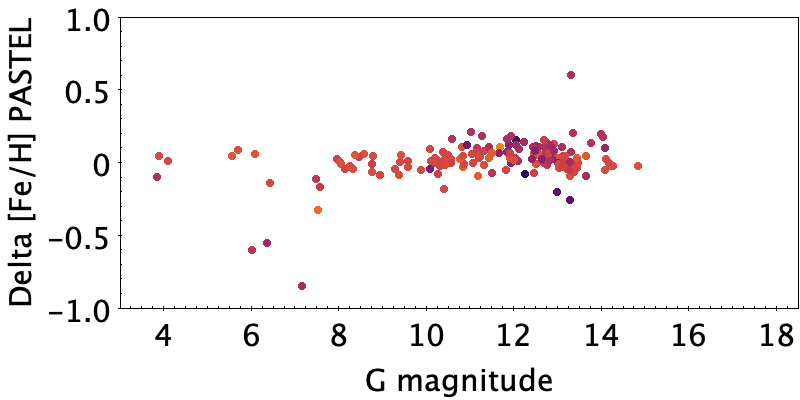}      
\includegraphics[width=0.48\columnwidth]{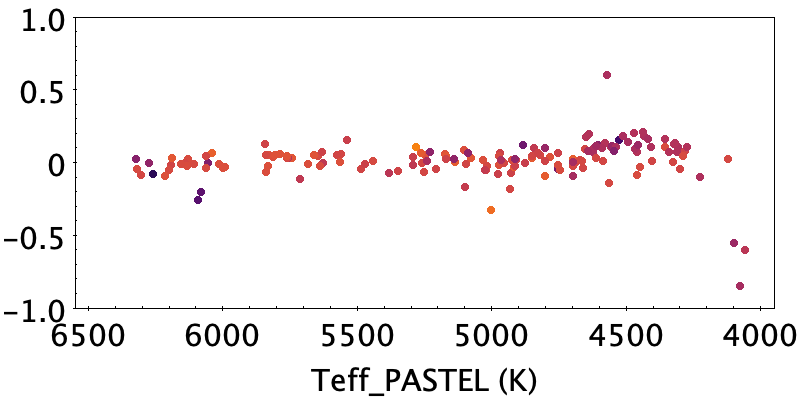}      
\includegraphics[width=0.48\columnwidth]{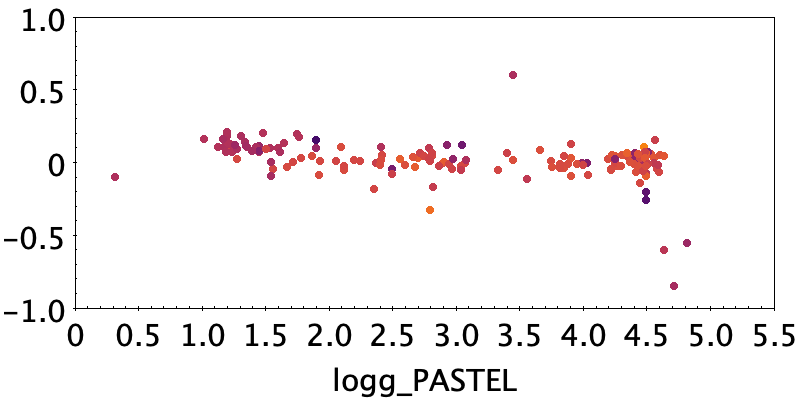}      
\includegraphics[width=0.48\columnwidth]{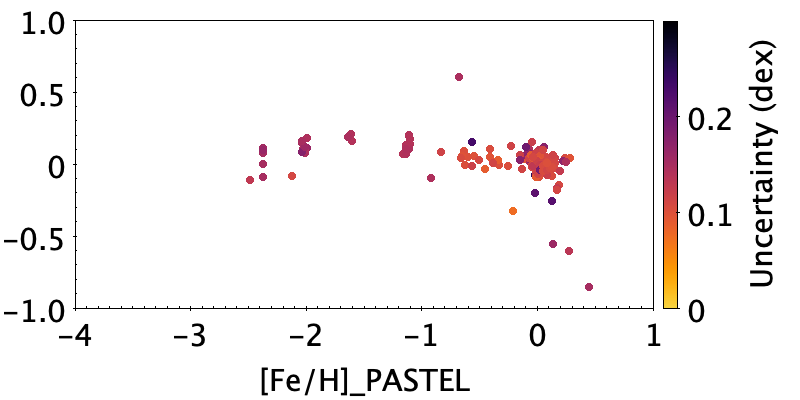}      

\includegraphics[width=0.48\columnwidth]{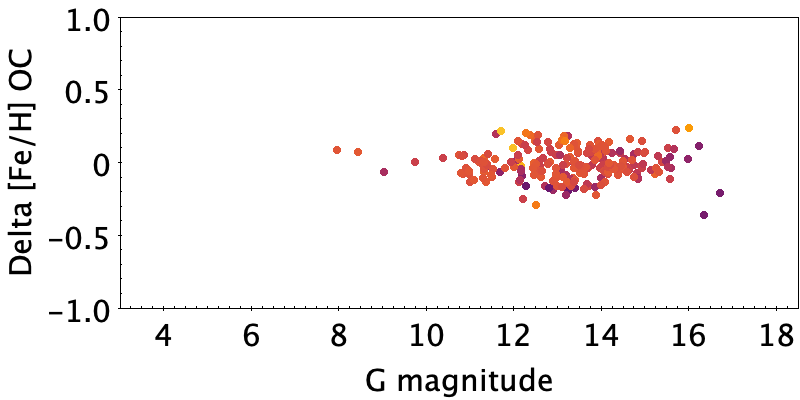}      
\includegraphics[width=0.48\columnwidth]{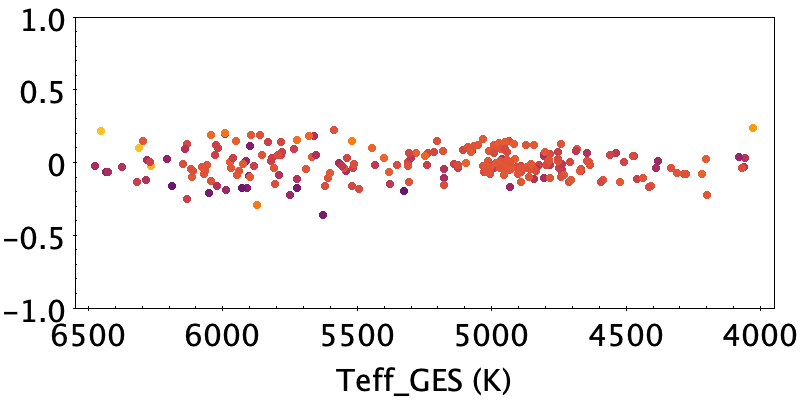}      
\includegraphics[width=0.48\columnwidth]{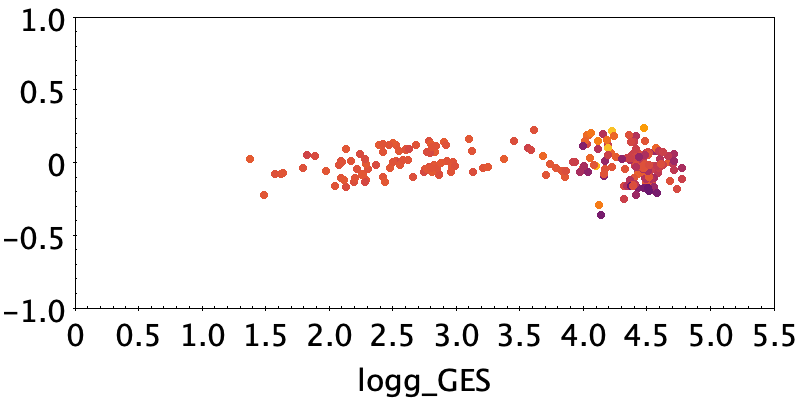}      
\includegraphics[width=0.48\columnwidth]{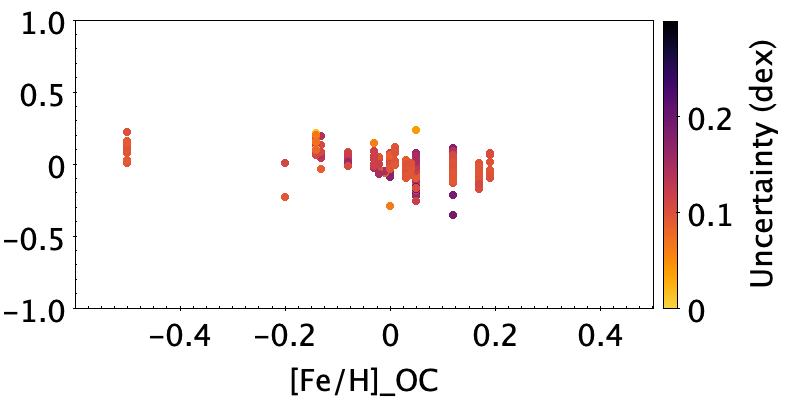}      

\includegraphics[width=0.48\columnwidth]{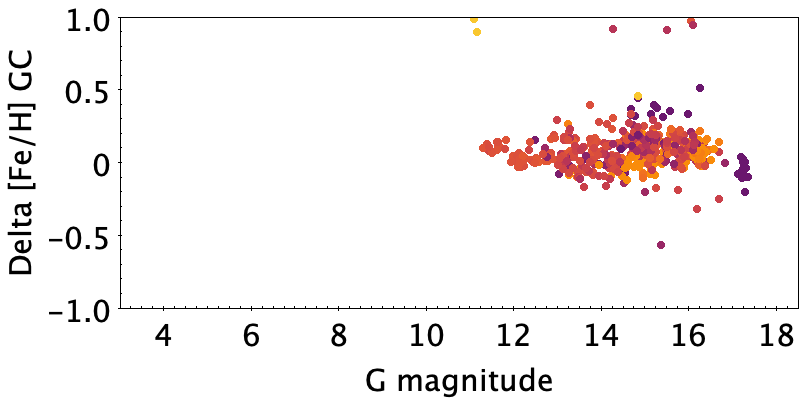}      
\includegraphics[width=0.48\columnwidth]{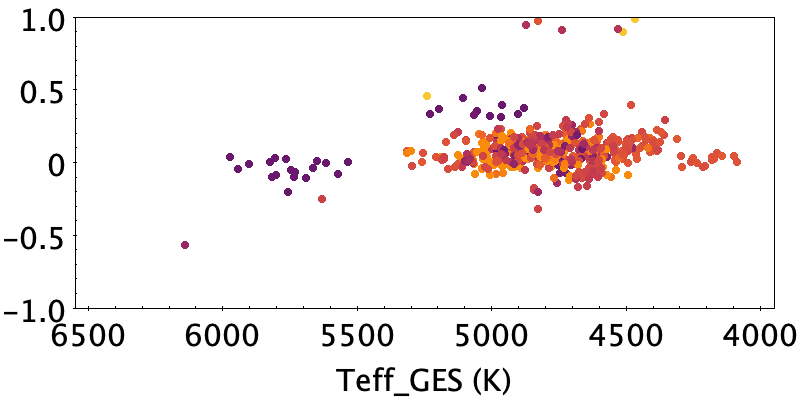}      
\includegraphics[width=0.48\columnwidth]{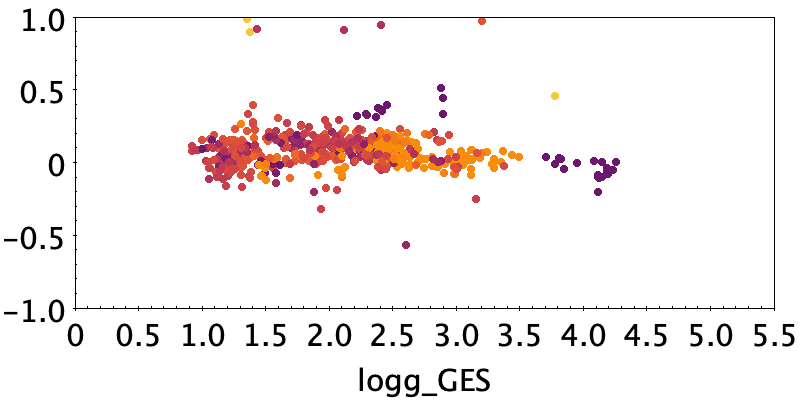}      
\includegraphics[width=0.48\columnwidth]{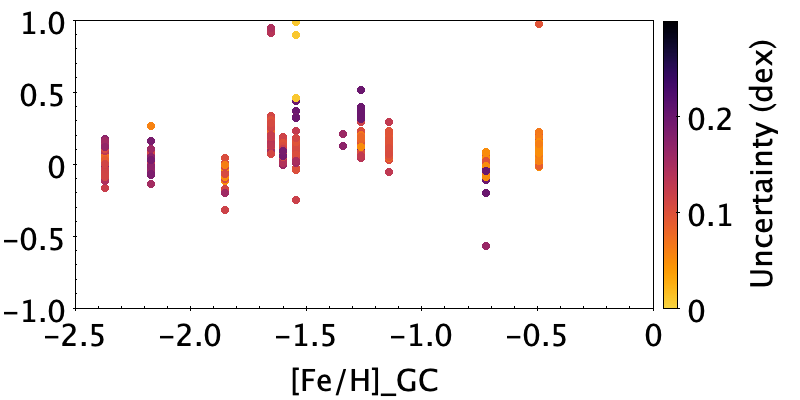}      
\caption{Same as Fig. \ref{f:apo_ref} for \ges.}
\label{f:gesdr3_ref}
\end{figure*}

We found 162 stars in common between \pastel\ and \ges, 239 FGK cluster members in 18 OCs and 513 FGK cluster members in 11 GCs. The residuals are shown in Fig.~\ref{f:gesdr3_ref}  as a function of magnitude and of \tgm.

The comparison to \pastel\ gives a good agreement over the full range of parameters, with a low dispersion which does not reflect the large quoted uncertainties. The comparison to OCs is very clean with a remarkably low scatter among the members. Several OCs have a dispersion of MAD=0.02 dex or lower: NGC 2243 (17 stars), NGC 2264 (7 stars), NGC 2682 (16 stars), NGC 6633 (6 stars), NGC 6802 (9 stars). The most observed OC is NGC 2516 with 55 stars (MAD=0.06 dex). For the GCs with at least 5 members, the scatter ranges from MAD=0.03 dex to MAD=0.06 dex. NGC 104 has 111 members observed (MAD=0.03 dex).




\subsection{RAVE}
The latest and final version of \rave, DR6, contains 451\,783 unique stars \citep{rave, ste20}. \rave\ spectra have an average resolution of R=7\,500 and cover the IR Ca triplet region at 8410-8795 \AA. The MADERA pipeline derives APs by fitting the spectra to a grid of synthetic spectra built from MARCS models.
The best model is found from a combination of a decision-tree algorithm and a projection method \citep[see][for further details]{kor11}. APs are then calibrated on a sample of reference stars. \feh\ is then determined by the GAUGUIN pipeline which fits individual  lines on a pre-computed grid of synthetic spectra, interpolated to the MADERA APs. Errors on \feh\ are computed by combining in a quadratic sum the 
propagation of errors of the stellar atmospheric parameters  and
the internal error of GAUGUIN due to noise. The later internal error is the standard deviation, at a given S/N and for a given spectral line, of
500 measurements of \feh\ from noisy synthetic spectra of Sun-like and Arcturus-like stars.  We note that the individual uncertainties quoted in the survey represent the precision of the metallicities while the accuracy has been tested using synthetic spectra with noise with the conclusion that GAUGUIN-derived values intrinsically do not suffer from large systematics.  We selected the most reliable results for FGK stars with the flag algo\_conv\_madera=0 and with fe\_h\_chisq\_gauguin < 2.5 and fe\_h\_error\_gauguin < 0.3. The median \feh\  uncertainty for the corresponding 196,448 stars  is 0.15 dex. 

\begin{figure*}[ht!]
\centering
\includegraphics[width=0.48\columnwidth]{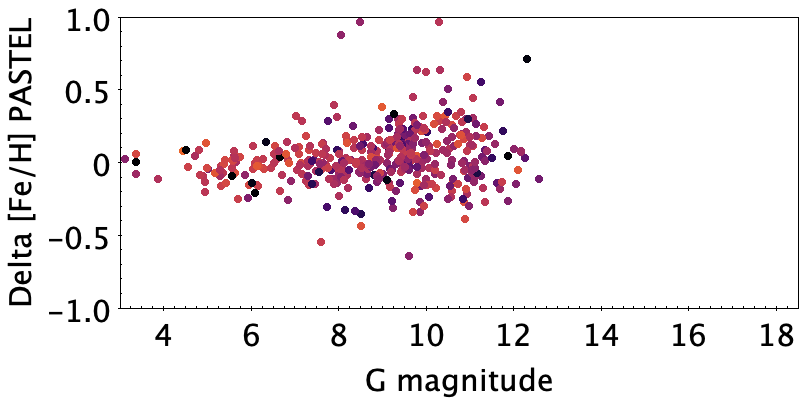}      
\includegraphics[width=0.48\columnwidth]{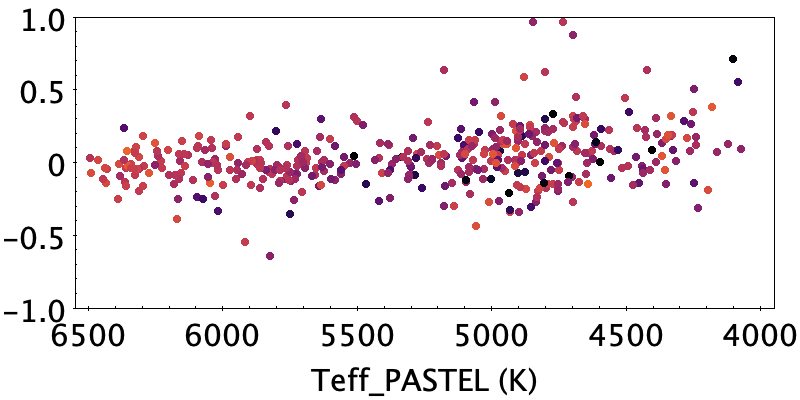}      
\includegraphics[width=0.48\columnwidth]{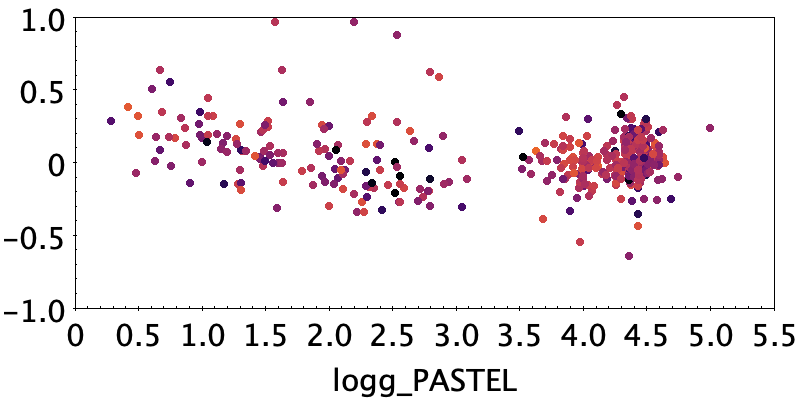}      
\includegraphics[width=0.48\columnwidth]{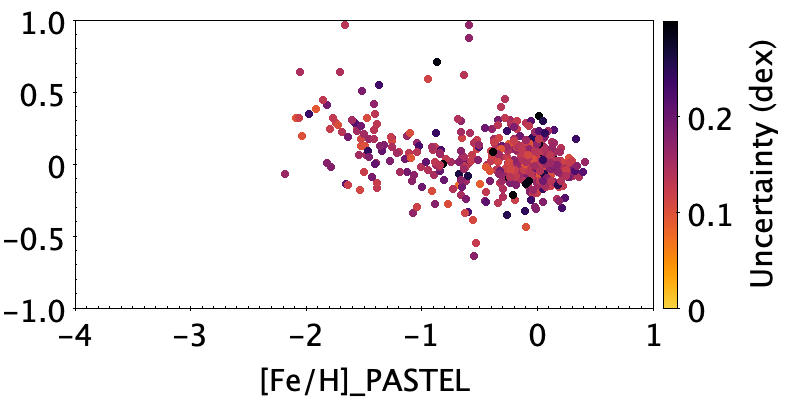}      

\includegraphics[width=0.48\columnwidth]{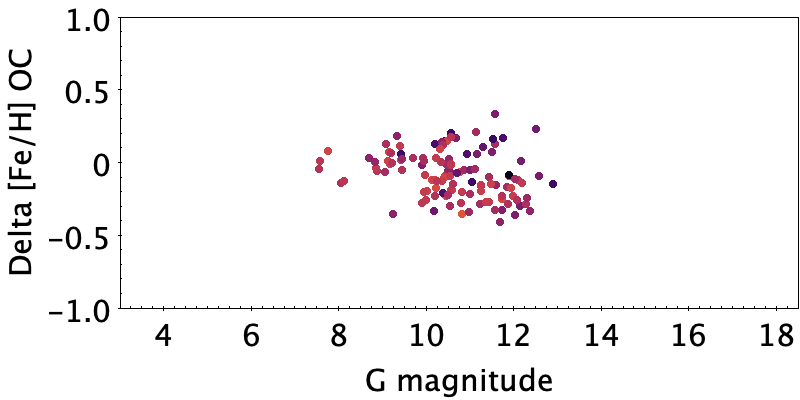}      
\includegraphics[width=0.48\columnwidth]{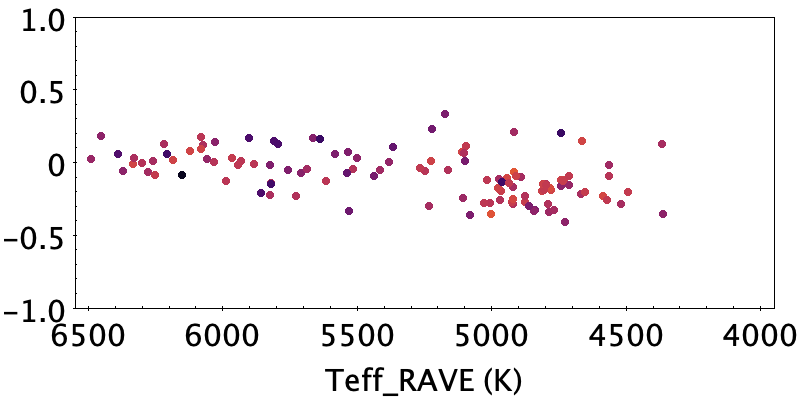}      
\includegraphics[width=0.48\columnwidth]{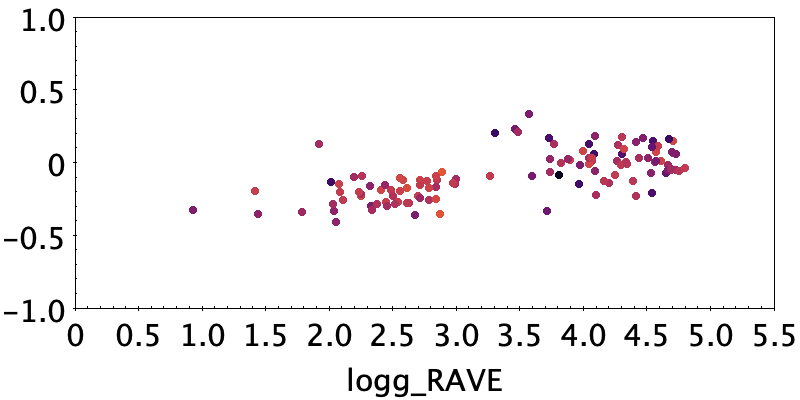}      
\includegraphics[width=0.48\columnwidth]{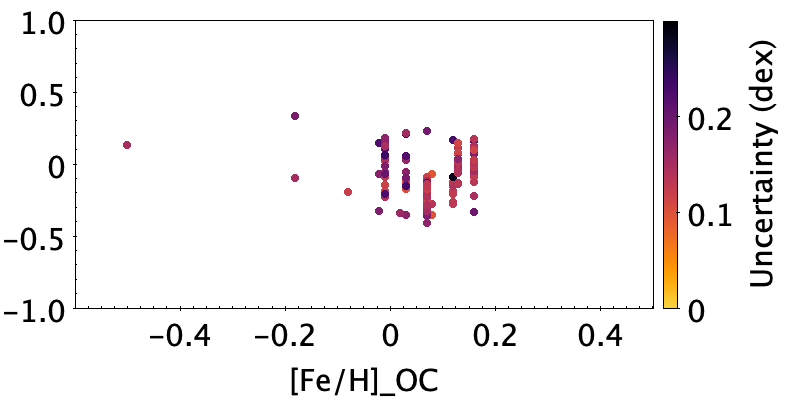}      

\includegraphics[width=0.48\columnwidth]{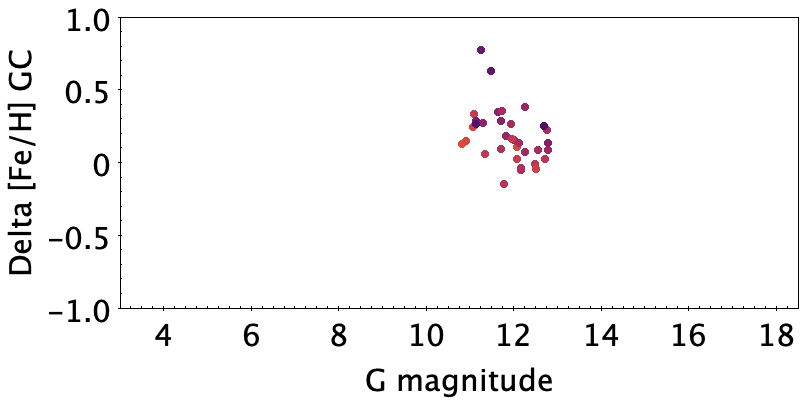}      
\includegraphics[width=0.48\columnwidth]{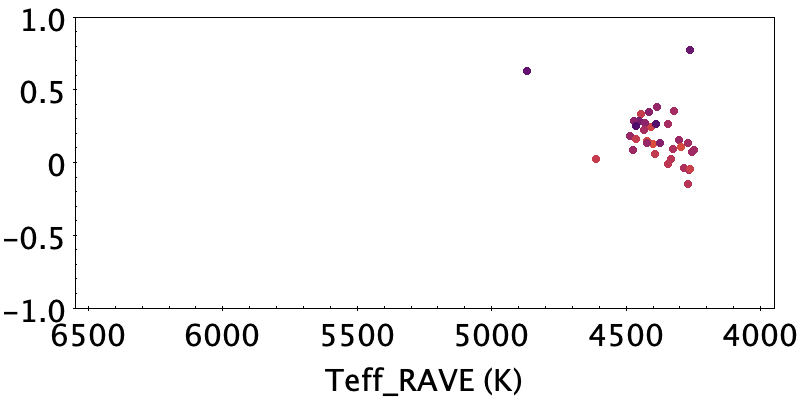}      
\includegraphics[width=0.48\columnwidth]{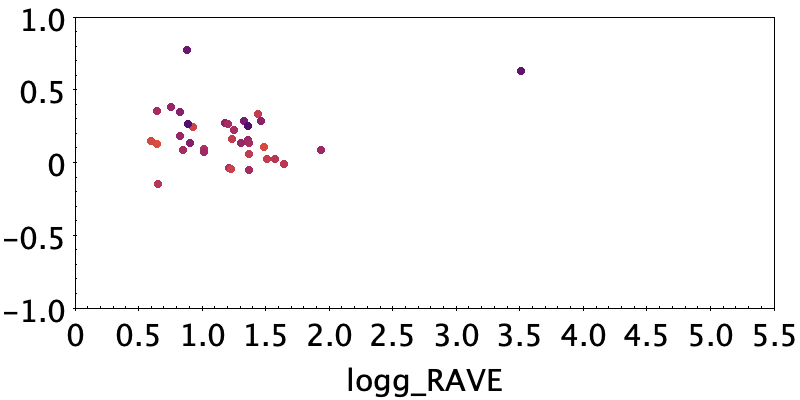}      
\includegraphics[width=0.48\columnwidth]{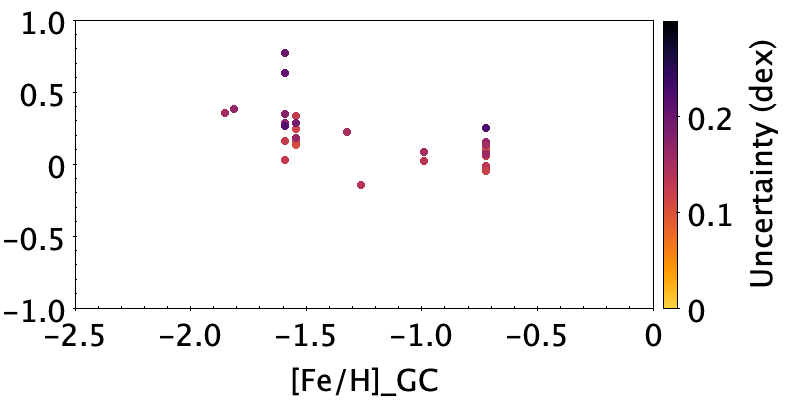}      
\caption{Same as Fig. \ref{f:apo_ref} for \rave.}
\label{f:rave_ref}
\end{figure*}

There are 427 stars in common between \pastel\ and this \rave\ sample, 119 FGK cluster members in 16 OCs and 35 FGK cluster members in 8 GCs. The residuals are shown in Fig.~\ref{f:rave_ref}  as a function of magnitude and of \tgm.

In the comparison to \pastel\ we note a positive offset of about +0.1 dex or larger for the coolest stars (\teff$<$5000 K), the most evolved giants (\logg$<$2) and the most metal-poor stars (\feh$<$-1.5). The OC residuals show that the metallicity of giants is systematically underestimated compared to that of dwarfs. The dispersion among OC members ranges from 0.04 to 0.07 dex for the 6 OCs with at least 5 members. There are only three GCs with at least five members which show dispersions from 0.045 to 0.08 dex.

\subsection{RAVE-CNN}
\cite{gui20} provide another version of \rave\ APs based on Convolutional Neural Networks (CNN) trained with a set of 3904 stars with high quality APs from \apo\ {\small DR16}. \rave\ data is complemented by 2MASS and ALL\_WISE photometry, and by Gaia DR2 photometry and parallaxes. The parameters
are averaged over 80 CNN runs while the errors correspond to the  dispersion of the runs. Repeat observations show that these internal errors are realistic.
We considered only FGK stars with fe\_h\_flag\_cnn = 0 and fe\_h\_error\_cnn <= 0.3.  The median \feh\  uncertainty for the corresponding 381\,681 stars  is 0.045 dex.

\begin{figure*}[ht!]
\centering
\includegraphics[width=0.48\columnwidth]{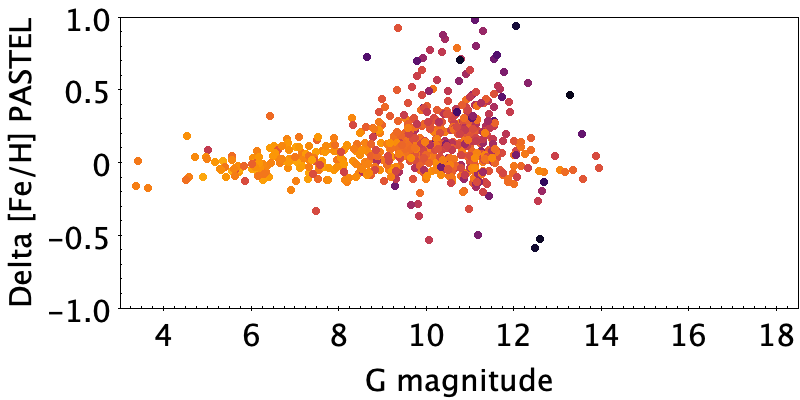}      
\includegraphics[width=0.48\columnwidth]{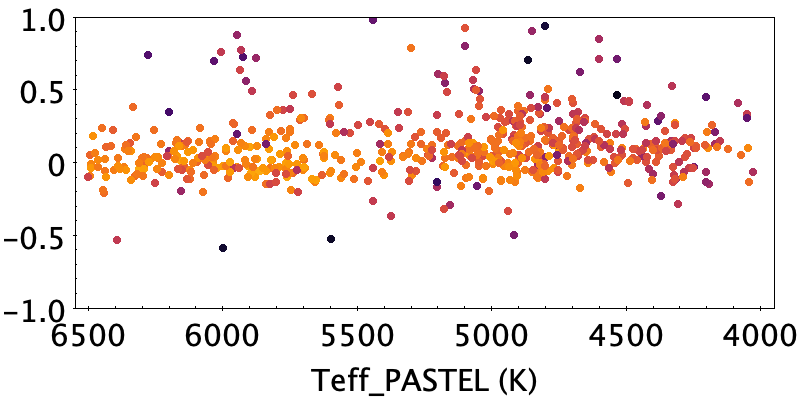}      
\includegraphics[width=0.48\columnwidth]{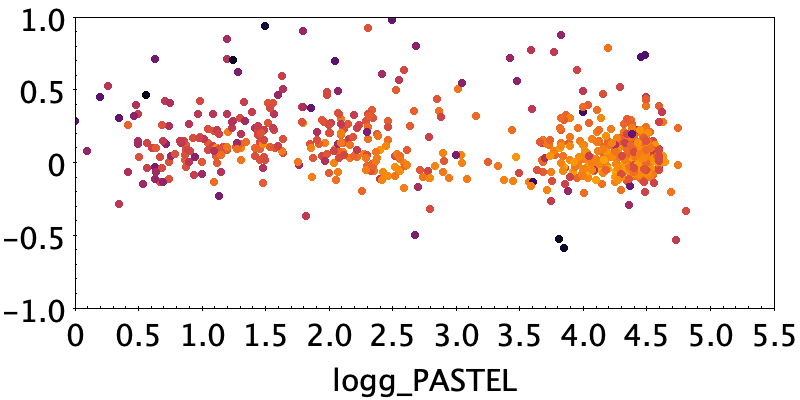}      
\includegraphics[width=0.48\columnwidth]{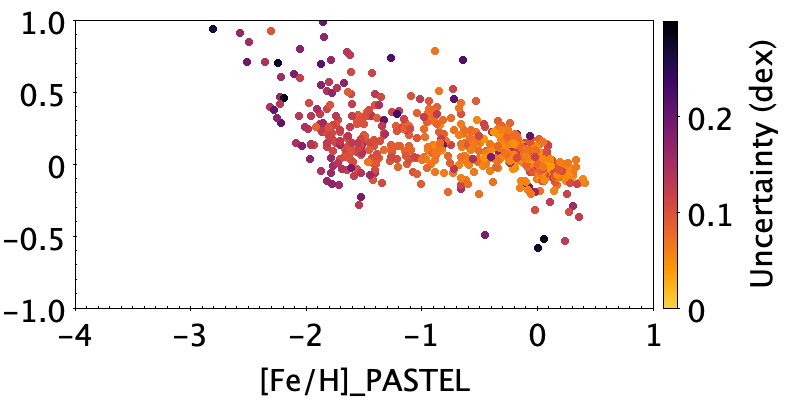}      

\includegraphics[width=0.48\columnwidth]{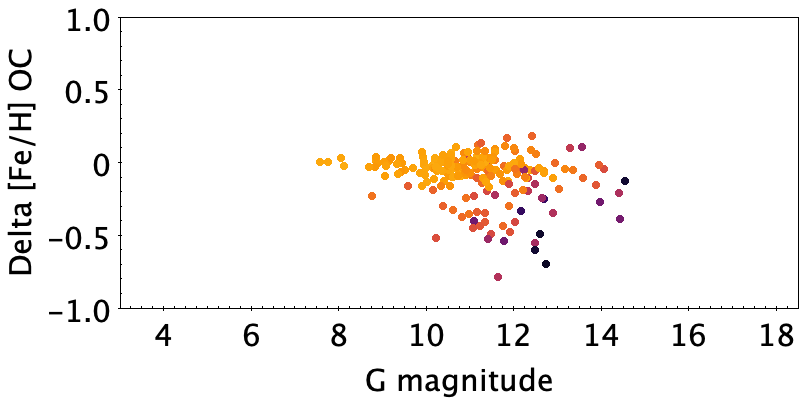}      
\includegraphics[width=0.48\columnwidth]{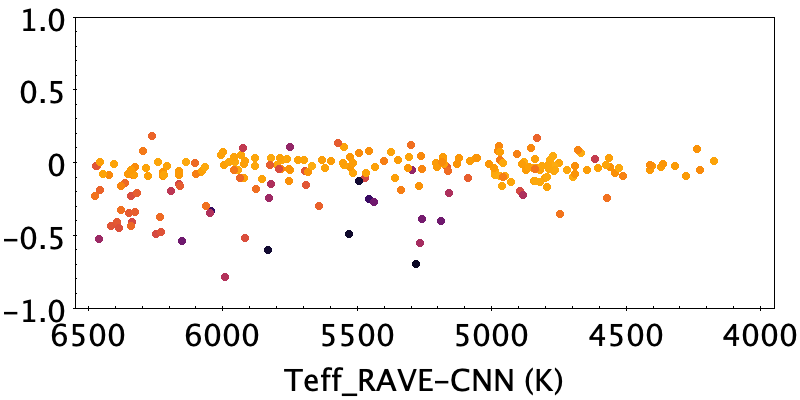}      
\includegraphics[width=0.48\columnwidth]{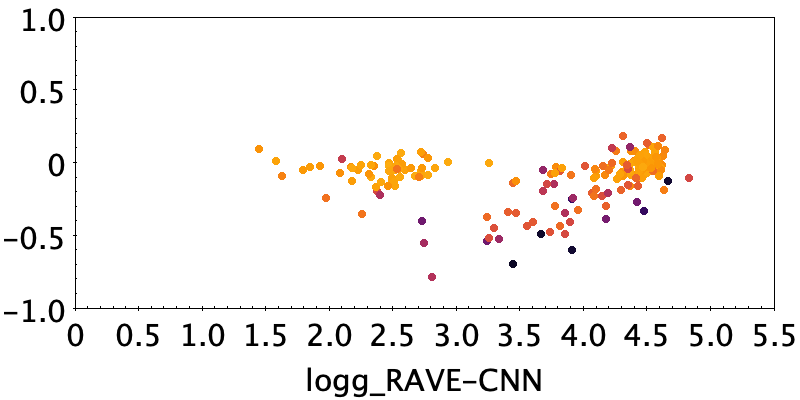}      
\includegraphics[width=0.48\columnwidth]{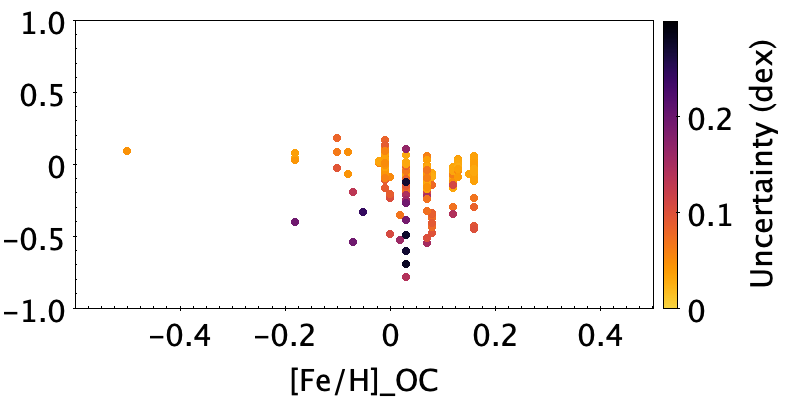}      

\includegraphics[width=0.48\columnwidth]{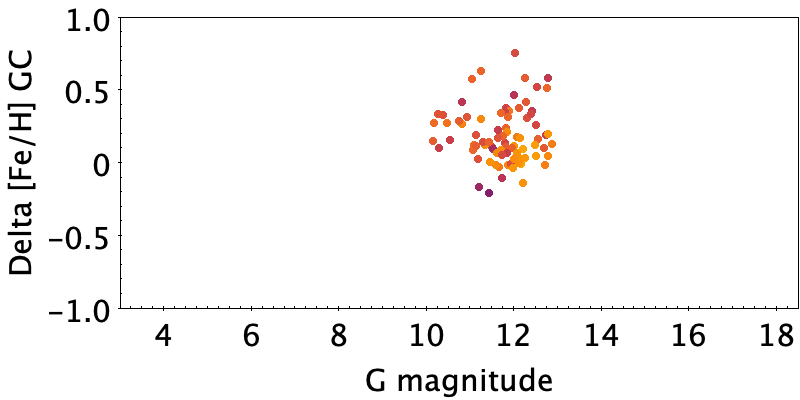}      
\includegraphics[width=0.48\columnwidth]{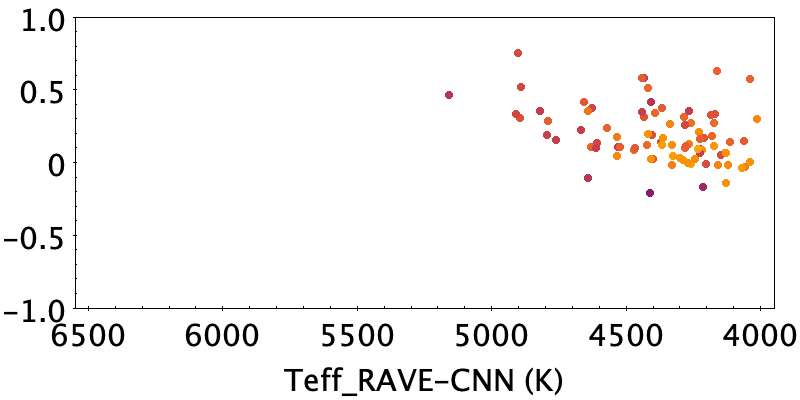}      
\includegraphics[width=0.48\columnwidth]{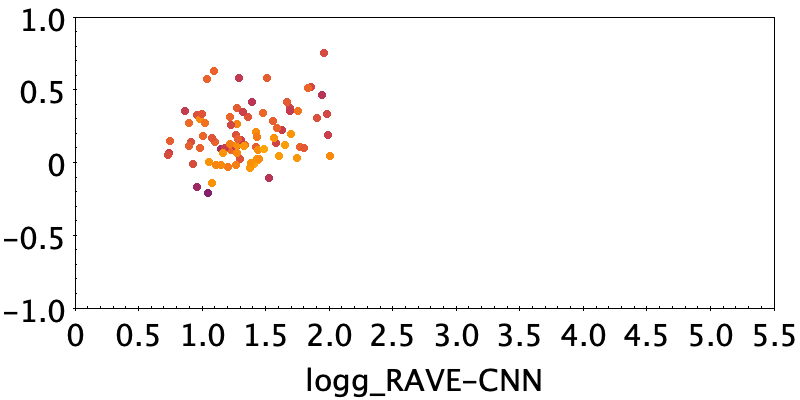}      
\includegraphics[width=0.48\columnwidth]{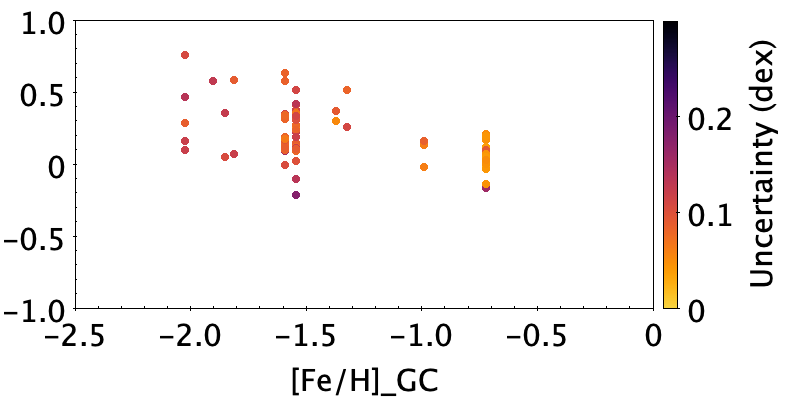}      
\caption{Same as Fig. \ref{f:apo_ref} for \rave-CNN.}
\label{f:cnn_ref}
\end{figure*}

There are 666 stars in common between \pastel\ and this \rave-CNN sample, 216 FGK cluster members in 25 OCs and 85 FGK cluster members in 10 GCs. The residuals are shown in Fig.~\ref{f:cnn_ref}  as a function of magnitude and of \tgm.

In the comparison to \pastel\ we note a change in the residual distribution at Gmag=9. For fainter stars, the dispersion is higher with a positive offset. The most deviating stars have also the largest uncertainties. The offset reaches +0.18 dex for the metal-poor stars (\feh$<$-0.5 dex) while there is decreasing trend in the residuals versus \feh\ for the metal-rich stars, with a slope of -0.32 dex$^{-1}$. The dispersion among OC members ranges from 0.02 to 0.06 dex, with only one cluster showing a large dispersion of MAD=0.12 dex (NGC 2423, 12 stars). The dwarfs and giants do not behave similarly, the residuals of giants are smaller while there seems to be a trend with \logg\ for the dwarfs, as well as outliers which have large uncertainties in general. A significant positive offset for the most metal-poor GCs is visible, in agreement with the trend also observed for the metal-poor stars in \pastel. The dispersion for four GCs with at least five members can be as low as 0.05 dex (NGC 104, 26 stars) and as high as 0.18 dex for the most metal-poor cluster (NGC 6397, 5 stars). \cite{gui20} state that  {\it "the performance would improve a lot
if the size of the training sample was three to four times
larger, but this pilot study is limited by the current overlap
with APOGEE DR16"}. Indeed we note that the training sample has very few stars with \feh$<$-0.5 dex and almost none with \feh$<$-1.0 dex resulting in a poor parametrisation of halo stars.

\subsection{LAMOST}
The currently public  version available  in VizieR is LAMOST DR5 \citep{luo19}. LAMOST spectra have a resolution of R$\sim$1\,800 and cover the optical range 3690-9100\AA.  The \lamost\ stellar pipeline \citep{xia15} derives \tgm\ by matching the flux-calibrated spectra  to empirical templates from the MILES library \citep{san06}. Two algorithms are used successively, first a weighted mean of parameters of best-matching templates, then a $\chi^2$ minimisation to further improve the parameters. Errors of the final parameters are estimated by combining the random and systematic errors, and are functions of S/N and of \tgm. Random errors are estimated from repeat observations, while the systematic errors are derived by applying the pipeline to the MILES templates. We considered only FGK stars with e\_\_Fe\_H\_$\le$0.3.  The median \feh\  uncertainty for the corresponding 4\,539\,240 stars  is 0.08 dex. 

The cross-match of LAMOST with the reference catalogues has been performed with the equatorial J2000 coordinates and a radius of 3''.

\begin{figure*}[ht!]
\centering
\includegraphics[width=0.48\columnwidth]{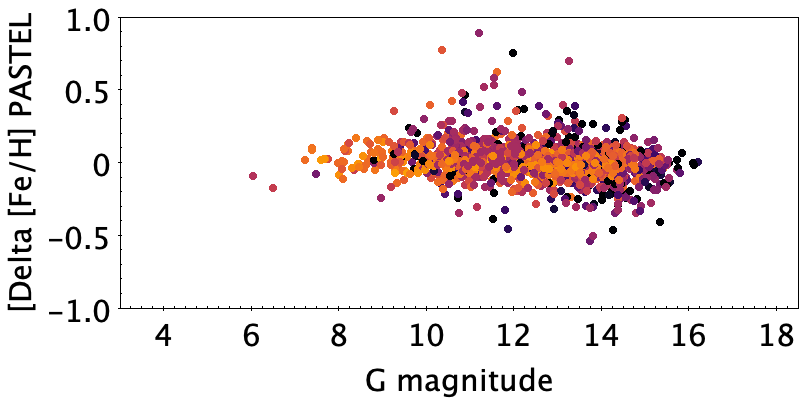}      
\includegraphics[width=0.48\columnwidth]{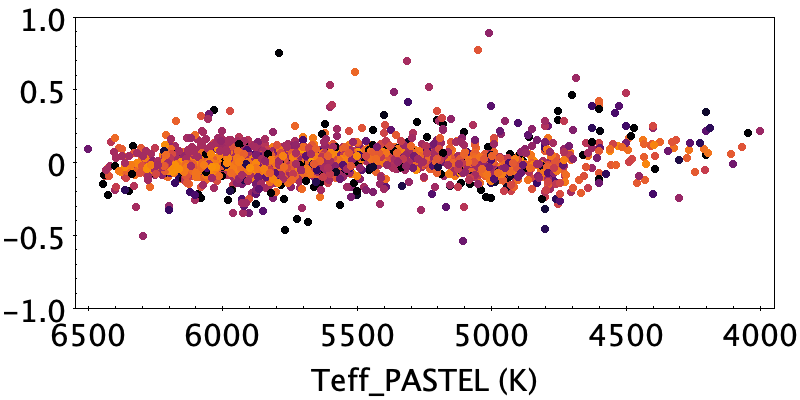}      
\includegraphics[width=0.48\columnwidth]{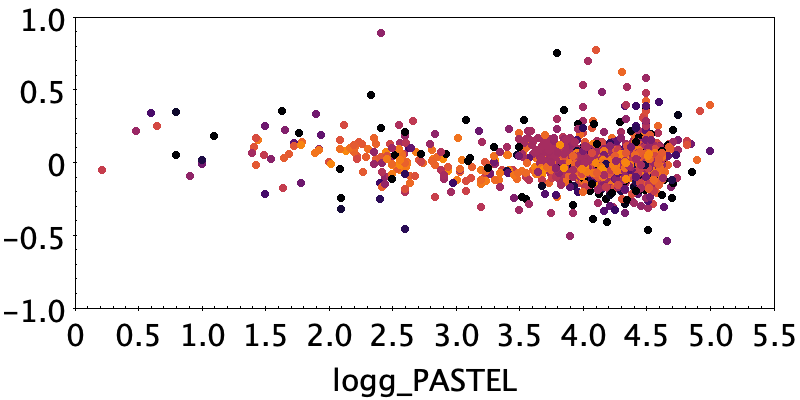}      
\includegraphics[width=0.48\columnwidth]{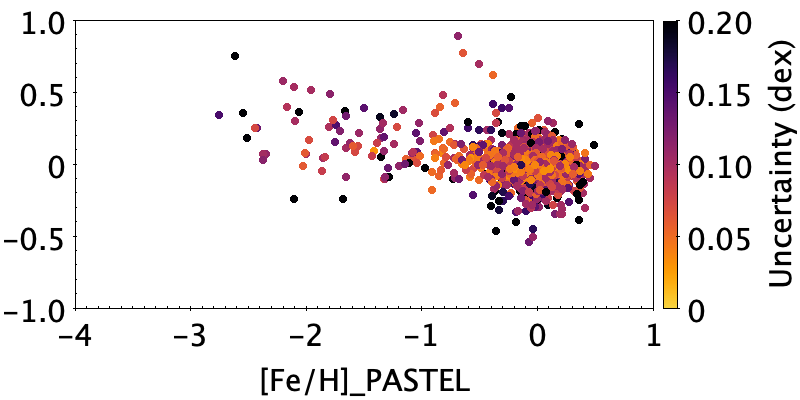}      

\includegraphics[width=0.48\columnwidth]{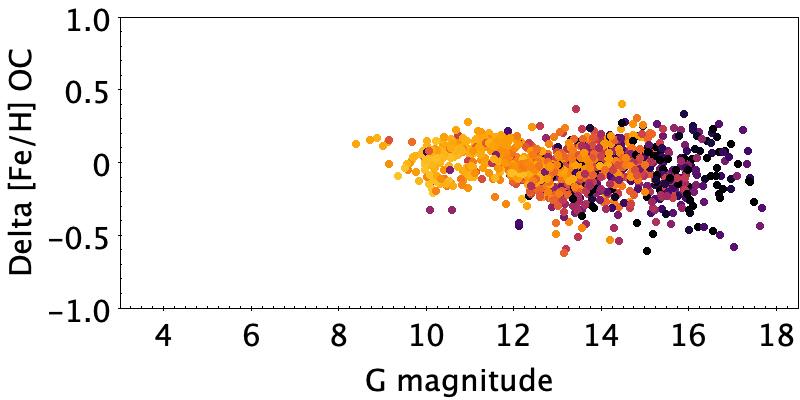}      
\includegraphics[width=0.48\columnwidth]{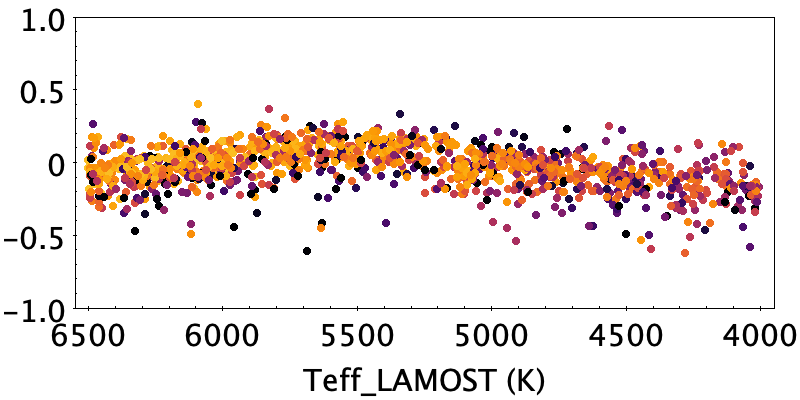}      
\includegraphics[width=0.48\columnwidth]{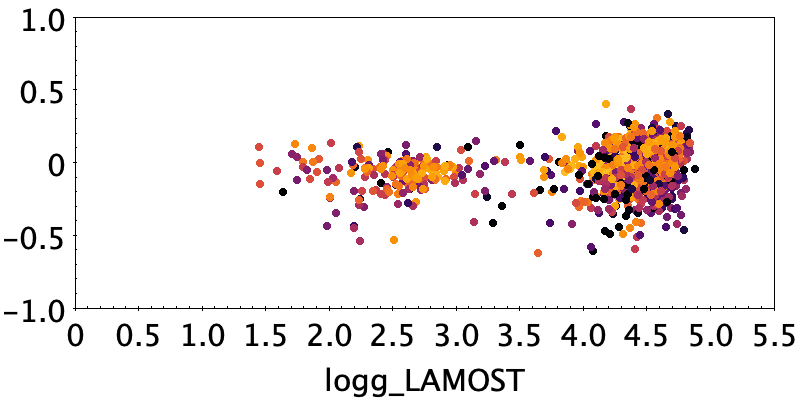}      
\includegraphics[width=0.48\columnwidth]{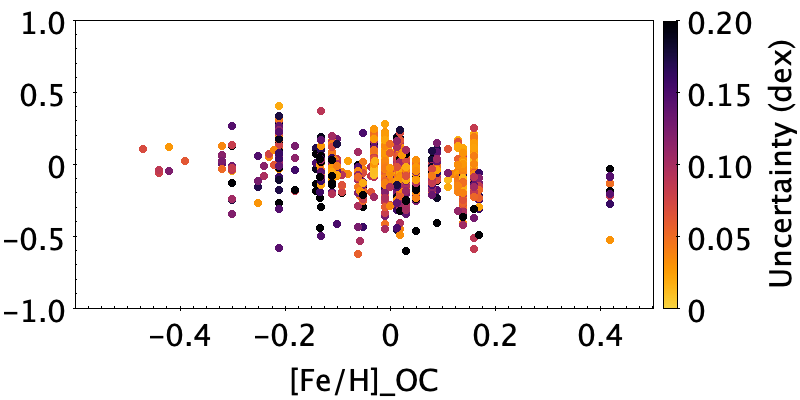}      

\includegraphics[width=0.48\columnwidth]{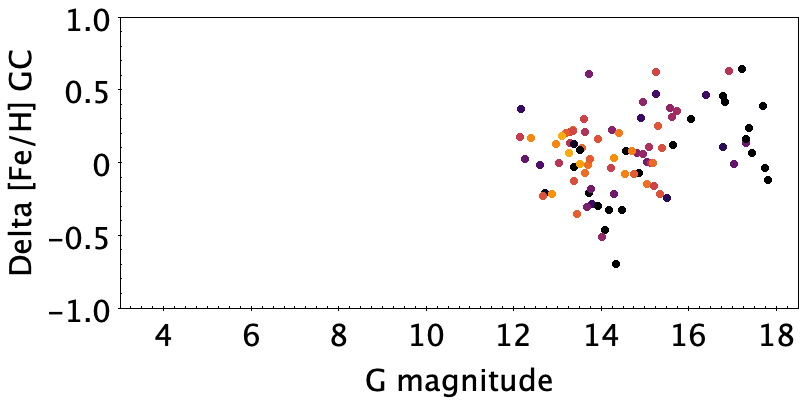}      
\includegraphics[width=0.48\columnwidth]{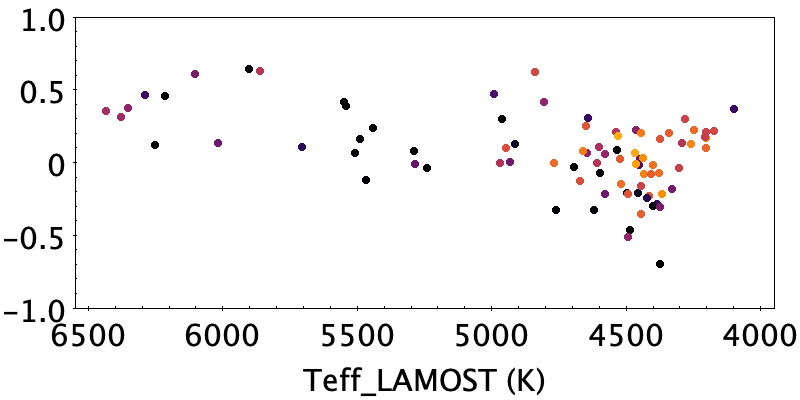}      
\includegraphics[width=0.48\columnwidth]{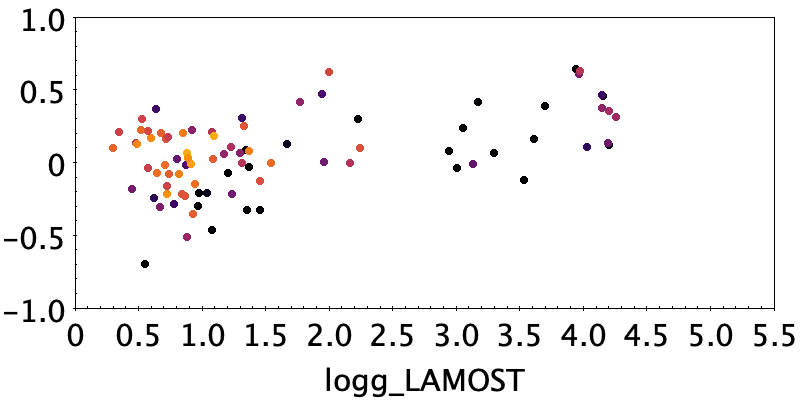}      
\includegraphics[width=0.48\columnwidth]{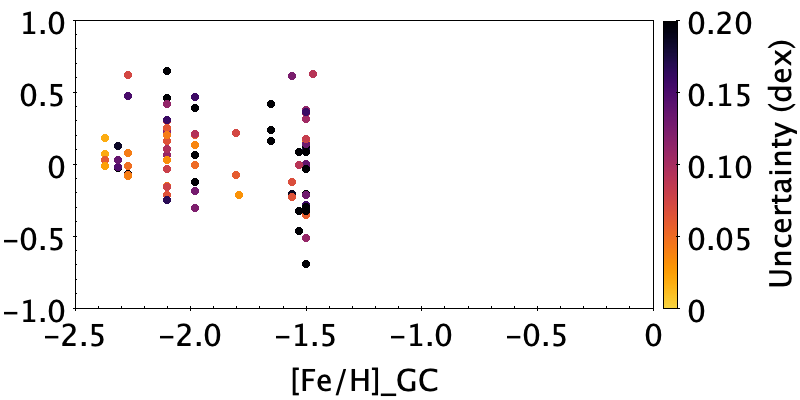}      
\caption{Same as Fig. \ref{f:apo_ref} for \lamost.}
\label{f:lamost_ref}
\end{figure*}

There are 1\,767 stars in common between \pastel\ and \lamost, 1\,175 FGK cluster members in 51 OCs and 87 FGK cluster members in 13 GCs. The residuals are shown in Fig.~\ref{f:lamost_ref}  as a function of magnitude and of \tgm.

There is a good overlap between \lamost\ and \pastel, mostly made of dwarfs with \feh$>$-0.5 dex. The few metal-poor stars show mainly positive residuals, like for the other surveys, indicating that \lamost\ tends to overestimate \feh\ compared to high-resolution analyses. The overall dispersion is low (MAD=0.055 dex) and the most deviating stars have also large uncertainties in general. There is also a good intersection with OCs, the majority having more than 15 members. The dispersion ranges from 0.01 to 0.14 dex with a median value of 0.05 dex. There is a marked oscillation of the residuals with \teff\ in OCs, which gives negative residuals at the extrema, similar to what we observed with \galah. Here the negative offset is more pronounced on the cool side. There are only four GCs with at least 5 members and only one of them has a dispersion lower than 0.1 dex (NGC 5053, MAD=0.06 dex, 7 stars). A positive offset is seen for the faintest stars, the hottest stars and dwarfs, all these stars also having the largest uncertainties.

\subsection{LAMOST PAYNE}
Another version of LAMOST DR5 has been released \citep{xia19} where APs and abundances have been determined with the method {\it Data-Driven Payne}, an hybrid approach which combines constraints from theoretical spectral models (ATLAS12) and training on 4557 stars from GALAH DR2 and on 15,000 stars from APOGEE DR14. 
The internal precision of \feh\ is deduced from the standard
deviation of  repeat observations at different S/N.  We adopt the recommended parameters, selecting the FGK stars with FEH\_FLAG = 1 (reliable). We apply in addition FEH\_ERR $\le$ 0.3. The median \feh\  uncertainty for the corresponding 5,967,849 stars  is 0.06 dex. 

Like for \lamost\ the cross-match with the reference catalogues has been performed with the equatorial J2000 coordinates and a radius of 3''. We note that the intersection with  the reference catalogues is larger than for \lamost.

\begin{figure*}[ht!]
\centering
\includegraphics[width=0.48\columnwidth]{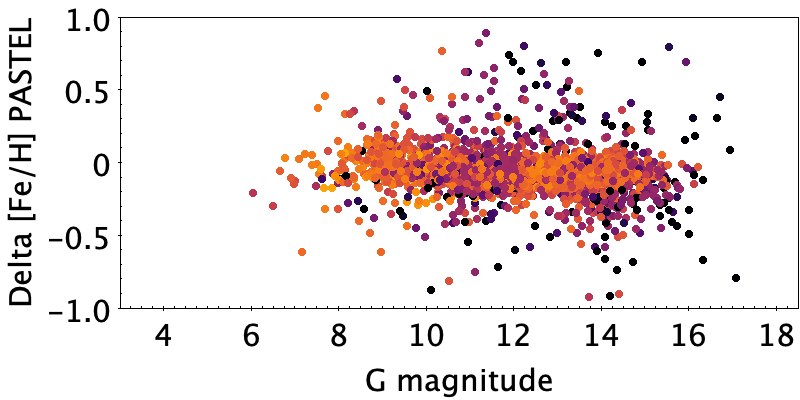}      
\includegraphics[width=0.48\columnwidth]{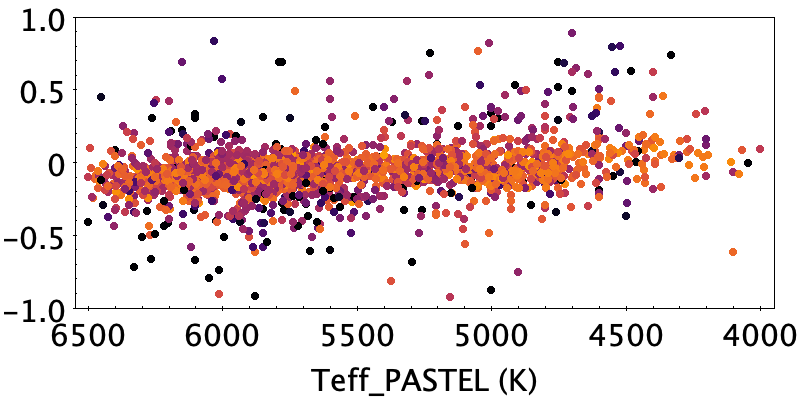}      
\includegraphics[width=0.48\columnwidth]{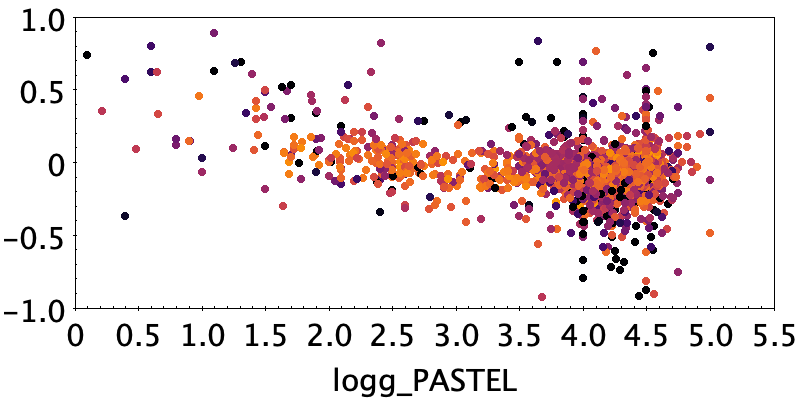}      
\includegraphics[width=0.48\columnwidth]{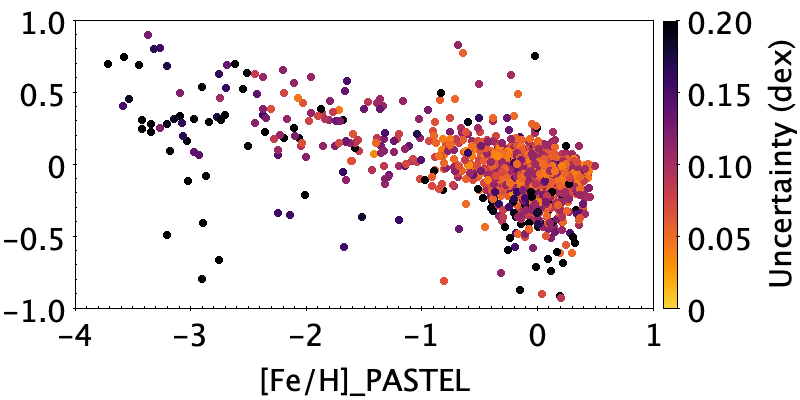}      

\includegraphics[width=0.48\columnwidth]{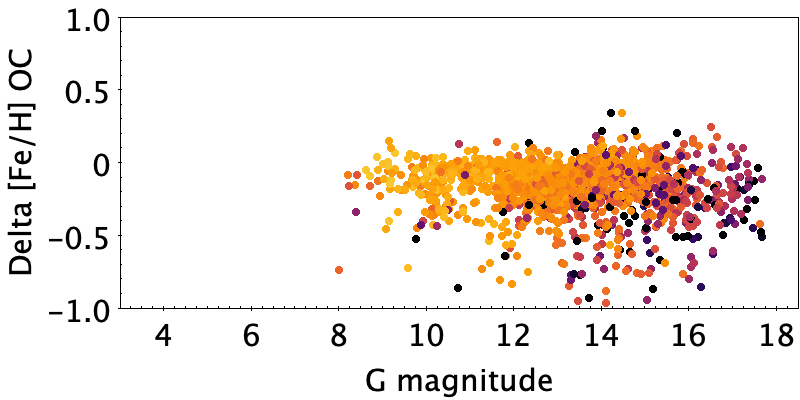}      
\includegraphics[width=0.48\columnwidth]{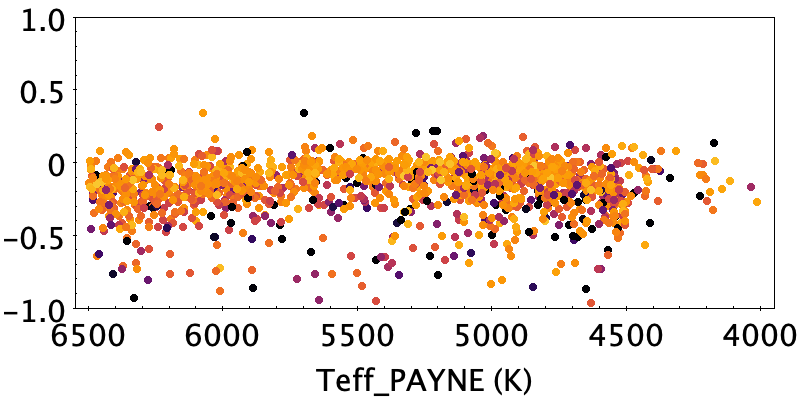}      
\includegraphics[width=0.48\columnwidth]{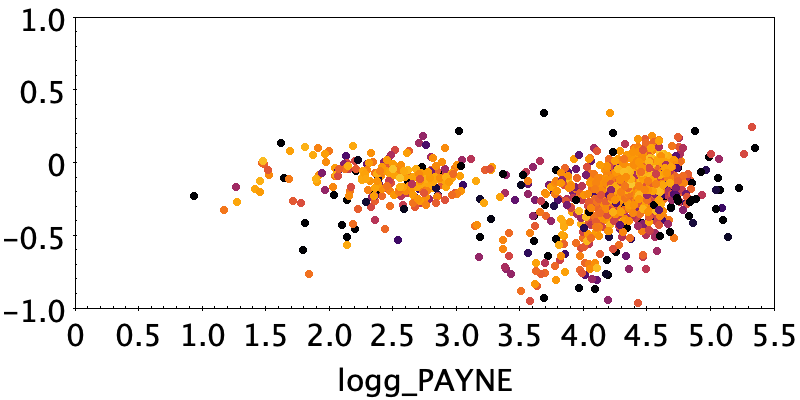}      
\includegraphics[width=0.48\columnwidth]{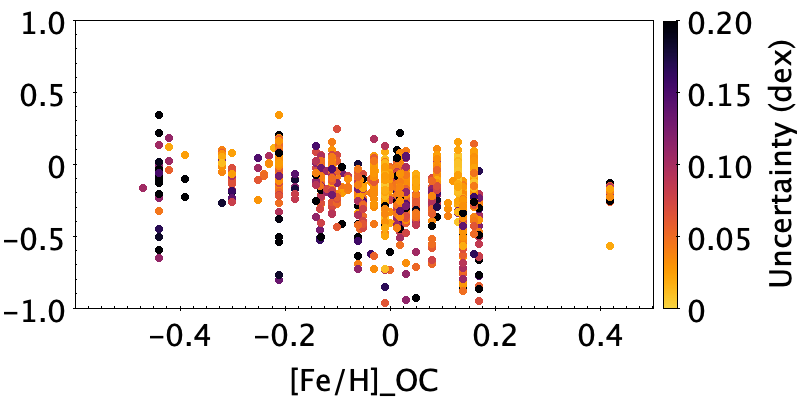}      

\includegraphics[width=0.48\columnwidth]{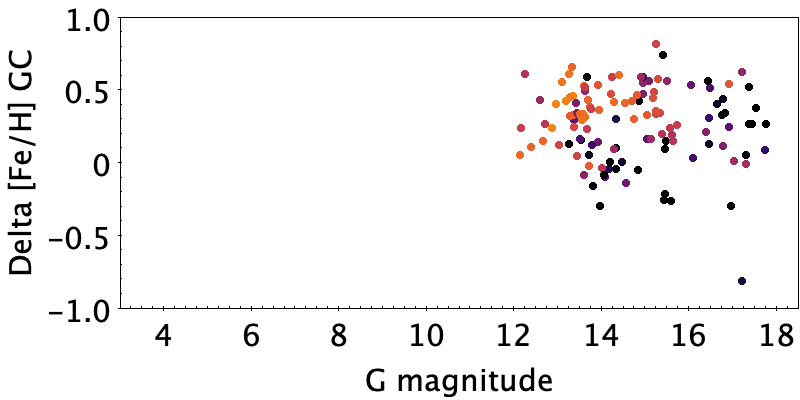}      
\includegraphics[width=0.48\columnwidth]{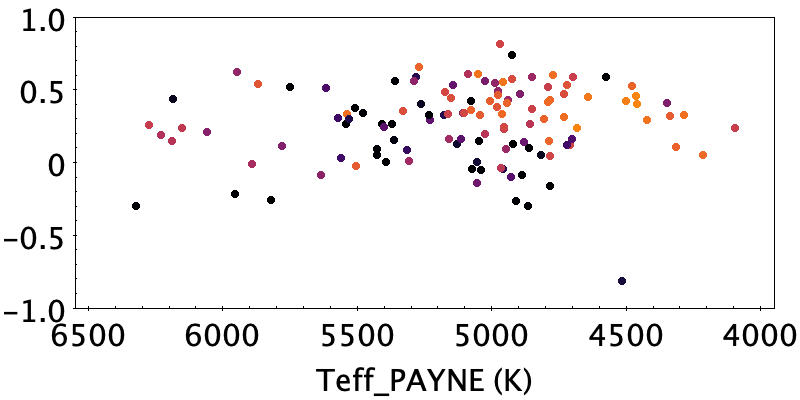}      
\includegraphics[width=0.48\columnwidth]{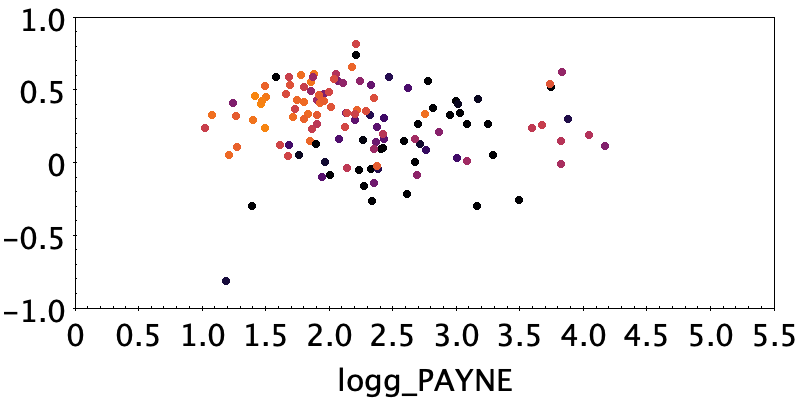}      
\includegraphics[width=0.48\columnwidth]{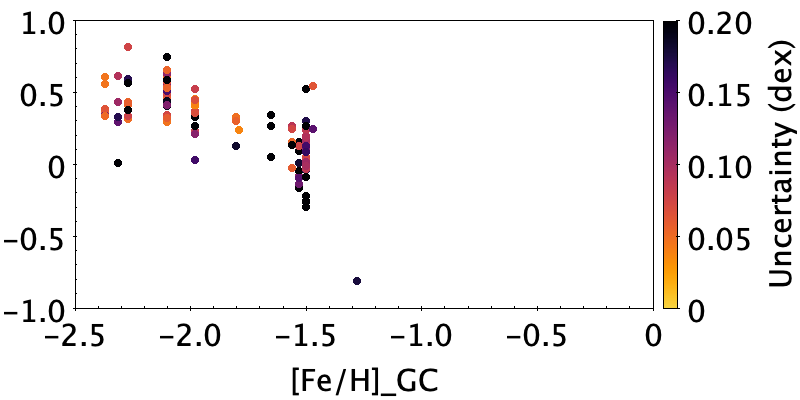}      
\caption{Same as Fig. \ref{f:apo_ref} for \lamost\ Data Driven Payne.}
\label{f:payne_ref}
\end{figure*}

There are 2243 stars in common between \pastel\ and \lamost\ Payne, 1470 FGK cluster members in 53 OCs and 123 FGK cluster members in 14 GCs. The residuals are shown in Fig.~\ref{f:payne_ref}  as a function of magnitude and of \tgm.

\lamost\ Payne behaves globally like \lamost\, with more outliers and larger systematics, inherited from the training sets. The largest deviations correspond quite well to the largest uncertainties. There are more metal-poor stars in common with \pastel\ but with a trend that gives a large positive offset of  +0.31 dex at \feh$<$-2 dex, a large dispersion which also corresponds to large uncertainties.  The OC residuals are significantly negative with a dispersion larger than with \lamost. Interestingly, the dwarfs and giants behave differently, with distributions of the residuals looking similar to those from \rave-CNN. Residuals of dwarfs are more dispersed and show a positive slope with \logg.   The dispersion among GC members is slightly lower than for \lamost, from 0.02 to 0.10 dex. There is a pronounced positive offset reaching +0.5 dex at the lowest metallicities. The source of these strong biaises is not clear since \lamost\ Payne uses two different training sets as well as constraints from synthetic spectra but we presume that the performances could be improved with a larger homogeneous training set covering better the full parameter space.

\subsection{SEGUE}
The Sloan Extension for Galactic Understanding and Exploration \citep[SEGUE, ][]{yan09} provides stellar spectra at R$\sim$2000 over the wavelength range 3800--9200 \AA\ for about 500,000 stars. The \segue\ Stellar Parameter Pipeline \citep[SSPP, ][]{lee08} uses multiple techniques to estimate \tgm, up to twelve methods for \feh, with a procedure which gives at the end a recommended value and its error that we adopt here. The SSPP was improved by \cite{smo11} who validated the results with OCs and GCs. They obtained a typical internal uncertainty of 0.05 dex on \feh, and a dispersion of 0.11 dex when the results are compared to high-resolution values. 

The cross-match of \segue\ with other catalogues based on equatorial J2000 coordinates was performed with a radius of 5''.

\begin{figure*}[ht!]
\centering
\includegraphics[width=0.48\columnwidth]{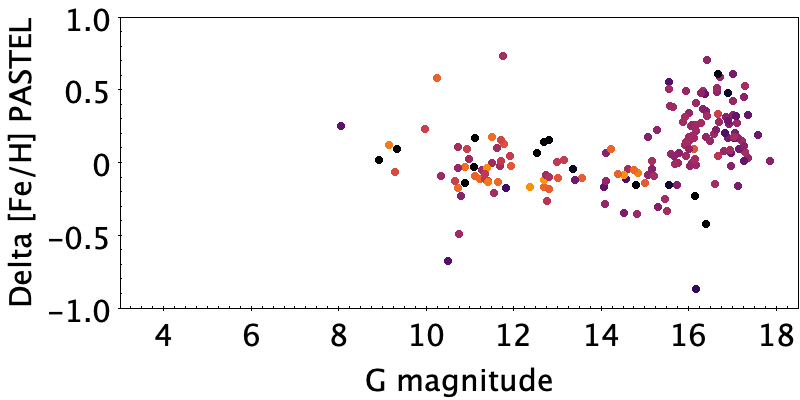}      
\includegraphics[width=0.48\columnwidth]{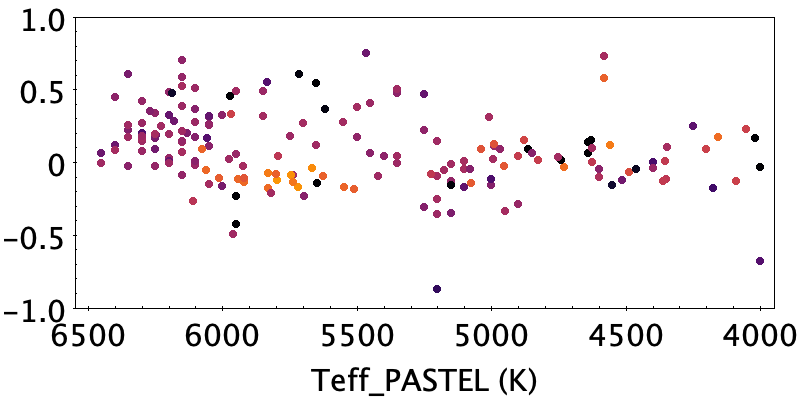}      
\includegraphics[width=0.48\columnwidth]{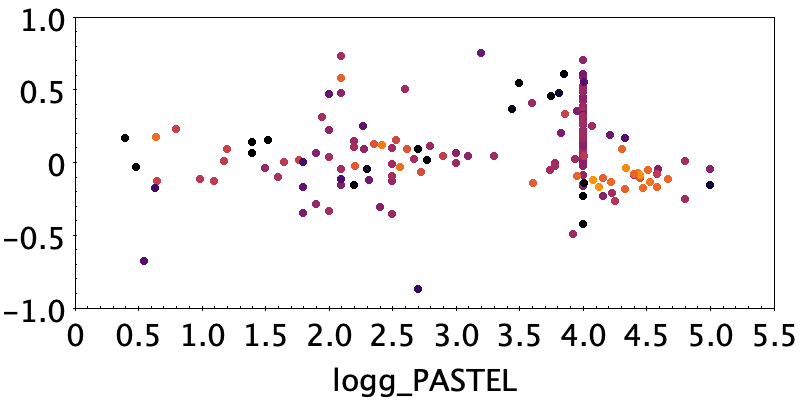}      
\includegraphics[width=0.48\columnwidth]{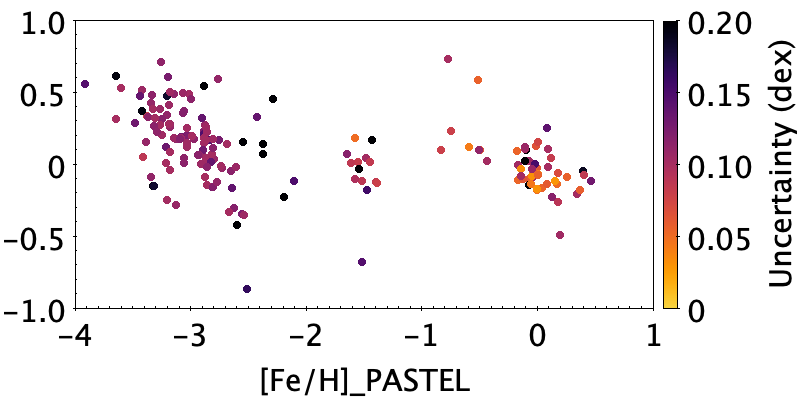}      

\includegraphics[width=0.48\columnwidth]{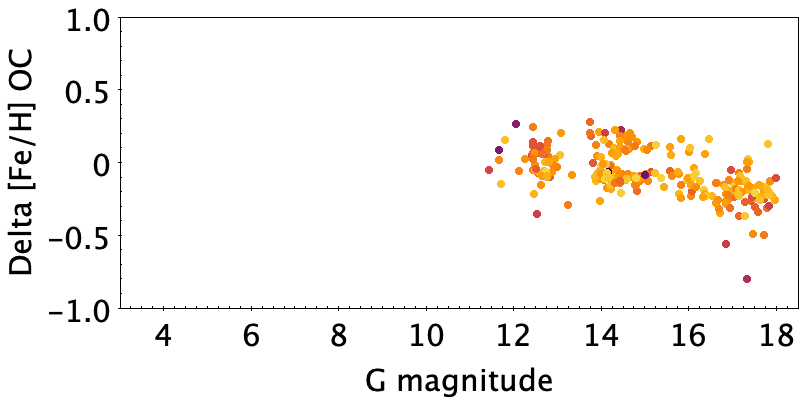}      
\includegraphics[width=0.48\columnwidth]{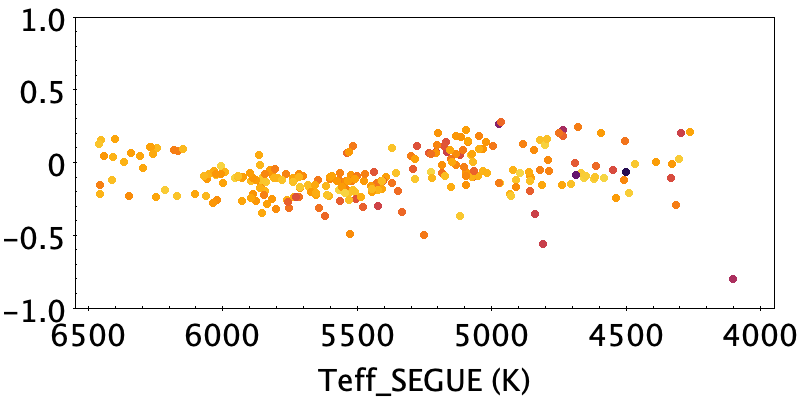}      
\includegraphics[width=0.48\columnwidth]{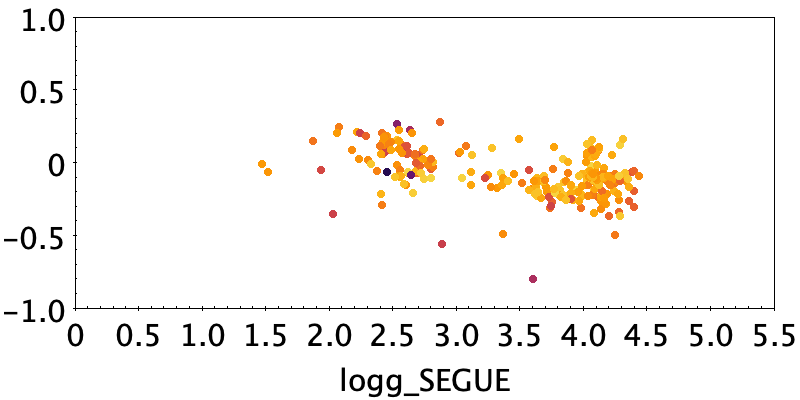}      
\includegraphics[width=0.48\columnwidth]{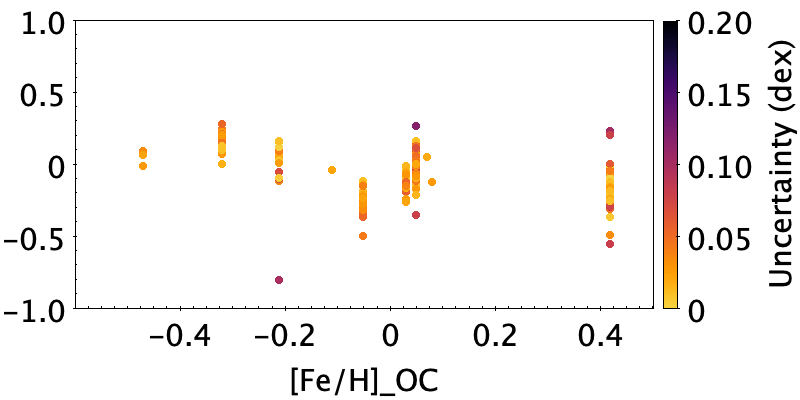}      

\includegraphics[width=0.48\columnwidth]{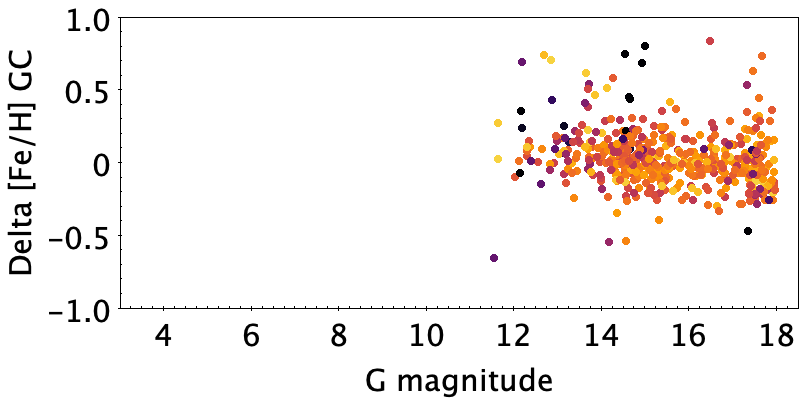}      
\includegraphics[width=0.48\columnwidth]{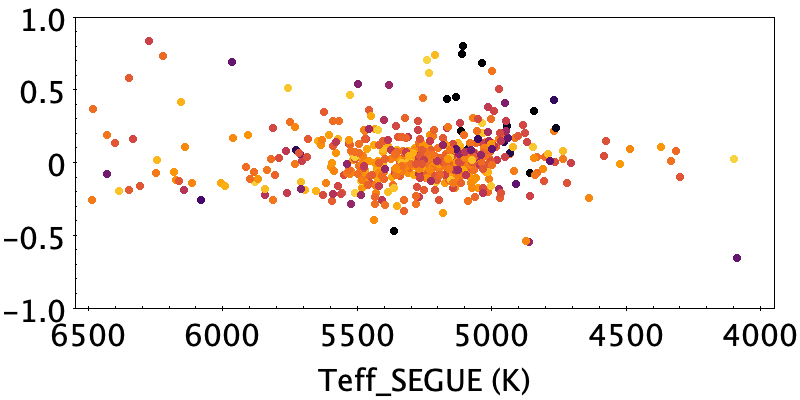}      
\includegraphics[width=0.48\columnwidth]{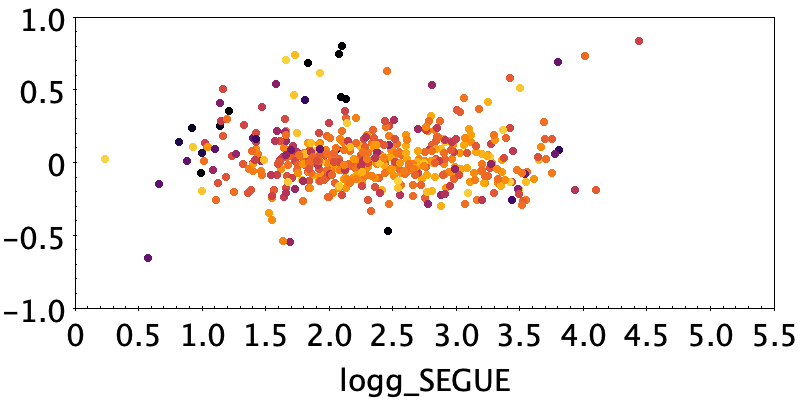}      
\includegraphics[width=0.48\columnwidth]{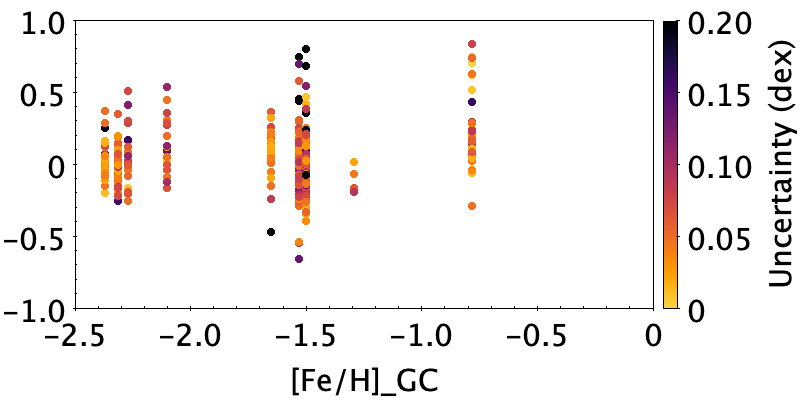}      
\caption{Same as Fig. \ref{f:apo_ref} for \segue. The multiple stars at \logg=4.0 in the \pastel\ plot come from one bibliographical reference, a high-resolution spectroscopic follow-up of extremely metal-poor stars from \segue\ \citep{aok13}, which assumed the surface gravity for all turn-off stars to be that value.}
\label{f:segue_ref}
\end{figure*}

We found 181 stars in common between \pastel\ and \segue, 266 FGK cluster members in 10 OCs and 509 FGK cluster members in 9 GCs. The residuals are shown in Fig.~\ref{f:segue_ref}  as a function of magnitude and of \tgm.

In the comparison to \pastel\ the faintest stars (G$\ge$13.5), which seem to be also the most metal-poor (\feh$<$-2 dex), show a positive offset of +0.18 dex. For the other common stars there is no offset and the dispersion is MAD=0.085 dex. For the six OCs with at least five members, the dispersion is remarkably low despite the resolution of the survey, from 0.02 dex (NGC 2682, 67 stars) to 0.06 dex (NGC 7789, 49 stars). Dwarfs, which seem to correspond to faint stars, tend to have lower metallicities than giants. The typical dispersion for GCs is around 0.08 dex. NGC 6205 is the most populated cluster for \segue\ with 186 members and has MAD=0.08 dex. NGC 5024 shows a poorer performance with MAD=0.16 dex (17 members). There is no trend of the residuals with G magnitude, \teff\ and \logg.

\section{Surveys vs \pastel}

All the comparisons to \pastel\ are summarised in Table \ref{t:stat_surveys_pastel} which gives the median offsets of the \feh\ residuals and their MAD.  When relevant,  trends have been evaluated through a simple linear fit. In general there is a good agreement for the metal-rich regime  (\feh$\ge$-0.5 dex) with negligible offsets, and typical dispersions of 0.04-0.06 dex for the higher resolution surveys (\apo, \galah, \ges) up to 0.10 dex for the other surveys. In this metallicity range, there is however a significant correlation of the residuals with \feh\ for \apo\ which is reproduced in \rave-CNN and \lamost-PAYNE, the surveys using \apo\ as a training set for their parametrisation methods. We note that these data driven methods give more outliers, larger offsets, dispersions and trends than the more classical methods used for \rave\ and \lamost. It is worth to note that \lamost\ and \segue\ show a remarkable precision, better than 0.1 dex, despite their low resolution.

For all the surveys, the metal-poor stars (\feh$<$-0.5 dex) have their metallicity overestimated compared to the high resolution high signal-to-noise (S/N) determinations listed in \pastel\ (also visible with GCs). Offsets range from +0.06 dex for \apo\ to +0.18 dex for \rave-CNN. This result needs to be investigated further owing to the important implication that it has for galactic studies. If this bias is confirmed in massive spectroscopy, it implies in particular that metallicity gradients in the Milky Way cannot be reliably estimated from surveys. Can \pastel\ be the source of this bias? This is not likely owing to the number of different papers (more than 1200) which have been considered in the average of APs, and the fact that the cross-match between \pastel\ and the surveys involves different samples of stars. This bias more likely results from the analysis pipelines of the surveys which are poorly constrained in the metal-poor range due to a lack of reference stars fitting the specific observing requirements. This highlights the need for surveys to observe metal-poor stars for calibration purposes.

Several surveys (\apo, \galah, \segue) seem to provide underestimated metallicities for stars with \feh$>$0 dex.  \rave-CNN and \lamost-PAYNE also show this behaviour, inherited from \apo. This trend is difficult to quantify due to the small extension of the metallicity range on the positive side.


Table \ref{t:stat_surveys_pastel} also provides the typical \feh\ uncertainties (median values) as given in each survey and in \pastel\ for the corresponding selection of stars. This allowed us to verify that their combination through a  quadratic sum is consistent with the dispersion of the residuals. In most cases the MAD of the residuals is lower than the total uncertainty of the catalogues which indicates that their precision is possibly better than expected. The most remarkable case is \ges\ which exhibits a small dispersion of MAD=0.04 dex for \feh$\ge$-0.5 dex while the quoted uncertainties have a median value of 0.10 dex. On the contrary \rave-CNN quotes small errors for the faint stars which are not consistent with the large dispersion of the residuals. A similar disagreement is seen for \lamost-PAYNE in the metal-poor regime. 

\begin{table*}[h]
  \centering 
  \caption{Summary of the \feh\ differences between the  surveys and \pastel.}
  \label{t:stat_surveys_pastel}
  \small
\begin{tabular}{| l | l | r c c | c c | l |}
\hline
 Survey  & Constraint & N             & MED &  MAD & $\sigma_{\rm Survey}$ & $\sigma_{\rm PASTEL} $ & slope \\  
\hline
APOGEE &   & 2155  & -0.01 & 0.05 &  0.01 & 0.05 & \\
APOGEE &   \feh$<$-0.5 & 308  & 0.06 & 0.07 &    0.01 & 0.07 &\\
APOGEE &    \feh$\ge$-0.5 & 1847  & -0.02 & 0.05 &   0.01 & 0.05 & -0.18  \\
\hline
GALAH &   &  232   & 0.00 & 0.07 &    0.055 & 0.05 & \\
GALAH &   \feh$<$-0.5 &  48  & 0.14 & 0.12 &    0.07 & 0.085 &  \\
GALAH &    \feh$\ge$-0.5&  184   & -0.02 & 0.06 &     0.05 & 0.05 &\\
GALAH &   \logg$<$3.8&  91   & 0.00 & 0.125 &    0.06 & 0.07 & \\
GALAH &   \logg$\ge$3.8&  141   & 0.00 & 0.05 &    0.05 & 0.05 &  \\
\hline
GES &   & 162    & 0.02    & 0.05 &    0.10 & 0.05 &\\
GES&  \feh$<$-0.5   & 44    & 0.09    & 0.04 &      0.10 & 0.10 &\\
GES&  \feh$\ge$-0.5   & 118    & 0.00    & 0.04 &     0.10 & 0.05 & \\
\hline
RAVE &   &  427   & 0.01 & 0.10 &   0.135 & 0.05 & \\
RAVE &  \teff$<$5000 K &  177   & 0.09 & 0.13 &    0.14 & 0.07 &\\
\hline
RAVE-CNN &   &  666   & 0.08 & 0.11 &   0.055 & 0.05 & \\
RAVE-CNN &  Gmag$> $9 & 150   & 0.13 & 0.19 &   0.07 & 0.06 & \\
RAVE-CNN &  \feh$<$-0.5 &  336   & 0.18 & 0.14 &    0.07 & 0.07 &\\
RAVE-CNN &  \feh$\ge$-0.5 &  330   & 0.01 & 0.07 &    0.04 & 0.05 &  -0.32  \\
\hline
LAMOST &   &  1767  & 0.00 & 0.055 &    0.03 & 0.05 &\\
LAMOST &   \feh$<$-0.5  &  136  & 0.08 & 0.095 &    0.025 & 0.07 &\\
\hline
LAMOST-PAYNE &   &  2243  & -0.07 & 0.07 &    0.03 & 0.05 & -0.17\\
LAMOST-PAYNE &   \feh$<$-0.5   &  246  & 0.15 & 0.16 &    0.05 & 0.08 &    \\
\hline
SEGUE &   &  181   & 0.07 & 0.16 &   0.04 & 0.10 & \\
SEGUE &  \feh$<$-2    &  111   & 0.17 & 0.16 &    0.05 & 0.10 &\\
SEGUE &   \feh$\ge$-2   &  70   & -0.045 & 0.085 &    0.03 & 0.07 &\\
\hline
\end{tabular}
\tablefoot{ Differences are measured with the median value MED and the dispersion MAD for the N common stars with various constraints. The typical (median) \feh\ uncertainty listed in each catalogue for the corresponding selected stars is also given. When relevant the slope of the linear fit on the distribution of the residuals vs \feh\ is provided. }
\end{table*}

\section{Surveys vs Open Clusters}

\begin{figure}[ht!]
\centering
\includegraphics[width=\columnwidth]{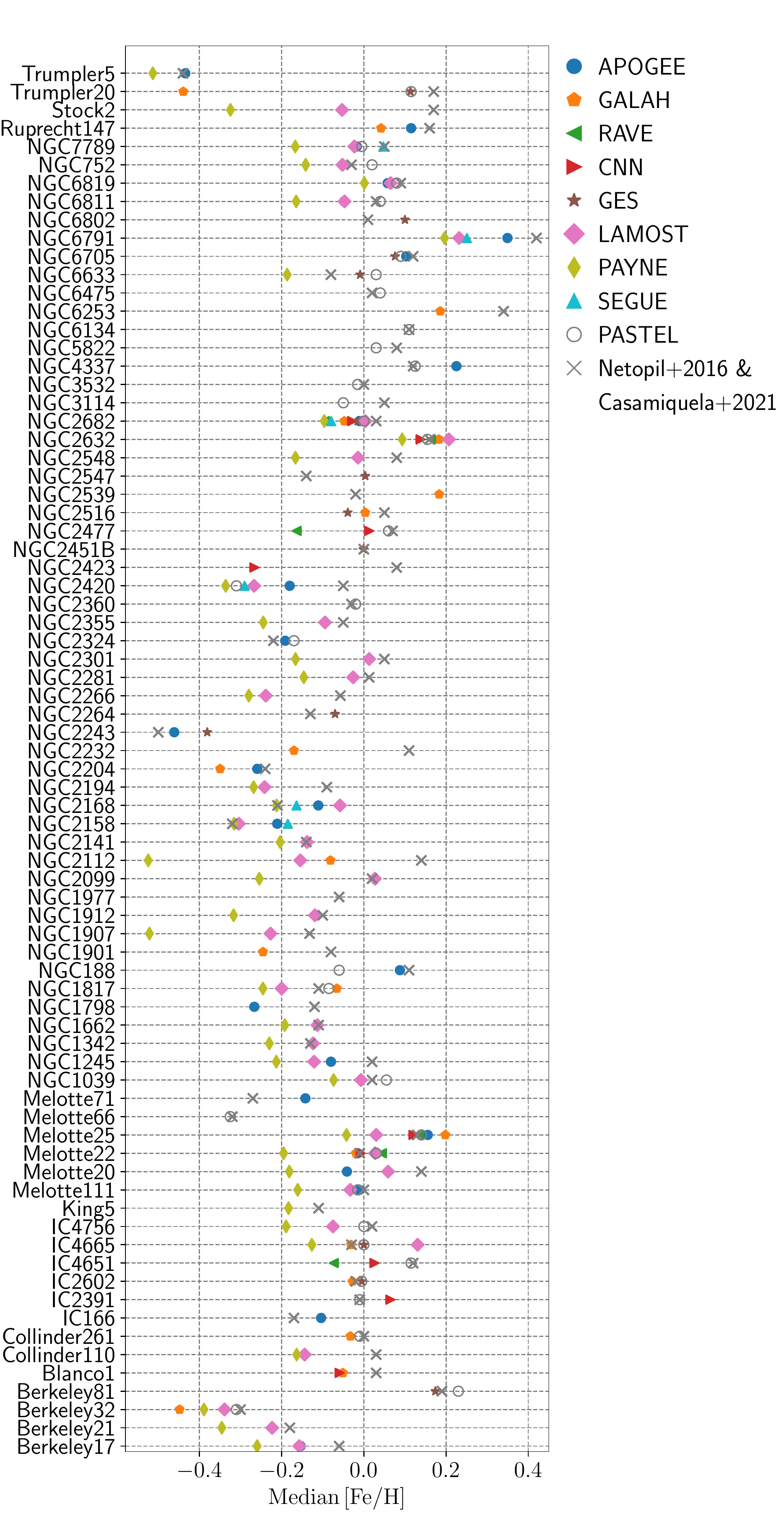}      
 \caption{Median \feh\ obtained by the different surveys and catalogues for OCs with at least 5 FGK members. }
\label{f:med_oc}
\end{figure}

The \feh\ residuals for OC members identified by each survey were shown in  the middle panels of Figures \ref{f:apo_ref} to \ref{f:gesdr3_ref} and commented in the previous sections. The \feh\ reference value for each cluster is adopted from \cite{net16} and \cite{cas21}. The offsets which are seen for some clusters and some surveys can be due to a systematic error in the survey or in the literature, or both. They are therefore difficult to interpret. More relevant is the agreement of the median \feh\ value obtained by different surveys for a given cluster. Fig. \ref{f:med_oc} represents the median \feh\ computed for the 76 OCs which have at least five members in one of the nine surveys, including \pastel. For the majority of clusters observed by several surveys, there is in general a good agreement, although there are a few clusters where the agreement is poor, with variations reaching more than 0.2 dex (e.g. Trumpler 20). The most metal-poor OCs (e.g. Trumpler 5, NGC 2243, Berkeley 32), have their low metallicity confirmed by several surveys. Similarly, the high metallicity of NGC 6791 and Berkeley 81 is found in good agreement by several surveys. It is very clear in Fig. \ref{f:med_oc} that the median \feh\ from \lamost-Payne systematically lies below the others, indicating that this survey has more metal-poor zero-point at the solar metallicity compared to all the other surveys. 

Remarkably NGC 2682 (M67) is part of the nine surveys while the Pleiades (Melotte 22)  and the Hyades (Melotte 25) appear in seven surveys. The results for these well observed clusters are presented in Table \ref{t:melotte}. For M67 the median \feh\ values vary from -0.10 to 0.0 dex with low dispersions from 0.02 to 0.06 dex. Interestingly the survey with the lowest spectroscopic resolution, \segue, exhibits the lowest scatter. We cannot exclude that the good performances in all the surveys are related to the fact that M67 is a common reference, used by most surveys for calibrations and validations. For the Pleiades all the surveys except \lamost-Payne agree well, with MED ranging from -0.02 to 0.04 dex, and MAD ranging from 0.03 to 0.09 dex. For the Hyades, there are larger differences for the median \feh\ due to lower values from \lamost\ and \lamost-Payne. The Hyades is known to be a metal-rich OC \citep[e.g. ][]{cas20} which is confirmed by \pastel, \apo, \galah, \rave\ and \rave-CNN with MED between +0.12 and +0.20 dex and MAD between 0.02 to 0.06 dex.

It is also informative to compare, for each catalogue, the dispersion among members of a given cluster and the typical uncertainties quoted in that catalogue, which we expect to be consistent.  This is verified in most cases with a few notable exceptions. We note that \ges\ and \rave\ provide \feh\ precisions which look pessimistic owing to the low dispersion obtained within OCs. On the contrary, \lamost-Payne gives small \feh\ errors for the Pleiades and Hyades members which do not correspond to the significant dispersion observed for these two clusters.

\begin{table*}[h]
  \centering 
  \caption{Metallicity of M67, the Pleiades and the Hyades in the different surveys. }
  \label{t:melotte}
\begin{tabular}{| l | r c c c | r c c c | r c c c |}
\hline
   & \multicolumn{4}{ c |}{M67} & \multicolumn{4}{ c |}{Pleiades} & \multicolumn{4}{  c |}{Hyades} \\
Survey  & N             & MED &  MAD & $\sigma$ & N             & MED &  MAD &$\sigma$ &   N             & MED &  MAD & $\sigma$   \\  
\hline
PASTEL  & 91 &   0.0 &  0.04 & 0.05 &14  &  0.03  & 0.055  & 0.05 & 55 & 0.14 & 0.03 & 0.04\\
APOGEE  & 178 & -0.01 & 0.03 &0.01& 180 & -0.015 & 0.03  & 0.01 & 7 & 0.155 & 0.04 & 0.01 \\
GALAH   & 130 & -0.05 & 0.05 & 0.06& 42  & -0.02  & 0.07  & 0.06  & 23 & 0.20 & 0.06 & 0.04\\
GES     & 16 &  -0.015 & 0.02 & 0.10 &     &        &       & &    &       &    &  \\
RAVE    & 8 &   -0.09 & 0.04 & 0.16& 15  &  0.04  & 0.07   & 0.18& 16 & 0.13 & 0.06 & 0.13\\
CNN     & 34 &  -0.03 & 0.06 & 0.08& 25  & -0.01  & 0.05   & 0.04 & 20 & 0.12 & 0.02 & 0.04\\
LAMOST  & 93 &    0.0 & 0.04 &0.03 & 148 &  0.03  & 0.09   & 0.03 & 26 & 0.03 & 0.07 & 0.03\\
PAYNE   & 99 &  -0.10 & 0.05 & 0.03& 184 & -0.195 & 0.14   &0.03 & 38 & -0.04 & 0.18 & 0.02\\
SEGUE   & 67 &  -0.08 & 0.02 &   0.02  &        &        &   & &       &      & &\\
\hline
\end{tabular}
\tablefoot{ The number of members (N), median \feh\ (MED) and median absolute deviation are provided. The $\sigma$ column gives the typical (median) uncertainty of the N stars as provided in the corresponding catalogue.}
\end{table*}

\begin{figure*}[ht!]
\centering
\includegraphics[width=0.6\columnwidth]{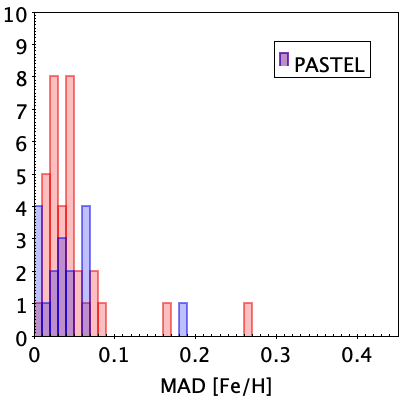}      
\includegraphics[width=0.6\columnwidth]{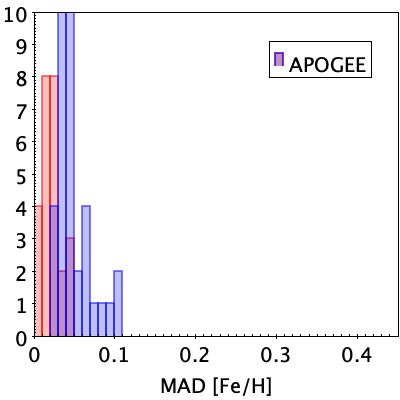}      
\includegraphics[width=0.6\columnwidth]{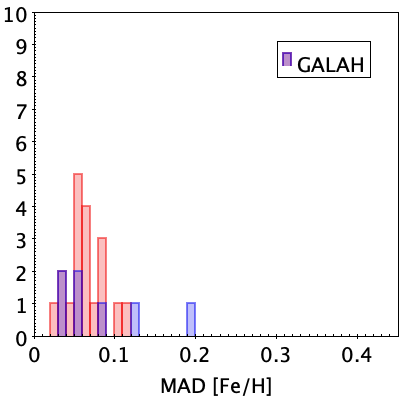}      
\includegraphics[width=0.6\columnwidth]{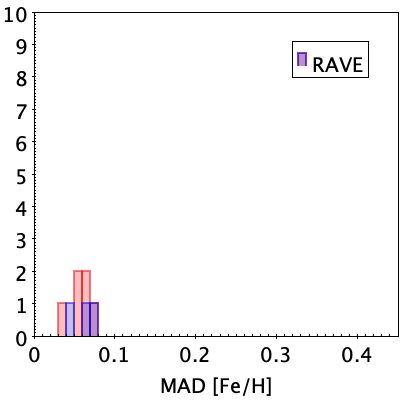}      
\includegraphics[width=0.6\columnwidth]{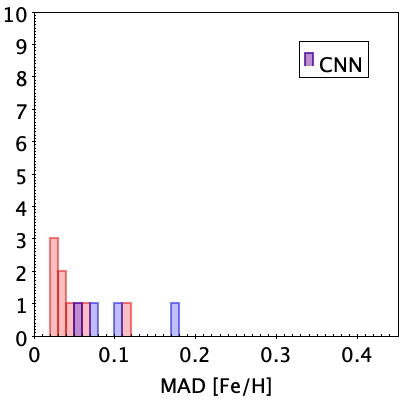}      
\includegraphics[width=0.6\columnwidth]{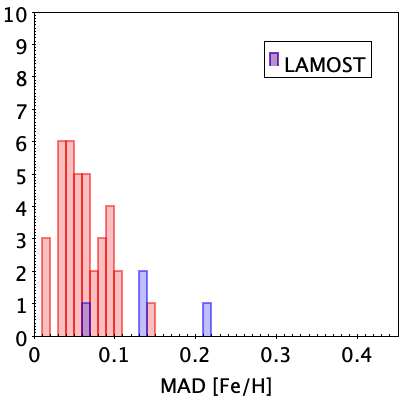}      
\includegraphics[width=0.6\columnwidth]{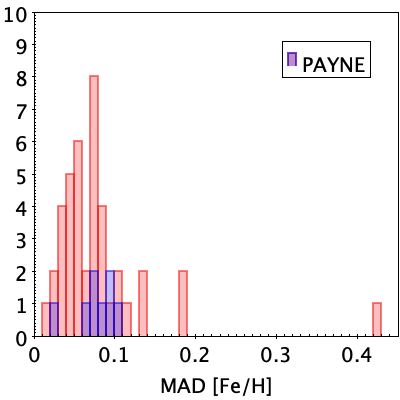}      
\includegraphics[width=0.6\columnwidth]{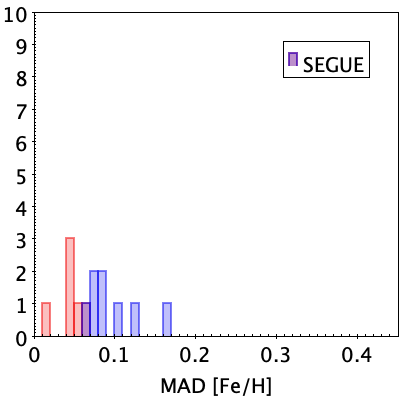}      
\includegraphics[width=0.6\columnwidth]{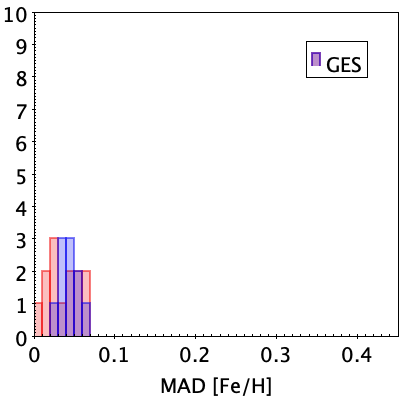}      
\caption{Histogram of the dispersion (MAD) of \feh\ obtained by the different surveys for OCs (in red) and GCs (in blue) with at least 5 FGK members. }
\label{f:mad_clusters}
\end{figure*}

Figure \ref{f:mad_clusters} shows the histogram of the MAD \feh\ for each survey.  \apo\ is clearly the survey which has the highest consistency among FGK members of OCs. This cannot directly be attributed to the ASCAP pipeline which does not use OCs for calibrations contrary to previous APOGEE releases, according to \cite{jon20}. \apo\ has observed 25 OCs in common with our reference sample and with at least five members and they have all a dispersion (MAD) lower than 0.05. The best performance is reached for Trumpler 5 (10 members, MED=-0.43 dex, MAD=0.007 dex), then NGC 2324  (6 members, MED=-0.19 dex, MAD=0.008 dex)  and NGC 1798 (9 members, MED=-0.27 dex, MAD=0.008 dex). It is worth to note that APOGEE has observed 128 OCs in total, reported in DR16, that \cite{don20} used to measure the radial metallicity gradient in the galactic disc.

\section{Surveys vs Globular Clusters}

\begin{figure}[ht!]
\centering
\includegraphics[width=0.9\columnwidth]{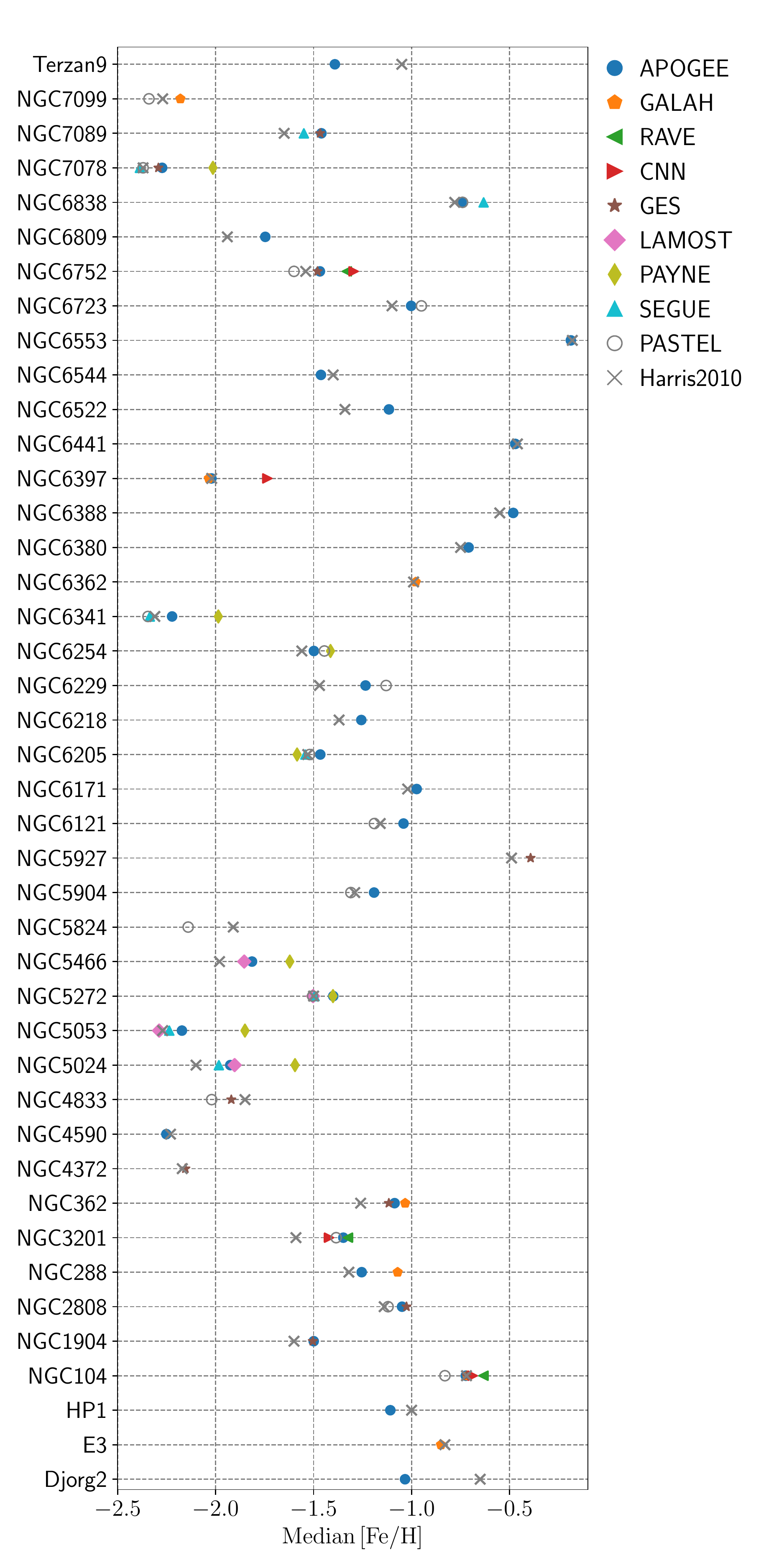}      
\caption{Median \feh\ obtained by the different surveys and catalogues for GCs with at least 5 FGK members. }
\label{f:med_gc}
\end{figure}

\begin{figure}[ht!]
\centering
\includegraphics[width=\columnwidth]{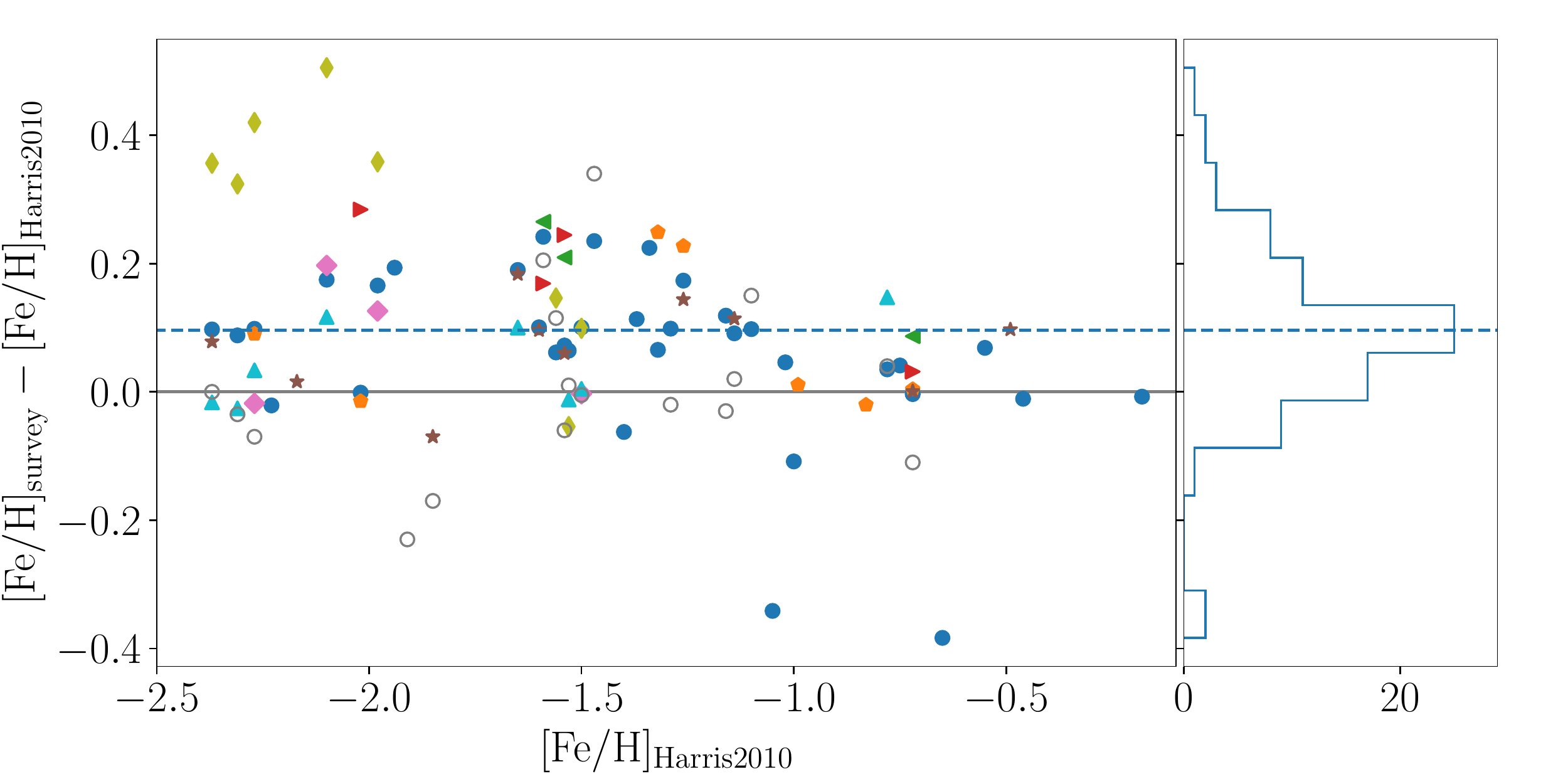}      
\caption{\feh\ residuals for 42 GCs with at least 5 FGK members. The dot colours are the same as in Fig. \ref{f:med_gc}. The histogram, represented without the \pastel\ values, shows a median offset of 0.096 dex for the surveys with respect to \cite{har10}.}
\label{f:correlation_gc}
\end{figure}

The bottom panels of Figures \ref{f:apo_ref} to \ref{f:gesdr3_ref} show the \feh\ residuals for GC members identified by each survey. The \feh\ reference values are adopted from \cite{har10} and do not represent the most up-to-date metallicities for those clusters. Nevertheless, for the 42 GCs with at least five members, the median \feh\ from the surveys show an excellent agreement  with the Harris metallicities, as shown in Fig. \ref{f:med_gc}, although slightly more metal-rich in general. When plotted all together, the \feh\ residuals are centered on  $\sim$0.1 dex (Fig. \ref{f:correlation_gc}). The tendency for the surveys to overestimate \feh\ of the metal-poor stars has already been noticed in the comparison to \pastel\ and is confirmed here in a different way with GCs.  Like for OCs, \lamost-Payne stands apart from the other surveys. In the previous section,  \lamost-Payne was systematically underestimating the median metallicity of clusters, here in the metal-poor regime it overestimates it.

The mildly metal-poor cluster NGC 104 has been observed by six surveys while NGC 5272 and NGC 7078 appear in five of them. The corresponding median \feh\ values are detailed in Table \ref{t:ngc}. For NGC 104 they vary from -0.63 dex with \rave\ to -0.83 dex with \pastel, a difference still compatible at the level of uncertainties of the two catalogues, 0.15 and 0.06 dex respectively. For this cluster there is a high level of consistency within each survey with dispersions a ranging from 0.03 to 0.06 dex. The agreement between surveys is also very good for NGC 5272 with median \feh\ ranging form -1.50 to -1.40 dex, although the dispersions are larger, in particular for \lamost\ (MAD=0.21 dex). For NGC 7078 the median \feh\ from \lamost-PAYNE lies well above the others despite the smallest internal dispersion (MAD=0.02 dex for 7 stars). Like for OCs, we note that the dispersion of the residuals is lower than the quoted uncertainties for \rave\ and \ges, while they are in a better agreement for the other surveys.

\begin{table*}[h]
  \centering 
  \caption{Like Table \ref{t:melotte} for the GCs NGC 104, NGC 5272, NGC 7078.}
  \label{t:ngc}
\begin{tabular}{| l | r c c c | r c c c| r c c c|}
\hline
   & \multicolumn{4}{ c |}{NGC 104} & \multicolumn{4}{ c |}{NGC 5272} & \multicolumn{4}{  c |}{NGC 7078} \\
Survey  & N             & MED &  MAD & $\sigma$ & N             & MED &  MAD& $\sigma$ &  N             & MED &  MAD& $\sigma$  \\  
\hline
PASTEL & 18 & -0.83 & 0.03   & 0.06 & 8 & -1.505 & 0.04     & 0.07 & 38 & -2.37 & 0.01 & 0.10\\
APOGEE & 186 & -0.72 & 0.03  & 0.01& 152 & -1.40 & 0.07    & 0.02 & 34 & -2.27 & 0.05 & 0.025\\ 
GALAH  & 253 & -0.72 & 0.05  & 0.07 &    &     &      &  &    &     &         &  \\    
GES  & 111 & -0.72 & 0.03  & 0.09&    &     &          & & 39 & -2.29 & 0.05 & 0.13 \\         
RAVE   & 14 & -0.63 & 0.045  & 0.15&    &     &       & &    &     &       &    \\   
CNN    & 26 & -0.69 & 0.05   & 0.05&    &     &       &  &    &     &        &  \\    
LAMOST &    &     &          && 23 & -1.50 & 0.21     & 0.13 &    &     &     &  \\
PAYNE  &    &     &           && 34 & -1.40 & 0.09  & 0.145 & 7 & -2.01 & 0.02 & 0.05\\
SEGUE  &    &     &           && 126 & -1.495 & 0.10 & 0.05 & 39 & -2.39 & 0.06  & 0.04 \\
\hline
\end{tabular}
\end{table*}

Figure \ref{f:mad_clusters} shows the histogram of the MAD \feh\ for each survey.  \apo\ has observed the largest number of clusters: 35 GCs with at least five members.  They have dispersions (MAD) ranging from 0.02 to 0.10 dex (median  0.04 dex). The lowest scatter (MAD=0.02 dex) is reached for  NGC 6441 (6 members, MED=-0.47 dex), NGC  6553 (7 members, MED=-0.19 dex)  and NGC 6723 (7 members, MED=-1.00 dex).


\section{Surveys vs \apo}

\begin{figure*}[ht!]
\centering
\includegraphics[width=0.48\columnwidth]{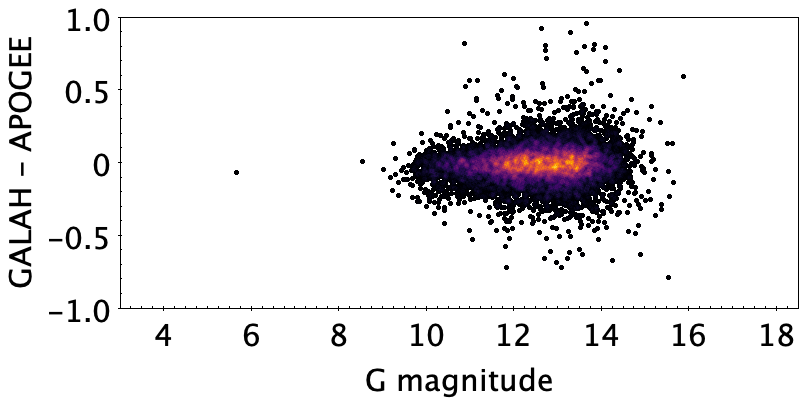}      
\includegraphics[width=0.48\columnwidth]{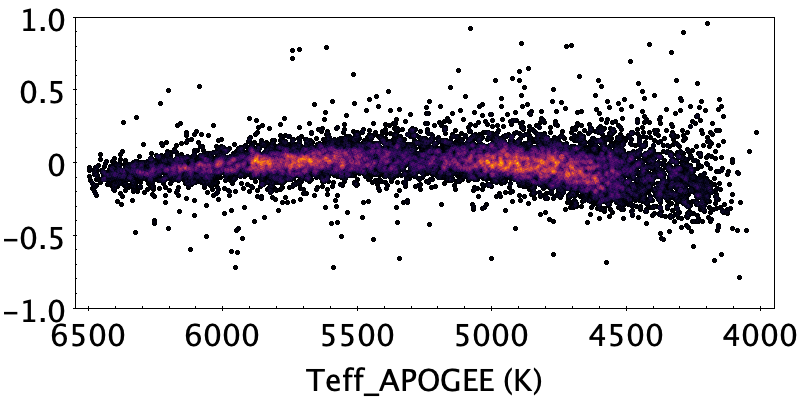}      
\includegraphics[width=0.48\columnwidth]{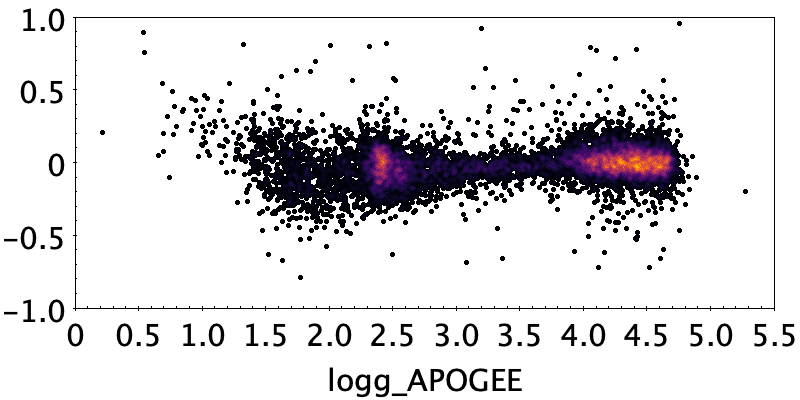}      
\includegraphics[width=0.48\columnwidth]{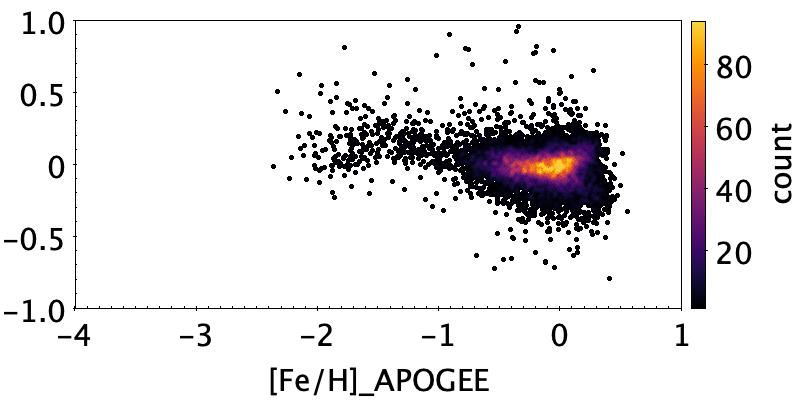}      

\includegraphics[width=0.48\columnwidth]{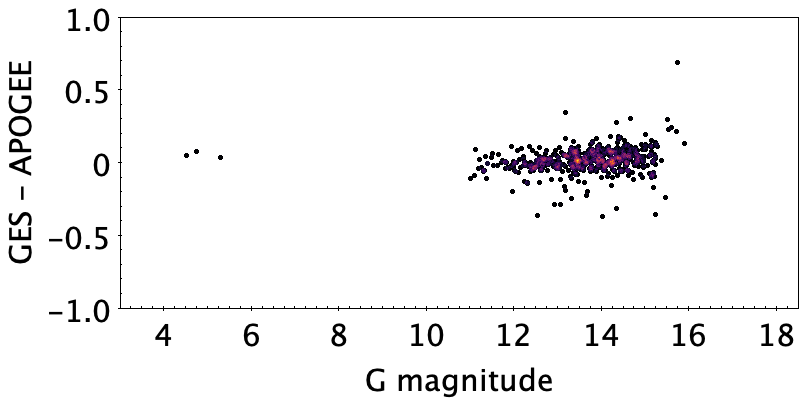}      
\includegraphics[width=0.48\columnwidth]{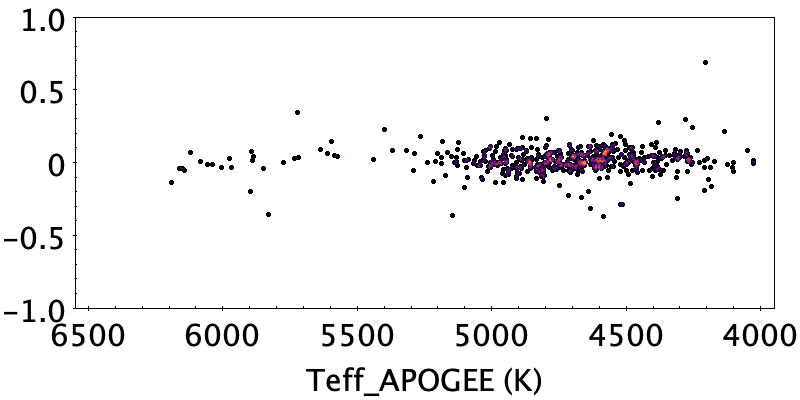}      
\includegraphics[width=0.48\columnwidth]{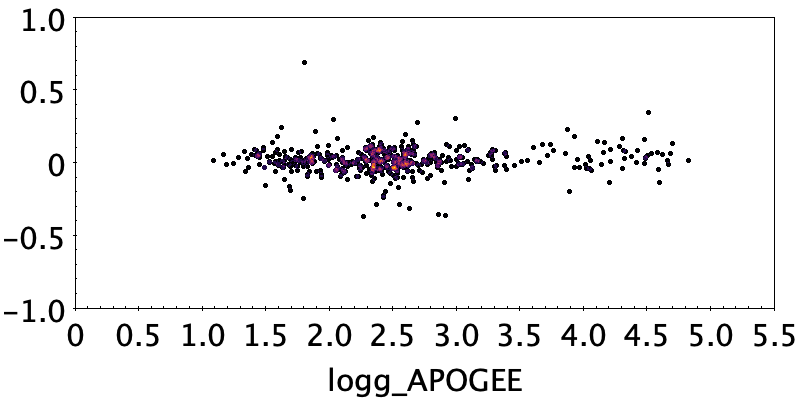}      
\includegraphics[width=0.48\columnwidth]{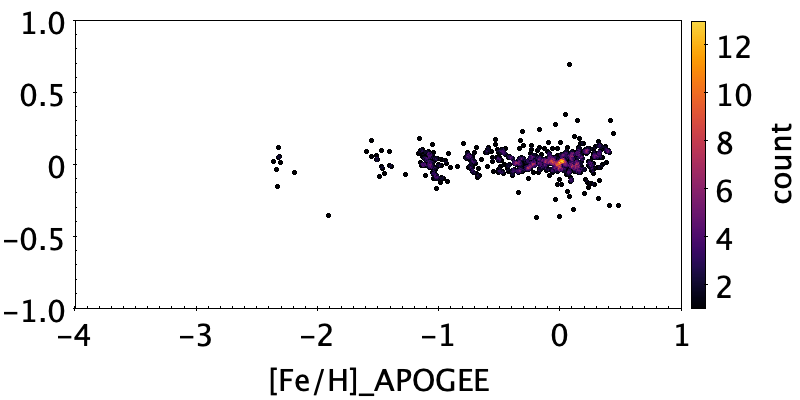}      

\includegraphics[width=0.48\columnwidth]{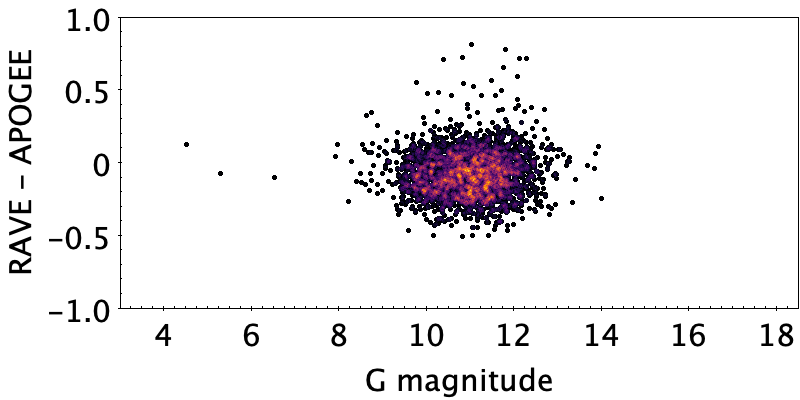}      
\includegraphics[width=0.48\columnwidth]{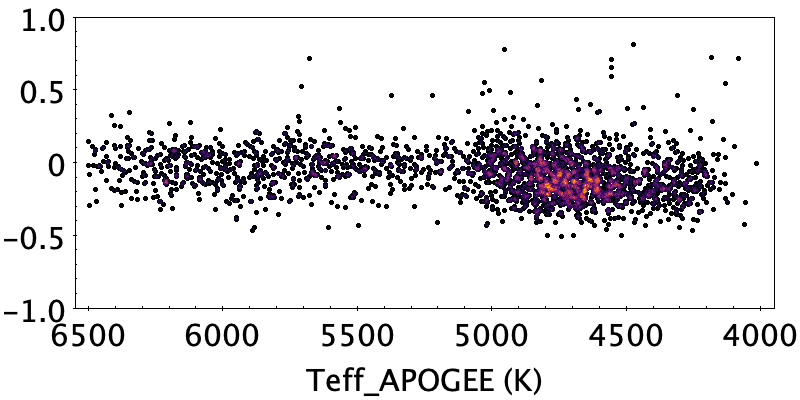}      
\includegraphics[width=0.48\columnwidth]{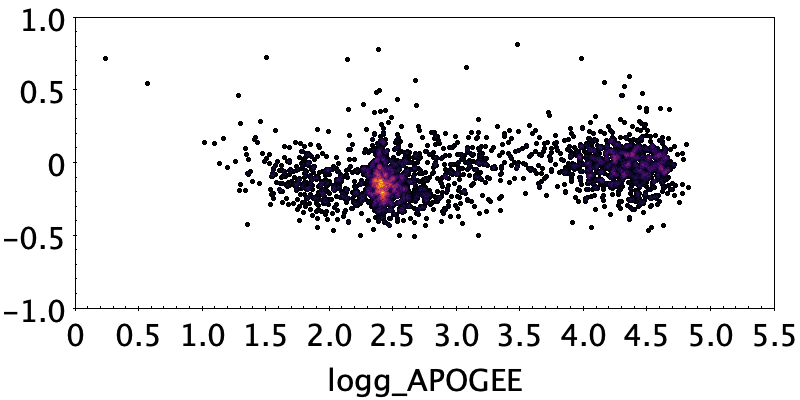}      
\includegraphics[width=0.48\columnwidth]{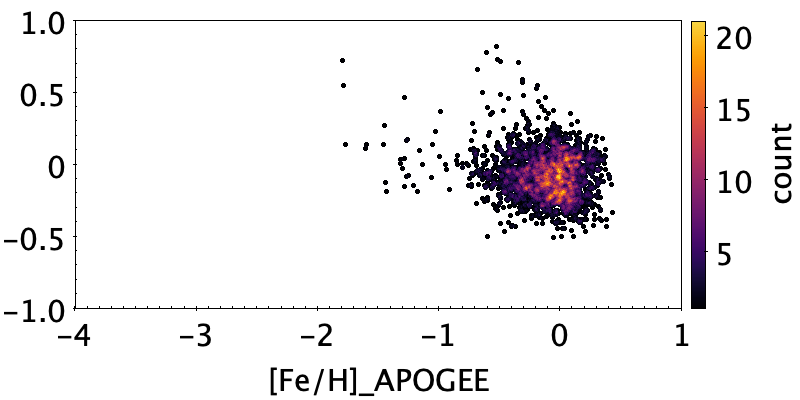}      

\includegraphics[width=0.48\columnwidth]{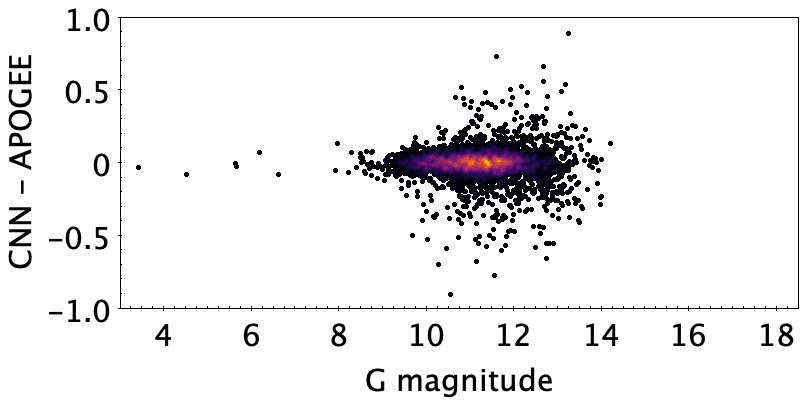}      
\includegraphics[width=0.48\columnwidth]{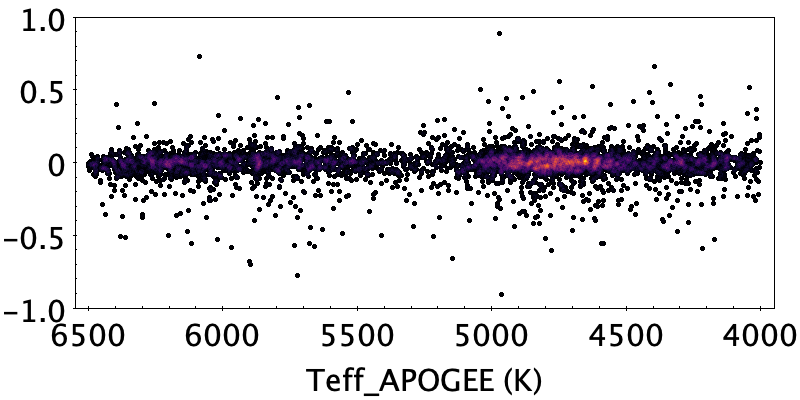}      
\includegraphics[width=0.48\columnwidth]{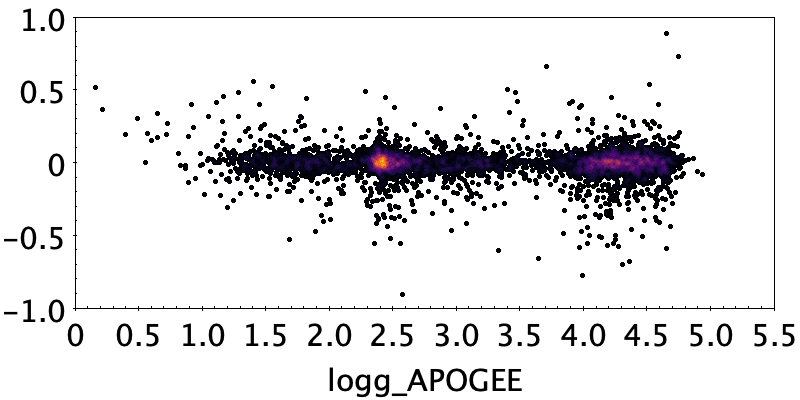}      
\includegraphics[width=0.48\columnwidth]{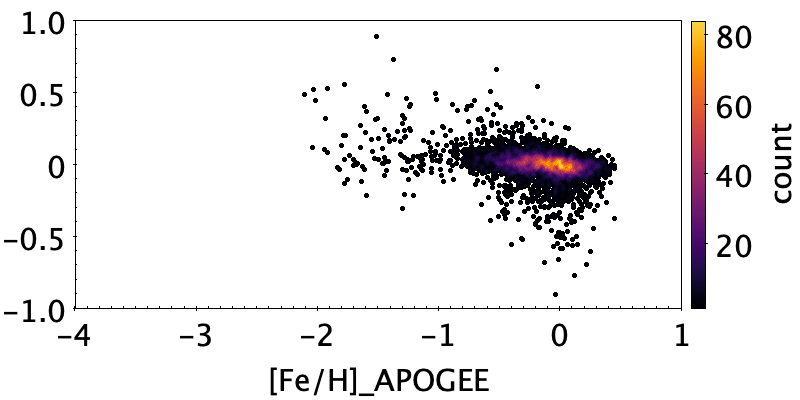}      

\includegraphics[width=0.48\columnwidth]{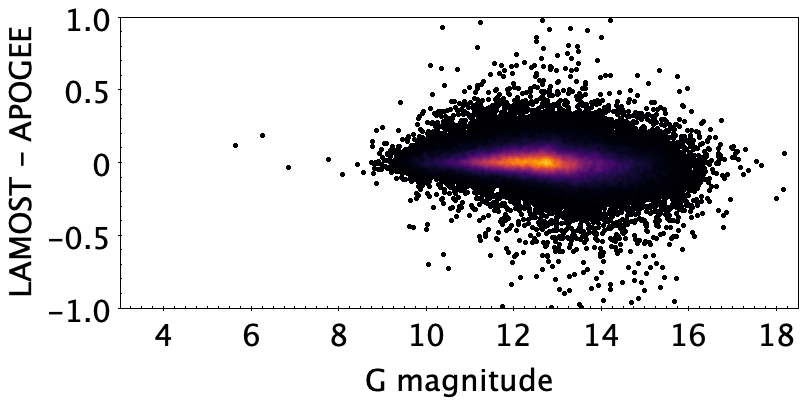}      
\includegraphics[width=0.48\columnwidth]{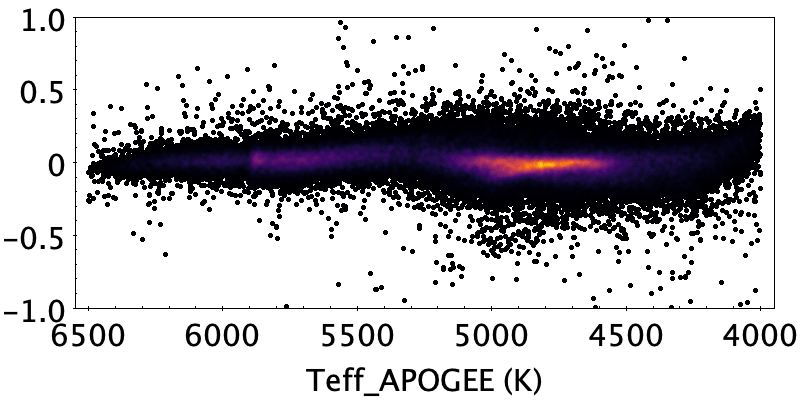}      
\includegraphics[width=0.48\columnwidth]{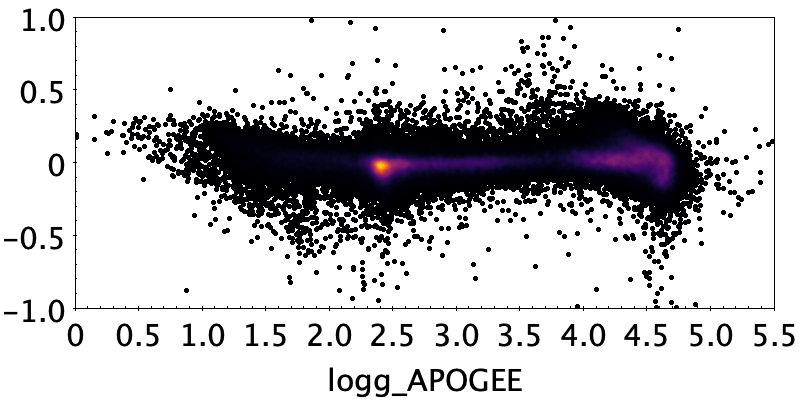}      
\includegraphics[width=0.48\columnwidth]{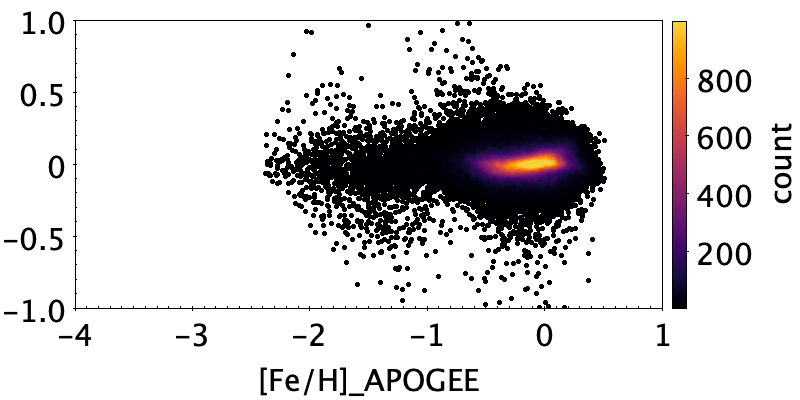}      

\includegraphics[width=0.48\columnwidth]{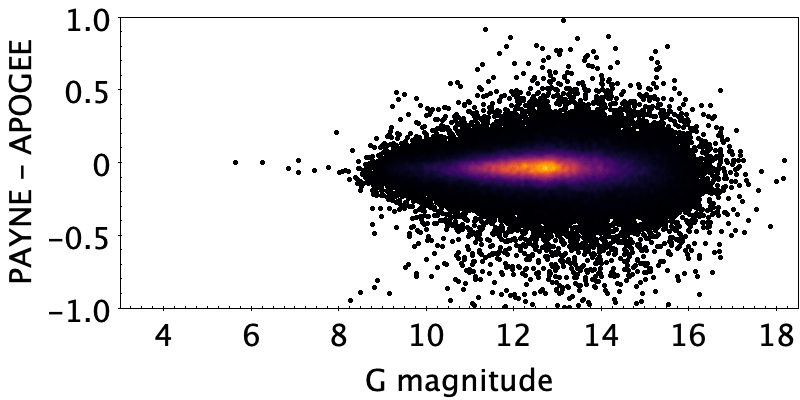}      
\includegraphics[width=0.48\columnwidth]{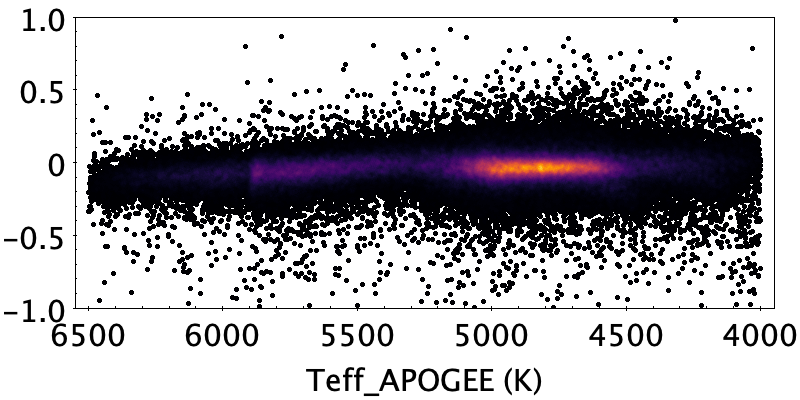}      
\includegraphics[width=0.48\columnwidth]{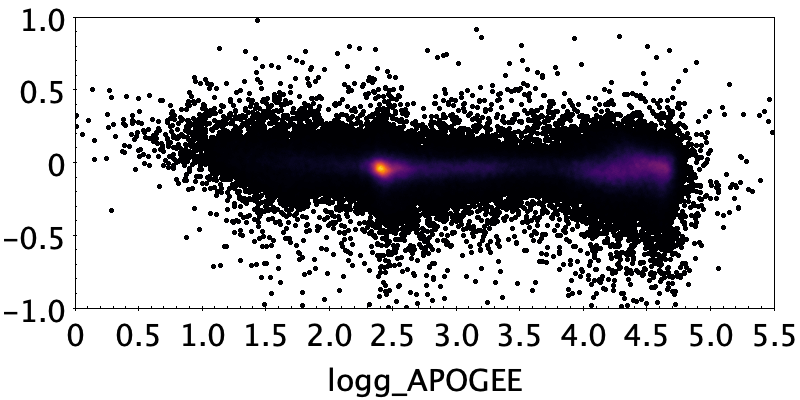}      
\includegraphics[width=0.48\columnwidth]{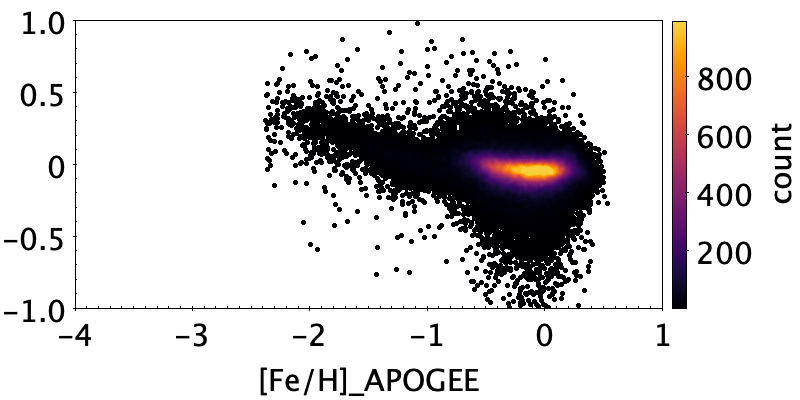}      

\includegraphics[width=0.48\columnwidth]{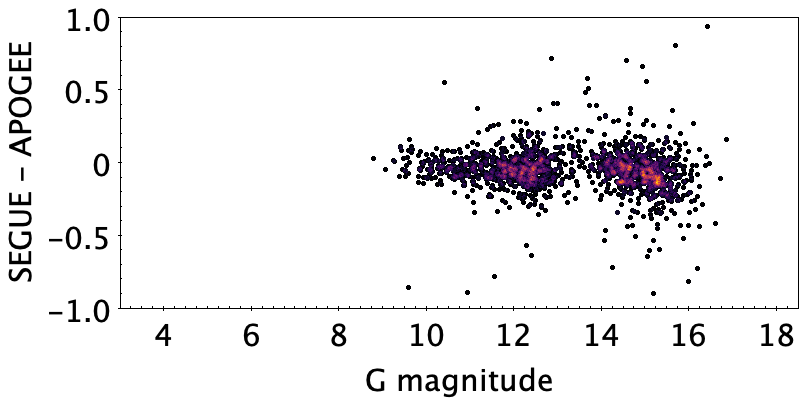}      
\includegraphics[width=0.48\columnwidth]{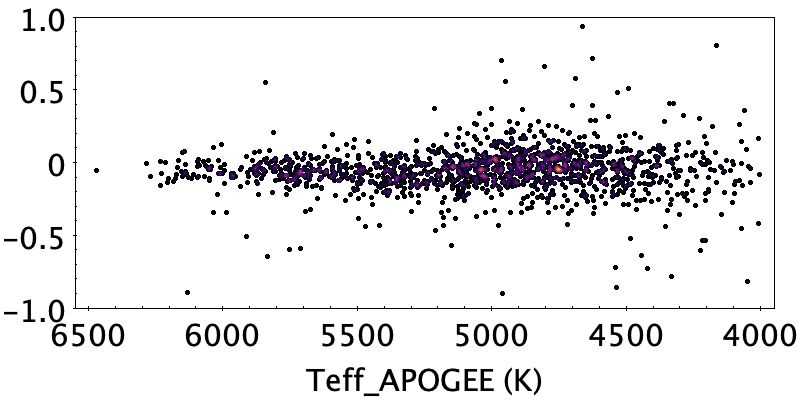}      
\includegraphics[width=0.48\columnwidth]{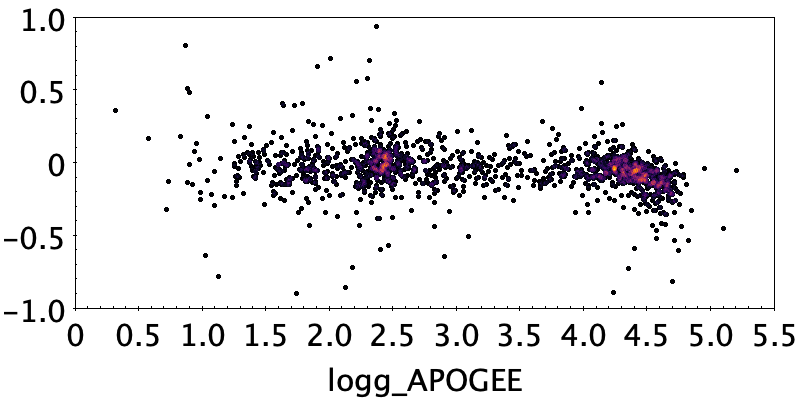}      
\includegraphics[width=0.48\columnwidth]{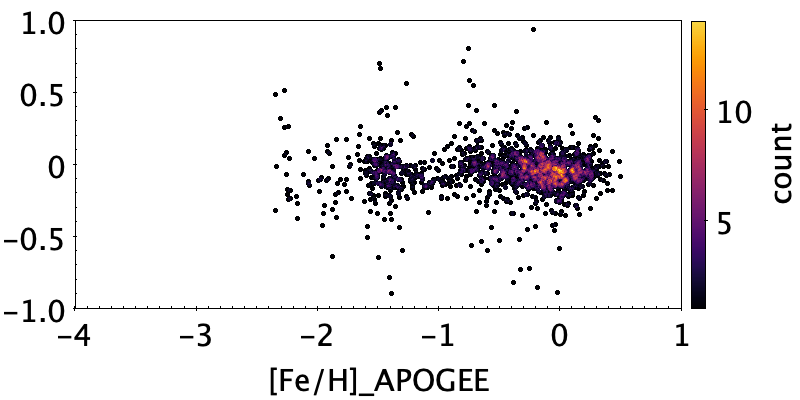}

\caption{\feh\ difference between the different surveys and \apo\ (surveys minus \apo) versus magnitude, \teff, \logg\ and \feh\ from \apo.  From the top to the bottom: \galah, \ges, \rave, \rave-CNN, \lamost, \lamost-PAYNE and \segue. The colour is scaled on the counts.}
\label{f:surveys_apo}
\end{figure*}

In the previous sections, we have shown that \apo\ performs very well, providing in particular a low \feh\  dispersion among members of OCs and GCs. It is thus relevant to use it as a reference catalogue to test the other surveys against it, in order to strengthen the statistics with more common stars. Here we therefore compare the \feh\ determinations from various surveys to those from \apo\ without considering \pastel. The residuals are shown in Fig. \ref{f:surveys_apo} and the MED and MAD are presented in Table \ref{t:stat_surveys_apo}, together with the uncertainties of the catalogues for the common stars. The intersection between the surveys and \apo\ varies from a few hundreds to nearly 96\,000 for \lamost-Payne. In general there is a good agreement of the \feh\ determinations with flat distributions of the residuals, well centered on zero. The lowest scatter (0.04 dex) is reached by \lamost\ and the \ges\ which implies that these surveys have precisions at this level. This confirms the findings of the previous sections. For \lamost\ this precision is a remarkable owing to its low resolution. For the \ges\ it implies that the quoted uncertainties are too pessimistic. The largest offset and dispersion are for \rave\ (MAD=-0.08 dex, MAD=0.10 dex). No trend is visible with the G magnitude. There are a slight trend  in \teff\ for \galah. The largest effects depend on \logg\ but they occur essentially the edges of the \logg\ range where the density of stars is lower.  Compared to \apo, \rave\ clearly underestimates the metallicity of giants. We note that there is no star in common between \apo\ and the other surveys more metal-poor than \feh=-2.5 dex. A striking feature is the positive offset for \galah\ for metal-poor stars, while \lamost, \segue\ and \ges\ have similar zero-points along the metallicity axis. 

 The \apo\ comparisons to \rave-CNN and \lamost-Payne are worth to be commented since  these two surveys use \apo\ for the training of their pipeline. The bulk of metallicities from \rave-CNN is in good agreement with  those from \apo\ with no offset and a low dispersion of 0.05 (this does not apply however on the small fraction of metal-poor stars which exhibit a positive offset). The systematics and precision look improved by comparison to \rave. This good performance is well explained by the fact that nearly all the stars in common between \rave-CNN and \apo\ were used for the training of the CNN pipeline.  What we see here is essentially the trained \feh\ versus the input \feh\ from \rave-CNN, already analysed by \cite{gui20}. For \lamost-PAYNE the distribution of the \feh\ residuals is very similar to that obtained with \lamost, with however a very pronounced linear trend  for the metal-poor stars. This suggests that the training set for \lamost-PAYNE (15\,000 stars from \apo\ DR14) was possibly too sparse in the metal-poor regime, resulting in these systematics.

\begin{table}[h]
  \centering 
  \caption{Summary of the \feh\ residuals of the surveys vs \apo}
  \label{t:stat_surveys_apo}
\begin{tabular}{| l | r c c | c c |}
\hline
 Survey  & N             & MED &  MAD & $\sigma_{\rm Survey}$ & $\sigma_{\rm APOGEE}$\\  
\hline
GALAH &   9\,331  & -0.01 & 0.06  & 0.08 & 0.01\\
GES &  491   & 0.01 & 0.04 & 0.08 & 0.01 \\
RAVE & 2\,269  & -0.08 & 0.10 & 0.15 & 0.008\\
CNN & 4\,498  & 0.0 & 0.04 & 0.04 & 0.009\\
LAMOST &  85\,726 & 0.0 & 0.04 & 0.04 & 0.01\\
PAYNE & 95\,924  & -0.04 & 0.05 & 0.04 & 0.01\\
SEGUE &   1\,427  & -0.05 & 0.07 & 0.03 & 0.01\\
 \hline
\end{tabular}
\end{table}

 The next step of this work will be to evaluate the agreement of abundance ratios from the different surveys. \apo\ DR16 provides abundances of 26 species
(C, C I, N, O, Na, Mg, Al, Si, P, S, K, Ca, Ti, Ti II, V, Cr, Mn, Fe, Co, Ni, Cu, Ge, Rb, Ce, Nd, Yb), while abundances of 30 species are given in \galah\ DR3 (Li, C,
O, Na, Mg, Al, Si, K, Ca, Sc, Ti, V, Cr, Mn, Co, Ni, Cu, Zn, Rb, Sr, Y, Zr, Mo, Ru, Ba, La, Ce, Nd, Sm, Eu) and 24 species in the current \ges\ public version (C, Li, N, O, Na, Mg, Al, S, Ca, Sc, Ti, Ti 2, V, Cr, Co, Ni, Zn, Y, Zr, Ba, La, Ce, Nd, Eu), offering interesting perspectives for comparisons. This is particularly relevant with the advent of Gaia DR3 that will publish chemical abundances (N, Mg, Si, S, Ca, Ti, Cr, Fe, FeII, Ni, Zr, Ce, Nd) for several millions of FGK stars \citep{rec16}.

 Another work for the future would be to combine the different surveys into a single homogenised catalogue. The agreement between metallicities from the different surveys is reasonable  enough in the metal-rich regime (\feh$\ge$-0.5) to attempt such a combination which would increase the sample size and sky coverage for galactic archeology studies. This implies however to develop a proper procedure that takes into account the differences between surveys to calibrate the metallicities onto a common scale. \cite{nan20} used the data driven approach {\it The Cannon} \citep{nes15,cas16} to combine metallicities and alpha-abundances from \apo\ DR16 and \galah\ DR3 in order to explore the radial and vertical gradients  and abundance trends in the Galactic disc. The stellar parameters of one survey were put on the scale of the other one and vice-versa, resulting in two catalogues that show some differences. These discrepancies reflect the difficulty to deal with an imperfect training set and with complex selection functions. Another on-going project is the Survey of Surveys that has already managed to homogenise radial velocities from different surveys \citep{tsa21} and plans to make a similar analysis with atmospheric parameters.

\section{Conclusion}
 We have assessed the \feh\ determinations of FGK stars in eight spectroscopic surveys by comparing them to independent sources built from the \pastel\ catalogue and clusters. We have tested the latest public versions of \apo, \galah, \rave, \lamost, \segue\ and the \ges, and the data-driven CNN and Payne versions of \rave\ and \lamost. 
 
  \pastel\ being a bibliographical catalogue, we have first adopted a weighted mean of the APs for each star. Then we have selected FGK stars with 4000$\le$\teff$\le$6500 K, which have a typical \feh\ uncertainty of  0.06 dex. We obtain a scatter (MAD) of 0.04 dex when OC members are considered, and 0.08 dex for GC members, the reference metallicity for clusters being adopted from the literature.  \pastel\ includes a number of metal-poor stars which allowed us  to probe the metal-poor regime in surveys. 
   
 To test the agreement between two sources of metallicity, we have used the median value of the residuals  to measure an eventual offset between them and the median absolute deviation of the residuals  which reflects the precision of both sources. We have looked for trends in the distribution of residuals vs G magnitude, and \tgm.  We have also checked whether the scatter of the residuals was consistent with the combined uncertainties quoted in the considered sources. 
 
 Our main conclusions are the following:
 \begin{itemize}
 \item in general all the surveys perform well in the metal-rich regime (\feh$\ge$-0.5 dex) with negligible offsets and dispersions lower than 0.10 dex whatever the resolution. This is verified with both \pastel\ and the OCs
 \item all the surveys overestimate \feh\ in the metal-poor regime (\feh$<$-0.5 dex) with offsets ranging from +0.06 to +0.18 dex on average. This is verified with both \pastel\ and the GCs
 \item the metallicities based on data-driven methodologies show offsets, dispersions and trends significantly larger than those obtained with classical methods on the same spectra. The biases of the training set are amplified. In addition the quoted uncertainties look too optimistic. 
 \item in most cases uncertainties of the surveys are consistent with the scatter observed in clusters. A notable exception is the \ges\ which has too pessimistic uncertainties.
 \item \apo\ has a typical precision better than 0.05 dex over the full metallicity range but systematically overestimates low metallicities and underestimates high metallicities with a linear trend
 \item \lamost\ performs as well as surveys of higher resolution
\end{itemize}
 
 Our investigation has highlighted biases at the extrema of the metallicity range where atmospheric models and automated pipelines are less constrained. The differences between surveys seriously hamper any attempt to simply combine them, in order to improve the statistics for instance. The combination of surveys requires an elaborated procedure of homogenisation of the iron abundances. We would like to encourage the builders of spectroscopic surveys to include common reference stars in their observing plans for calibration and validation purposes.  This would enlarge the intersection between surveys and facilitate their homogenisation into a common scale. Such stars should have APs  measured from high resolution, high S/N spectroscopy or belong to well studied clusters. The \pastel\ catalogue is a useful resource to search well studied stars over the whole metallicity range, in particular at its extrema. Having calibration targets in common between surveys is very useful to track systematic differences. For this purpose, the OC M67 and the GC NGC 104 are excellent targets. The Hyades, Berkeley 81 and NGC 6791 are interesting to observe to constrain the high metallicity regime, while the most metal-poor OCs  (Berkeley 32, NGC 2243, Trumpler 5) can constrain metallicities around \feh=-0.4 dex. Many GCs are available to anchor the lowest metallicities. The recent revision of memberships in clusters based on Gaia data provides reliable lists of targets. We want to mention the sample of Gaia FGK Benchmark Stars \citep{hei15, jof15, jof18} which has been built for the calibration of Gaia's APs to be delivered in DR3 and to serve as common reference between Gaia and the ground-based surveys. A new extended version is in preparation.

\begin{acknowledgements}
 We thank the anonymous referee for the useful comments
and suggestions that helped clarify the paper. CS and NB warmly thank Philippe Prugniel who made a significant contribution in the 2020 version of \pastel. The preparation of this work has made extensive use of Topcat \citep{2011ascl.soft01010T}, of the Simbad and VizieR databases at CDS, Strasbourg, France, and of NASA's Astrophysics Data System Bibliographic Services. 
\end{acknowledgements}

\bibliographystyle{aa}
\bibliography{feh_v2}

\end{document}